\documentclass[a4paper]{aa}
\usepackage[dvips]{graphicx}
\usepackage{natbib,amssymb,astro,color}
\usepackage[version=3]{mhchem}
\usepackage[normalem]{ulem}
\bibpunct{(}{)}{;}{a}{}{,}
\begin{document}
\def \tpeak{\ensuremath{T_{\rm mb}}}
\def\dveq{\ensuremath{\Delta v}}
\def\wtot{\ensuremath{W_{\rm tot}}}
\def\jlow{\ensuremath{J_{\rm l}}}
\def\jup{\ensuremath{J_{\rm u}}}
\def\colfig{}
\def\elec{\ensuremath{{\rm e}^-}}
\def\corr#1#2{#2}
\def\cref#1{(\ref{equ#1})}
\title{Nitrogen chemistry and depletion in starless cores\thanks{Based
    partly on observations carried out with the IRAM~30~m
    telescope. IRAM is supported by INSU-CNRS/MPG/IGN.}}

\author{%
P.~Hily--Blant\inst{1}
\and
M.~Walmsley\inst{2}
\and
G.~Pineau~des~For\^ets\inst{3,4}
\and
D.~Flower\inst{5}
}
\date{}
\institute{LAOG (UMR 5571) Observatoire de Grenoble, BP 53 F-38041 GRENOBLE Cedex 9
\and INAF, Osservatorio Astrofisico di Arcetri, Largo Enrico Fermi 5, I-50125 Firenze, Italy
\and IAS (UMR 8617), Universit\'{e} de Paris--Sud, F-91405 Orsay, France
\and LERMA (UMR 8112), Observatoire de Paris, 61 Avenue de l'Observatoire, F-75014, Paris, France
\and Physics Department, The University, Durham DH1 3LE, UK}

\authorrunning{Hily-Blant \etal}

\offprints{P. Hily-Blant, \email{pierre.hilyblant@obs.ujf-grenoble.fr}}

\titlerunning{Nitrogen chemistry in starless cores}

\abstract%
    {}
    {We investigated the chemistry of nitrogen--containing species,
      principally isotopomers of CN, HCN, and HNC, in a sample of
      pre-protostellar cores.}
    {We used the IRAM 30~m telescope to measure the emission in
      rotational and hyperfine transitions of CN, HCN, \thcn, H\thcn,
      HN\thc, and HC\fifn\ in L~1544, L~183, Oph~D, L~1517B,
      L~310. The observations were made along axial cuts through the
      dust emission peak, at a number of regularly--spaced offset
      positions. The observations were reduced and analyzed to obtain
      the column densities, using the measurements of the less
      abundant isotopic variants in order to minimize the consequences
      of finite optical depths in the lines. The observations were
      compared with the predictions of a free--fall gravitational
      collapse model, which incorporates a non-equilibrium treatment
      of the relevant chemistry.}
    {We found that CN, HCN, and HNC remain present in the gas phase at
      densities well above that at which CO depletes on to grains. The
      CN:HCN and the HNC:HCN abundance ratios are larger than unity in
      all the objects of our sample. Furthermore, there is no
      observational evidence for large variations of these ratios with
      increasing offset from the dust emission peak and hence with
      density. Whilst the differential freeze--out of CN and CO can be
      understood in terms of the current chemistry, the behaviour of
      the CN:HCN ratio is more difficult to explain. Models suggest
      that most nitrogen is not in the gas phase but may be locked in
      ices. Unambiguous conclusions require measurements of the rate
      coefficients of the key neutral--neutral reactions at low
      temperatures.}
    {}

\keywords{ISM: abundances, ISM: chemistry, ISM individual objects:
L~1544, L~183, Oph~D, L~1517B, L~310}

\maketitle
%

\section{Introduction}

\begin{table*}[t]
  \centering
  \caption{The sample of cores observed. Note that $\rho$Oph~D is also
    known as L~1696A.}
  \begin{tabular}{l c c c c c c c c c}\hline\hline
    Source & $\alpha_{2000}$ & $\delta_{2000}$ & $\Delta \alpha,\Delta \delta^a$ &
    $\vlsr^b$ & $n(\hh)^c$ & $S_{1.2\rm mm}$ & $N(\hh)^d$ & $\tkin^e$&References\\
    & & & \arcsec & \kms & $\dix{5}$~\ccc & \Mjysr & \dix{22}\cc & K \smallskip\\\hline
    L~183 & 15:54:08.80 & -02:52:44.0 &$ (0,0)$ & 2.37 & 20 & 18.3 & 7.2 & 7.0 & (1) \\
    L~1544 & 05:04:16.90 & +25:10:47.0 &$ (0,0)$ & 7.20 & 14 & 17.2 & 6.7 & 7.0 & (2) \\
    Oph~D & 16:28:30.40 & -24:18:29.0 &$ (0,0)$ & 3.35 & 8.5 & 14.6 & 5.7 & 6.0 & (3) \\
    L~1517B & 04:55:18.80 & +30:38:03.8 &$(-10,-20)$& 5.87 & 2.2 & 7.1 & 2.8 &10.0 & (4) \\
    L~310 & 18:07:11.90 & -18:21:35.0 &$( 30, 80)$& 6.70 & 0.9 & 7.5 & 2.9 & 9.0 & (5) \\\hline
    \label{tab:cores}
  \end{tabular}
  \begin{list}{}{}
    \scriptsize
  \item $^a$ Offsets of the continuum peak with respect to the
    reference position.
  \item $^b$ Systemic LSR velocity.
  \item $^c$ Peak particle number density.
  \item $^d$ Column density computed assuming $\tdust=8$~K and
    $\kappa_\nu=0.01\rm\,cm^2g^{-1}$. The values increase by 25\% if
    $\tdust=7$~K.
  \item $^e$ Gas kinetic temperature from the literature.
  \item (1) \cite{pagani2004,pagani2007}, (2)
    \cite{tafalla2002}, \cite{crapsi2007}, (3) \cite{ward1999},
    \cite{harju2008} (4) \cite{tafalla2004} (5) \cite{bacmann2002}
  \end{list}
\end{table*}

Observations of rotational transitions of molecules and radicals play
a key role in deriving information on solar--mass objects in the early
stages of gravitational collapse. The variation of the intensities of
the emission lines can be interpreted in terms of the chemical
reactions and gas--grain interactions occurring in the medium, and the
line profiles and frequency shifts in terms of the kinematics of the
collapsing gaseous material. Indeed, apart from infrared observations
of dust continuum emission, which yield no chemical or kinematical
information, measurements of radio transitions of molecules provide
the only means of probing the evolution of pre-protostellar cores.

An obstacle to the use of molecular line emission to study the early
stages of star formation is the propensity of some molecules to freeze
on to the grains at the low temperatures, $T \approx 10$~K, which
prevail. Observations of prestellar cores have shown that the
fractional abundances of the carbon--containing species, CO and CS,
decrease strongly towards the core centres, whereas the fractions of
the nitrogen--containing species, \nnhp\ and NH$_3$, either remain
constant or even increase towards the centre, where the density is
highest \citep{tafalla2002}. Differential freeze--out of the C- and
N-containing species on to the grains was the generally--accepted
explanation of these observational results. However, recent
observations of the NO radical have demonstrated that the real
situation is more complicated. A comparison of the profiles of NO and
\nnhp\ along cuts through the prestellar cores L~1544 and L~183
\cite[][ hereafter A07]{akyilmaz2007} has shown that the fractional
abundance of NO, unlike that of \nnhp, {\it decreases} towards the
centres of these cores (their centres being identified with the peak
of the dust emission). Thus, not all nitrogen--containing species
remain in the gas phase at densities $\gtrsim \dix{6}$~\ccc, which
prevail in the central regions. On the other hand, still more recent
observations of CN \citep{hilyblant2008cn} have shown that the
emission of this radical follows closely the dust emission in both
L~1544 and L~183. In so far as these two objects are representative of
their class, it appears that the adsorption process must somehow
differentiate between nitrogen--bearing species.

In the present work, we extended our observations of
nitrogen--containing species to include isotopomers of HCN. In
addition to L~1544 and L~183, we have studied Oph~D, L~1517B, and
L~310. The observations are described and analyzed in
Section~\ref{observations}. Sections~\ref{sec:results} and
\ref{sec:cdens} describe the observations and present estimates of
N-bearing species abundances. In Section~\ref{chemistry}, we consider
the chemical processes, including freeze--out on to the grains, which
determine the gas--phase abundances of key N-- and also C-- and
O--containing species in prestellar cores. Section~\ref{models}
summarizes the model that has been adopted of the early stages of the
collapse of the representative prestellar core L~1544. The fractional
chemical abundances predicted by the model are presented and the
corresponding column density profiles are compared with the
observations. Finally, in Section~\ref{conclusions}, we make our
concluding remarks.

\section{Observational procedures and data reduction}
\label{observations}

The observations were performed at the IRAM~30~m telescope in January
2008. The data have been reduced and anlayzed using the CLASS90
software \citep{IRAM_report_2005-1}. The instrumental setup was
identical to that used by \cite{hilyblant2008cn}: frequency--switching
spectra, with a frequency-throw of 7.8~MHz, were recorded by the VESPA
facility, with 20~kHz spectral resolution and 20 to 80~MHz
bandwidth. The half-power beam-width is calculated as
HPBW$[\arcsec]=2460/\nu[\rm GHz]$, that is 28\arcsec\ at 87~GHz and
22\arcsec\ at 113~GHz. Cross--like patterns with 20\arcsec\ spacing
were observed towards each source. The crosses were centered on the
dust emission peak, as determined from published continuum maps; all
offsets quoted in the present paper refer to the central positions
listed in Table~\ref{tab:cores}. Table~\ref{tab:summary} summarizes
the lines observed for each object. The amplitude calibration was
checked every 10~min, the pointing every hour, and the focus every two
hours, typically. Instrumental spectral effects were compensated by
subtracting polynomial baselines from each spectrum before
folding. More details on the data reduction procedures can be found in
Appendix~\ref{AppA}. All results (unless explicitly stated) have been
translated from the antenna temperature scale (\tant) to the main-beam
temperature scale $\tmb=\tant\times\feff/\beff$, with the values of
\beff\ listed in Table~\ref{tab:summary}; \feff\ is the forward
efficiency, and \beff\ is the beam efficiency.

\begin{figure*}[t]
  \centering
  \includegraphics[width=0.3\hsize,angle=-90]{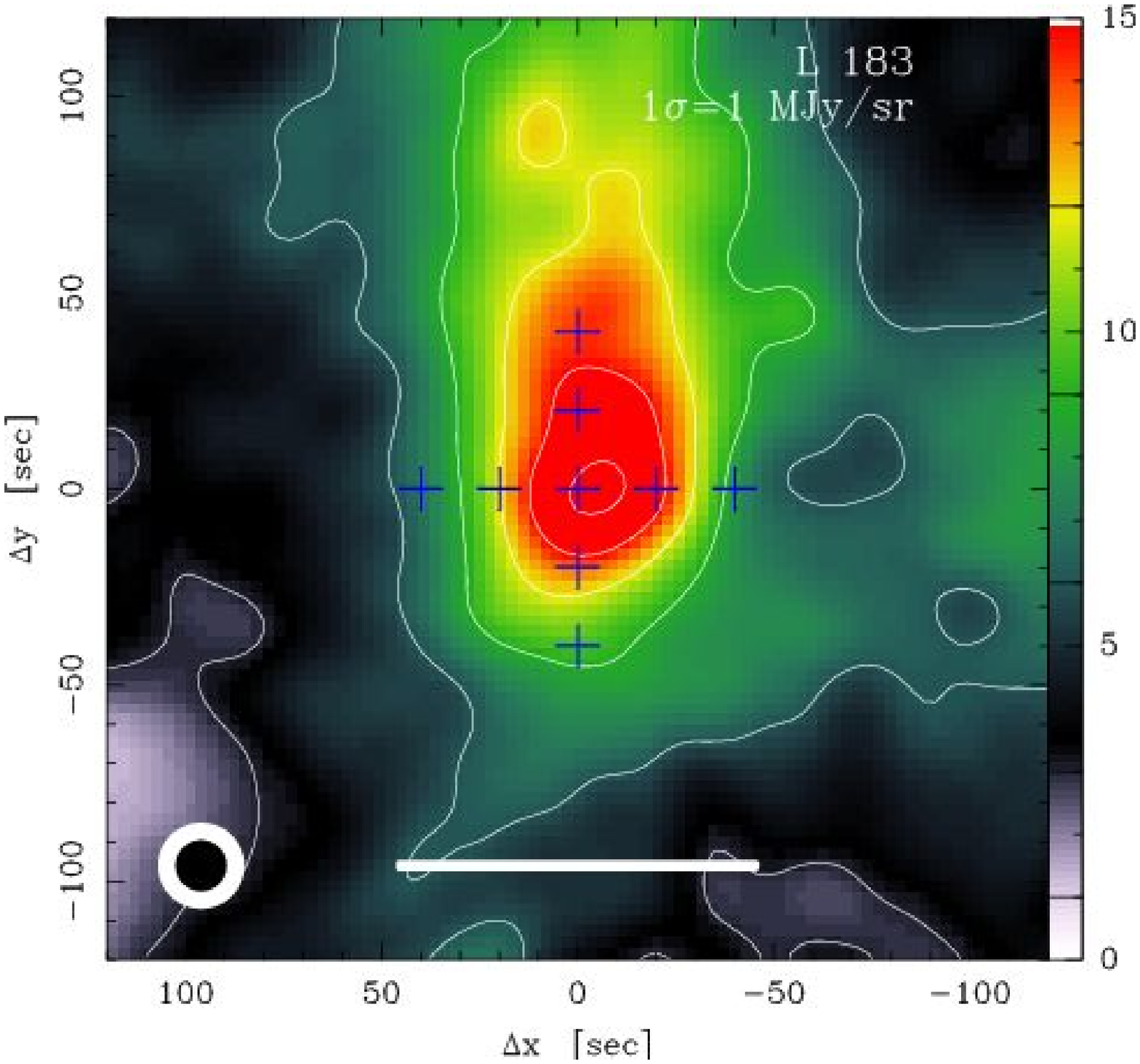}\hfill%
  \includegraphics[width=0.3\hsize,angle=-90]{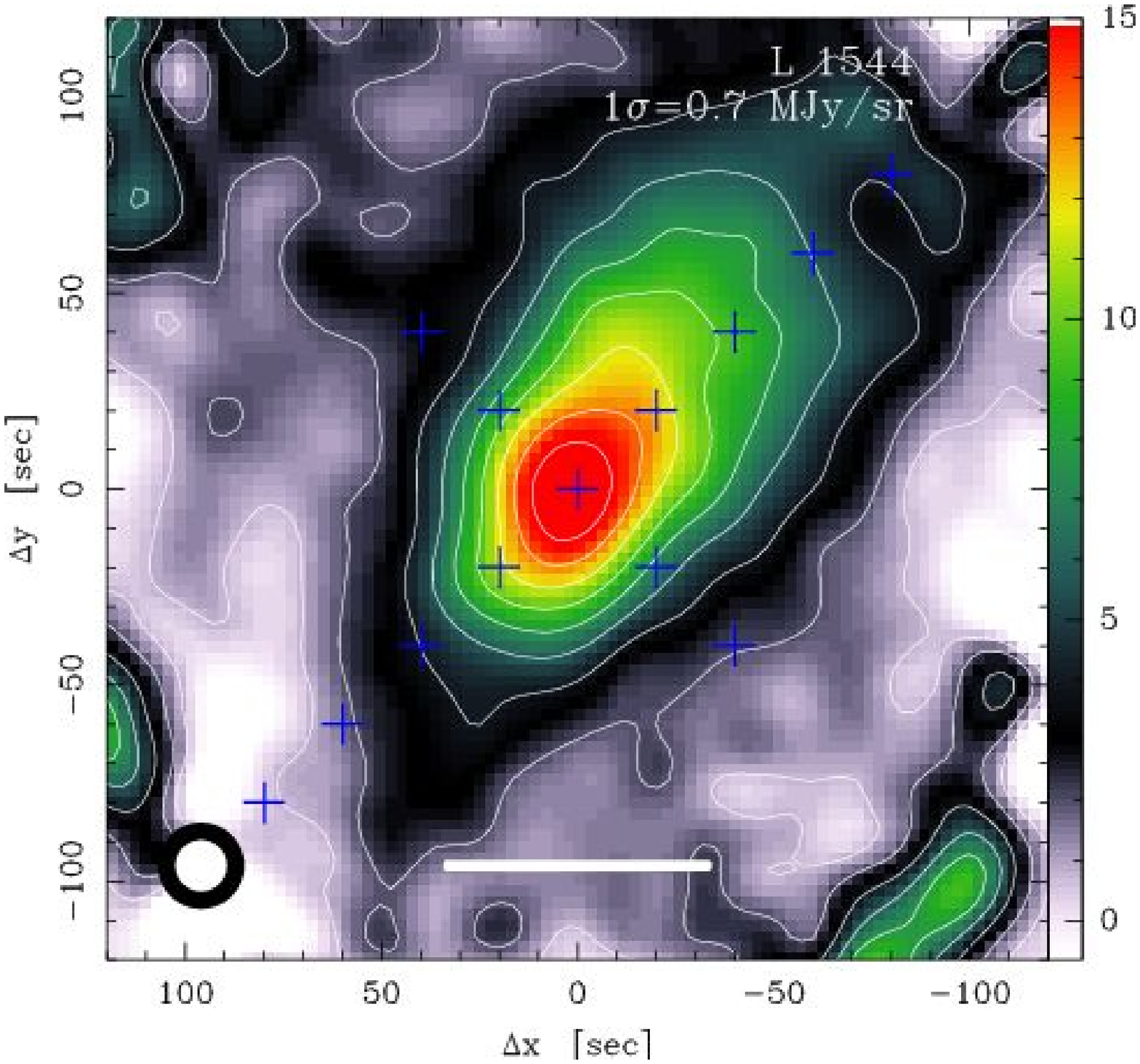}\hfill%
  \includegraphics[width=0.3\hsize,angle=-90]{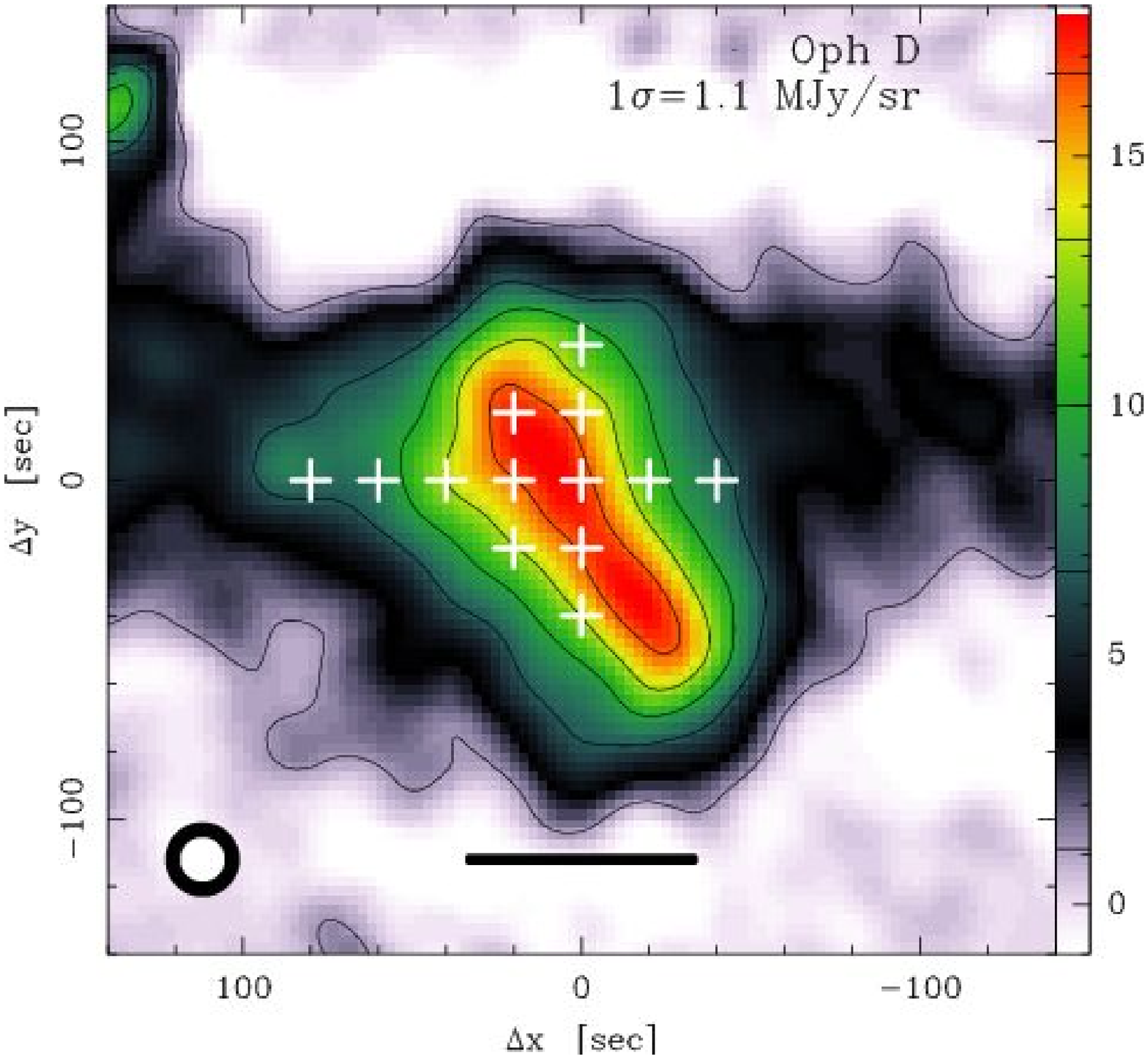}
  \includegraphics[width=0.3\hsize,angle=-90]{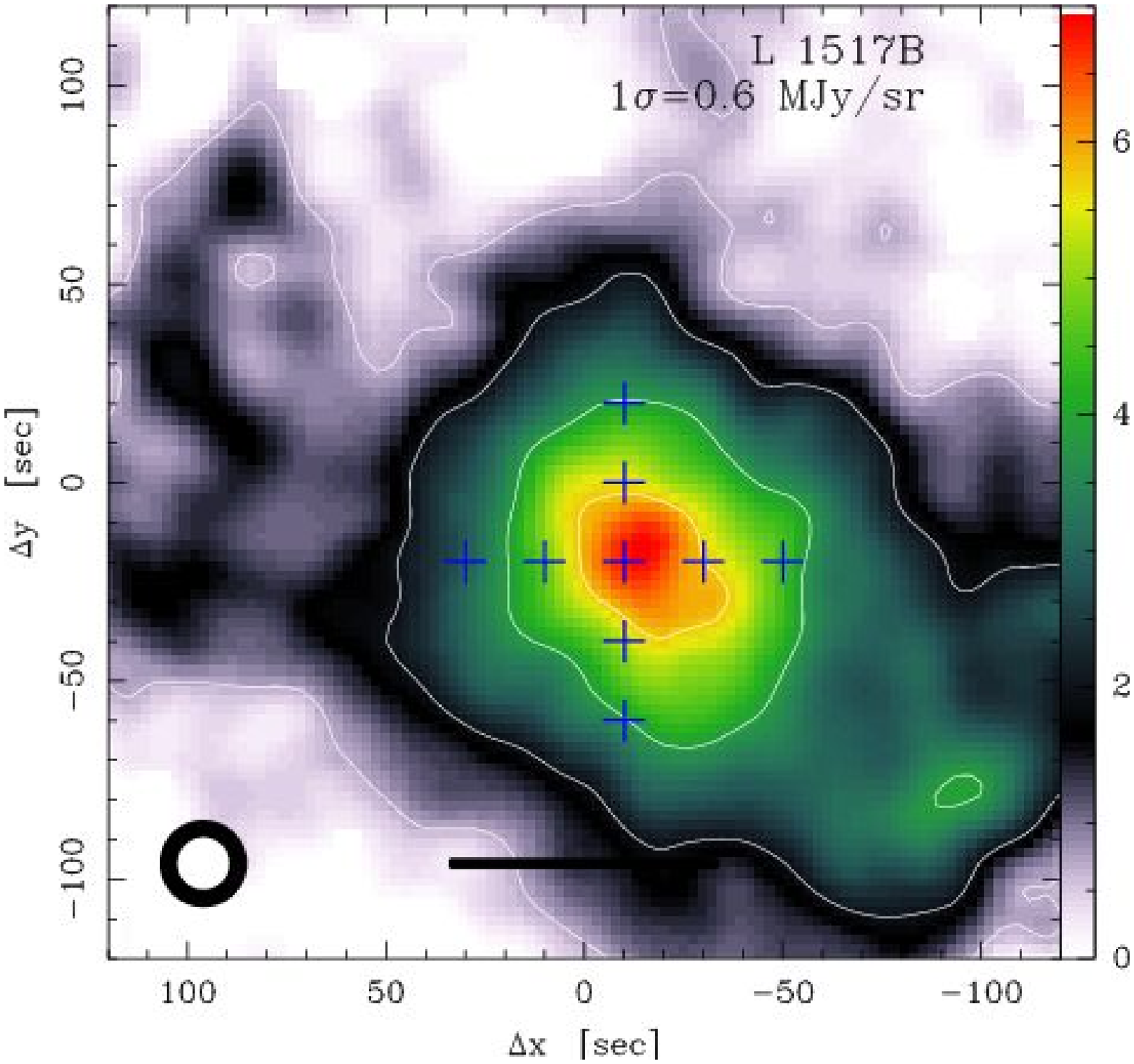}\hfill%
  \includegraphics[width=0.3\hsize,angle=-90]{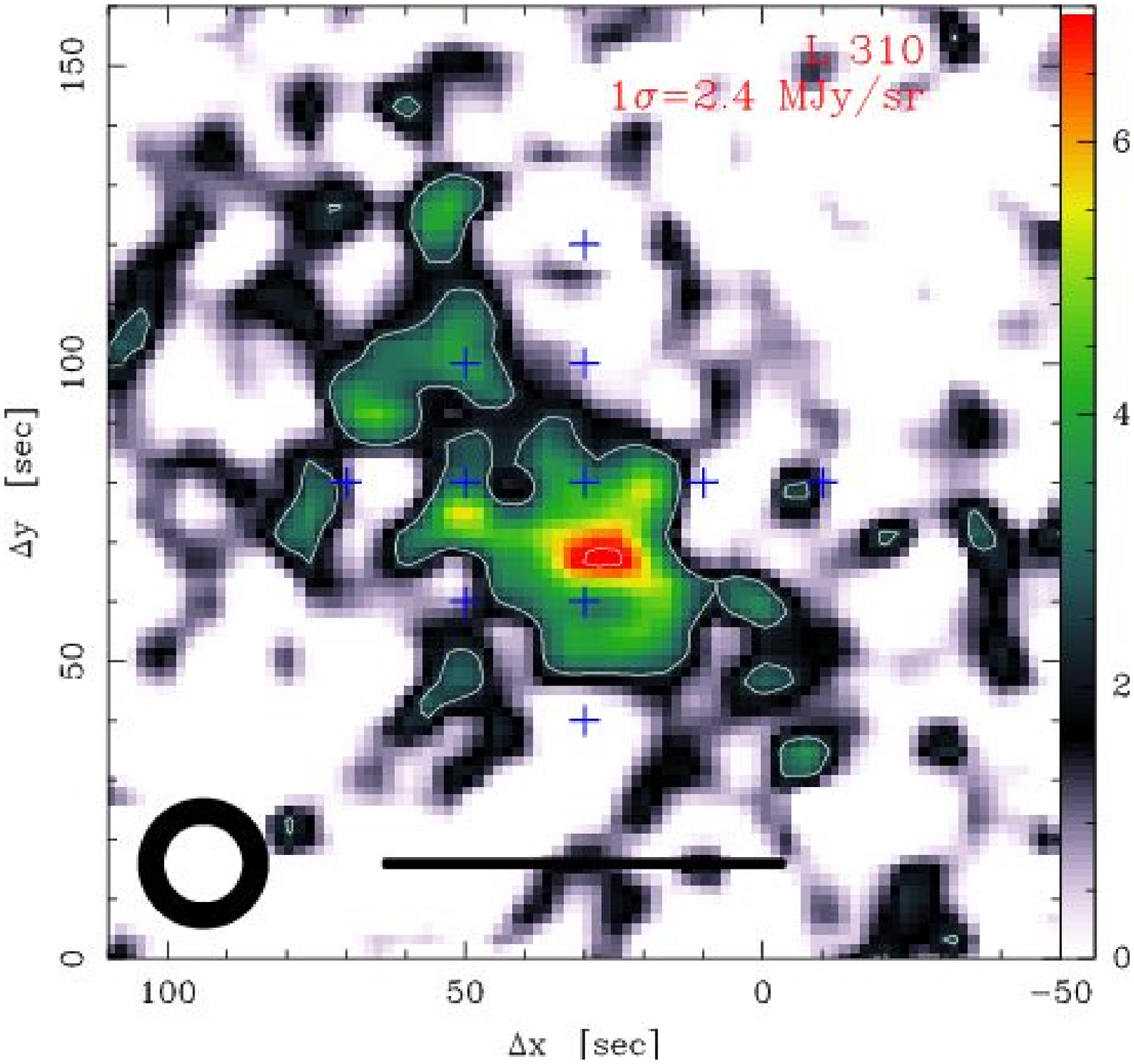}
  \caption{Continuum emission (\Mjysr) at 1.2mm with the locations of
    the line integrations (crosses). The HPBW at 1.2~mm and at the
    3~mm frequencies discussed in this paper are indicated. A linear
    scale of 0.05~pc is also shown, assuming a distance of 150~pc for
    all cores except L~183 (110~pc). Continuum maps for L~1544, Oph~D,
    L~183, L~1517B and L~310 are taken from \cite{ward1999},
    \cite{pagani2003}, \cite{tafalla2004} and \cite{bacmann2000}
    respectively. \colfig}
  \label{fig:dust}
\end{figure*}

\section{Observational results}
\label{sec:results}

All objects in our sample are pre-stellar, in the sense that none
shows \corr{unambiguous evidence of collapse}{signposts of embedded
  stellar objects}. Their peak particle number densities span more
than an order of magnitude (see Table~\ref{tab:cores}): the peak
densities decrease from approximately $\dix{6}$~\ccc\ (L~183, L~1544)
to $0.9\tdix{5}$~\ccc\ in L~310. All the objects have been extensively
observed, both in their lines and continuum (see
Fig.~\ref{fig:dust}). Our observations focused on the
nitrogen--bearing species CN, HCN, \thcn, H\thcn, HN\thc, and
HC\fifn. All these molecules present hyperfine structure
(HFS). However, in the cases of HN\thc\ and HC\fifn, the hyperfine
structures were not resolved in our 20~kHz resolution spectra. All
transitions are in the 3~mm band; the CN\jtwo\ line was observed in
parallel at 1.3~mm (see Table~\ref{tab:12cn}).

\subsection{Line properties}

\begin{figure*}
  \def\wa{0.33\hsize}
  \includegraphics[width=\wa,angle=0]{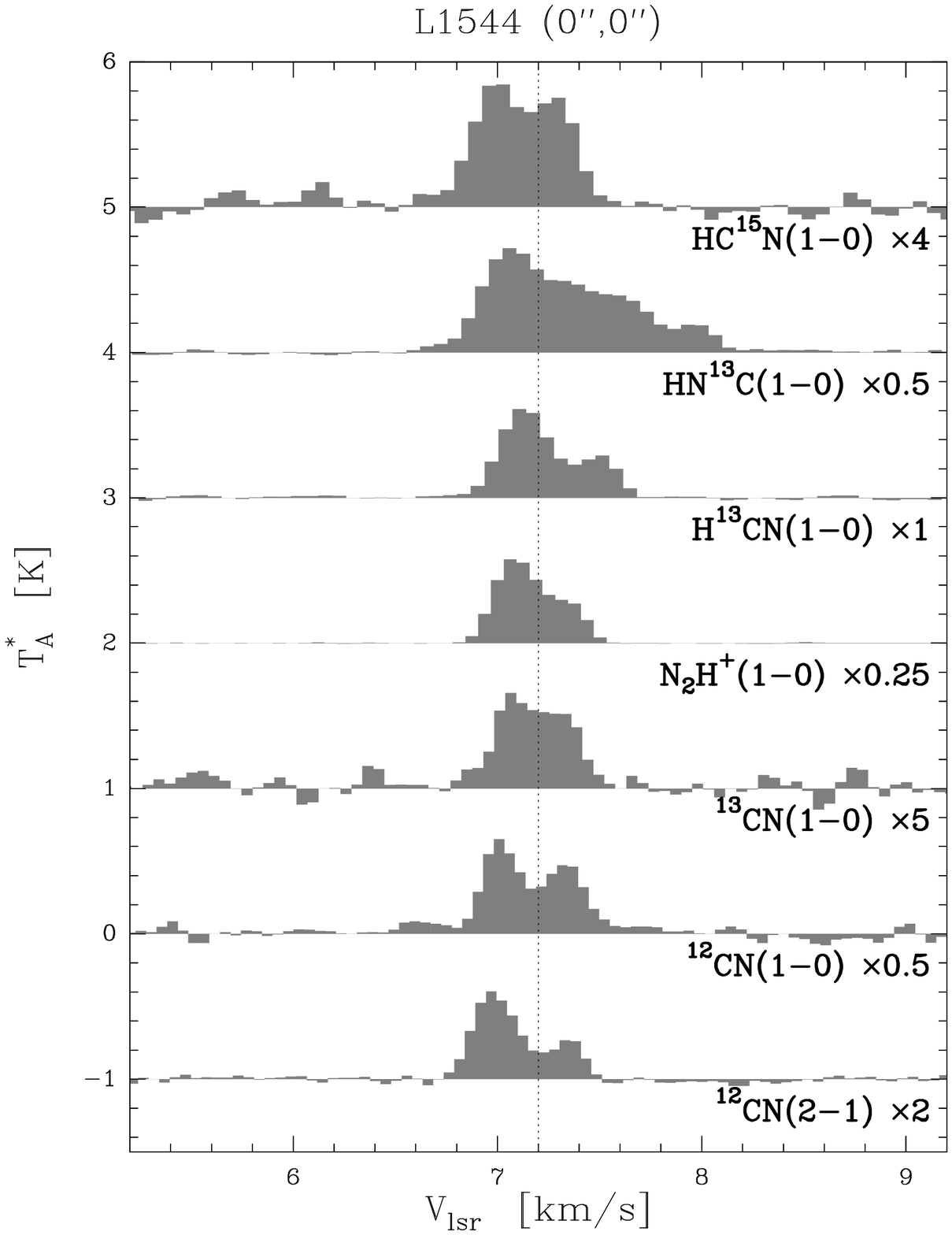}\hskip0.1\hsize
  \includegraphics[width=\wa,angle=0]{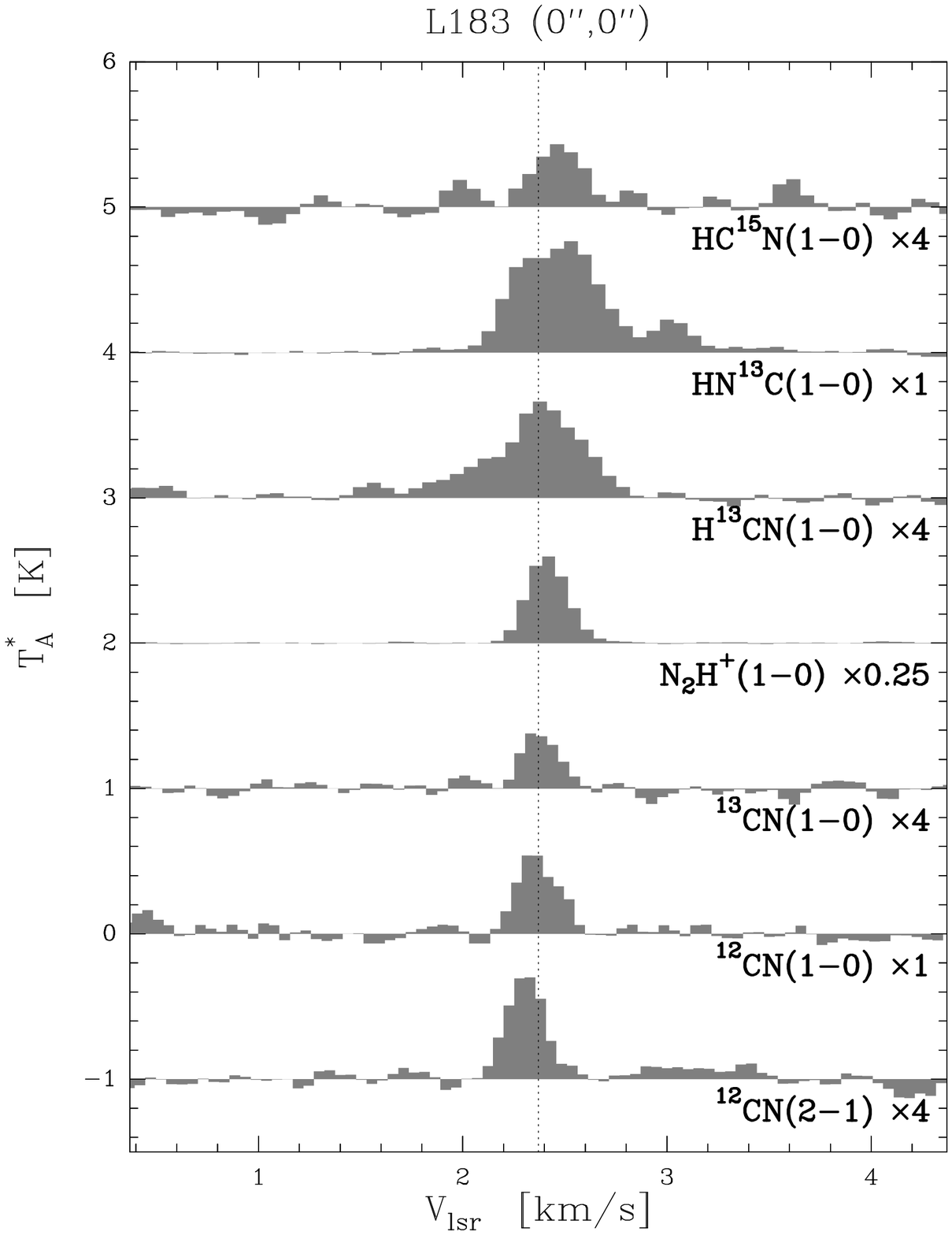}\\
  \includegraphics[width=\wa,angle=0]{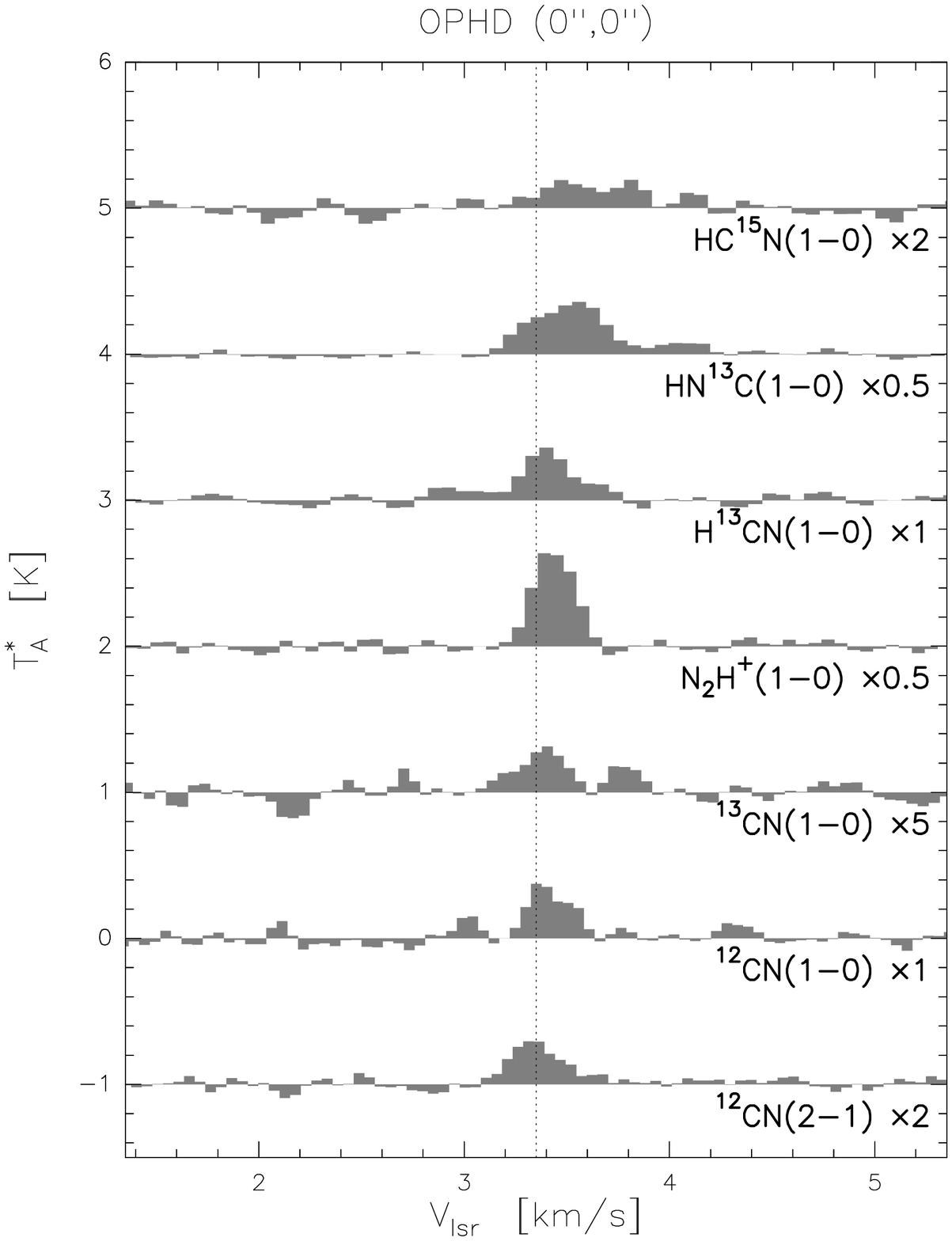}\hskip0.1\hsize
  \includegraphics[width=\wa,angle=0]{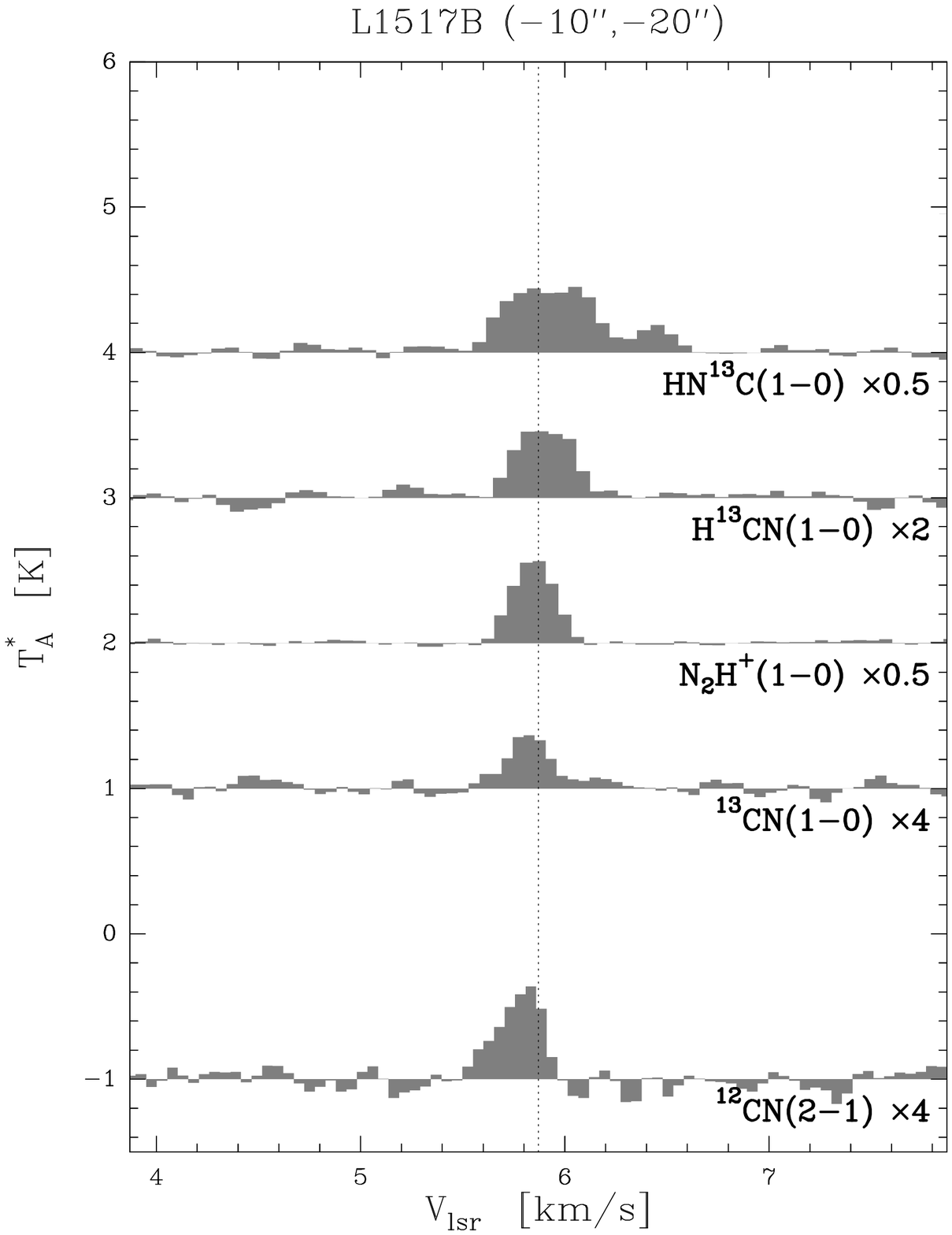}\\
  \includegraphics[width=\wa,angle=0]{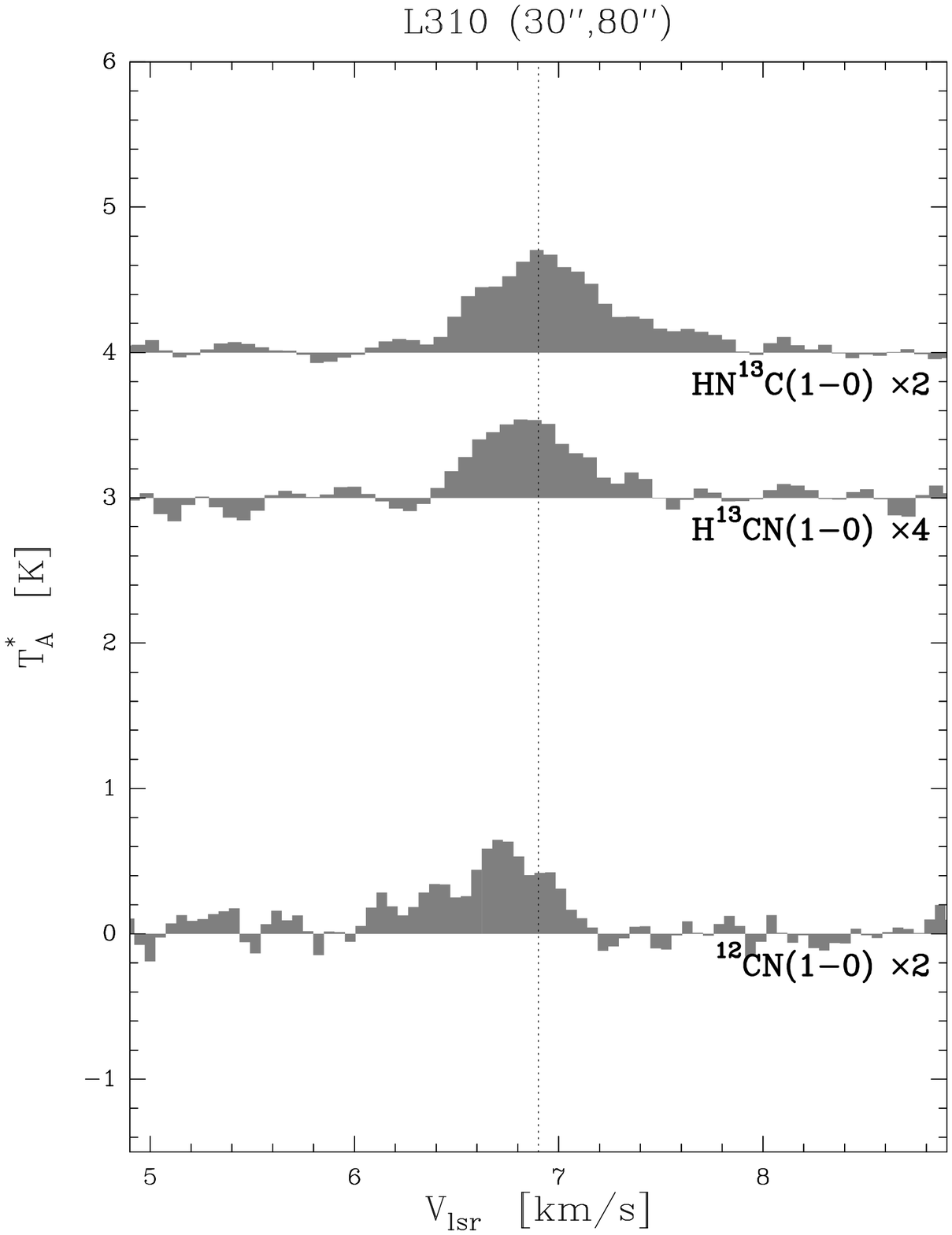}
  \caption{Comparison of the line profiles of different tracers; the
    spectra have been shifted vertically for clarity. For CN\jone,
    \jtwo, and \nnhp\jone, the weakest HFS components (at 113520.4315,
    226887.3520 and 93176.2650~MHz respectively) are shown. For each
    of the other lines, the strongest HFS component is considered:
    108780.2010, 86340.1840, 87090.8590, and 86054.9664~MHz for
    \thcn\jone, H\thcn, HN\thc\ and \hcfifn\jone\ respectively. The
    \nnhp\ spectrum is taken from A07. The spectra are for zero offset
    (cf. Table~\ref{tab:cores}). Towards L~1517B, the
    CN\jone\ spectrum is replaced by CN\jtwo\ and the \nnhp\ spectrum
    is taken from \cite{tafalla2002}. Towards Oph~D, the
    \nnhp\ spectrum at offset (0\arcsec, 0\arcsec) is taken from
    \cite{crapsi2005}.}
  \label{fig:profile-00}
\end{figure*}

All lines were observed successfully towards the three most centrally
peaked cores, L~183, L~1544, and Oph~D; the emission lines are shown
in Fig.~\ref{fig:profile-00}. For CN and \nnhp\jone, only the weakest
HFS components (at 113520.4315~MHz and 93176.2650~MHz) are shown. For
each of the other lines, the strongest HFS component is considered
instead: 108780.2010~MHz, 86340.1840~MHz, 87090.8590~MHz, and
86054.9664~MHz for \thcn\jone, H\thcn, HN\thc\ and
\hcfifn\jone\ respectively. The HN\thc\ hyperfine structure is not
fully resolved and the two hyperfine transitions of \hcfifn\ are not
resolved. The \thcn\ line is detected in the four densest objects,
L~183, L~1544, Oph~D and L~1517B.

Tables~\ref{tab:l183}-\ref{tab:l310} give the properties of all lines
towards all the positions in each source. The integrated intensities,
$W$, were derived from Gaussian fitting of a given HFS component (see
above). Several Gaussian components were fitted in some cases,
\eg\ for the known hyperfine structure of HN\thc, and also in the
obvious cases of multiple--component line profiles (L~1544). In such
cases, $W$ is the sum of the integrated intensities of each velocity
component. The peak temperature \tpeak\ is the maximum intensity over
the line. Given that the lines are, in general, not Gaussian, the
linewidth is estimated as the equivalent width, $\dveq=W/\tpeak$ and
the statistical uncertainty is obtained by propagating the errors. For
non-detections, upper limits on the integrated intensity were obtained
by fitting a Gaussian at a fixed position. Upper limits on the
intensity are at the 3$\sigma$ level while those on the integrated
intensity are at the 5$\sigma$ level.

Towards L~1544, all resolved lines show two clear peaks, with a dip
centred at 7.20~\kms. These two peaks are seen in several tracers
including \htco\ by \cite{hirota2003} who concluded that there are two
distinct velocity components along the line of sight
\citep{tafalla1998}. Owing to their double--peak profiles, lines
towards L~1544 have the largest integrated intensities of all the
lines that we observed.

The comparison of the \thcn\ linewidths shows that the lines towards
L~183 are the narrowest \corr{}{with full widths at half maximum}
(FWHM) $\approx0.20$~\kms. The H\thcn\ line in this source exhibits a
blue wing and the profile can be well fitted by two Gaussian
components with FWHM of 0.38 and 0.85~\kms; this blue wing is not
evident in any other tracer. Towards Oph~D and L~1517B, the linewidth
is larger by a factor of 2 to 3, although the comparison with L~1544
is rendered difficult by the double--peak line profiles. In several
tracers, \eg\ H\thcn, L~310 displays the largest linewidth
($<0.8$~\kms) but small integrated intensities. In all the sources,
\hcfifn\ has been detected, and the properties of the line, averaged
over all offset positions, are listed in Table~\ref{tab:hc15n}. The
peak and integrated intensities decrease as the peak density
decreases. Again, L~1544 seems to be an exception, owing to the
double--peak line profile. The FWHM are comparable ($\approx0.4~\kms$)
for all sources; but, once again, the FWHM is significantly larger (by
a factor 2) in L~310 than in the other sources.
\begin{table*}
  \centering
  \caption{Properties of the average \hcfifn\jone\ profiles and
    average fractional abundance towards the observed cores (main beam
    temperature scale).}
  \begin{tabular}{l r c c c c c c}\hline\hline
    Source & $\aver{\tmb}$ & $\aver{W}$ & $\aver{v_0}$ & \aver{\rm FWHM} & 
    $\aver{N(\hcfifn)}$ & $\aver{N(\hh)}_{20\arcsec}\,^b$ & $\aver{N}/\aver{2N(\hh)}\,^d$ \\
    \smallskip\\
    & mK & m\kkms &\kms & \kms & \tdix{10}\,\cc & \tdix{22}\,\cc & $\times\dix{-12}$ \smallskip\\\hline
    L~183 & 125 $\pm$ 15 & 47 $\pm$ 4 &2.49 & 0.35 $\pm$ 0.04 & 8.9$\pm$0.8 & 5.6 & 0.8\\
    L~1544$^a$ & 170 $\pm$ 12 & 85 $\pm$ 5 &7.20 & 0.48 $\pm$ 0.05 &16.1$\pm$1.0 & 4.3 & 1.9\\
    Oph~D & 102 $\pm$ 15 & 47 $\pm$ 6 &3.44 & 0.43 $\pm$ 0.08 & 8.9$\pm$1.1 & 5.6 & 0.8\\
    L~1517B & 82 $\pm$ 15 & 29 $\pm$ 3 &5.69 & 0.33 $\pm$ 0.04 & 5.5$\pm$0.6 & 2.0 & 1.4\\
    L~310 & 20 $\pm$ 10 & 21 $\pm$ 3 &6.69 & 0.98 $\pm$ 0.16 & 4.0$\pm$0.6 & 0.8 & 2.5\\
    \hline
  \end{tabular}
  \label{tab:hc15n}
  \begin{list}{}{}
    \scriptsize
  \item $^a$ Using a two--component Gaussian fit. The FWHM is taken to be the
    equivalent linewidth, in this case; $v_0=7.2$~\kms\ is adopted
    from \cite{hilyblant2008cn}.
  \item $^b$ Column density of \hh\ derived from the dust emission
    smoothed to 20\arcsec, assuming $\tdust=8$~K (see
    Table~\ref{tab:cores}). The total (statistical and systematic)
    uncertainty is 30\%.
  \item $^d$ The uncertainty is typically 40\%.
  \end{list}
\end{table*}

\subsection{Line ratios}

The ratios of total integrated intensities \wtot\ for some line
combinations in each source, are shown in
Fig.~\ref{fig:ratio-cnhcn-1}. Under the optically thin assumption,
these ratios reflect the relative abundances. The ratios CN/HCN or
\thcn/H\thcn\ are constant to within a factor of 2 across all the
cores and vary between about 0.5 to 5 from source to source. Towards
L~1517B, the \thcn/H\thcn\ ratio appears to decrease towards the
centre. Significant also is the fact that the H\thcn/HN\thc\ is
constant and of similar magnitude (0.2--0.8) in all sources,
independent of the central density.

\begin{figure*}
  \centering \def\wa{0.45\hsize}
  \includegraphics[width=\wa,angle=-90]{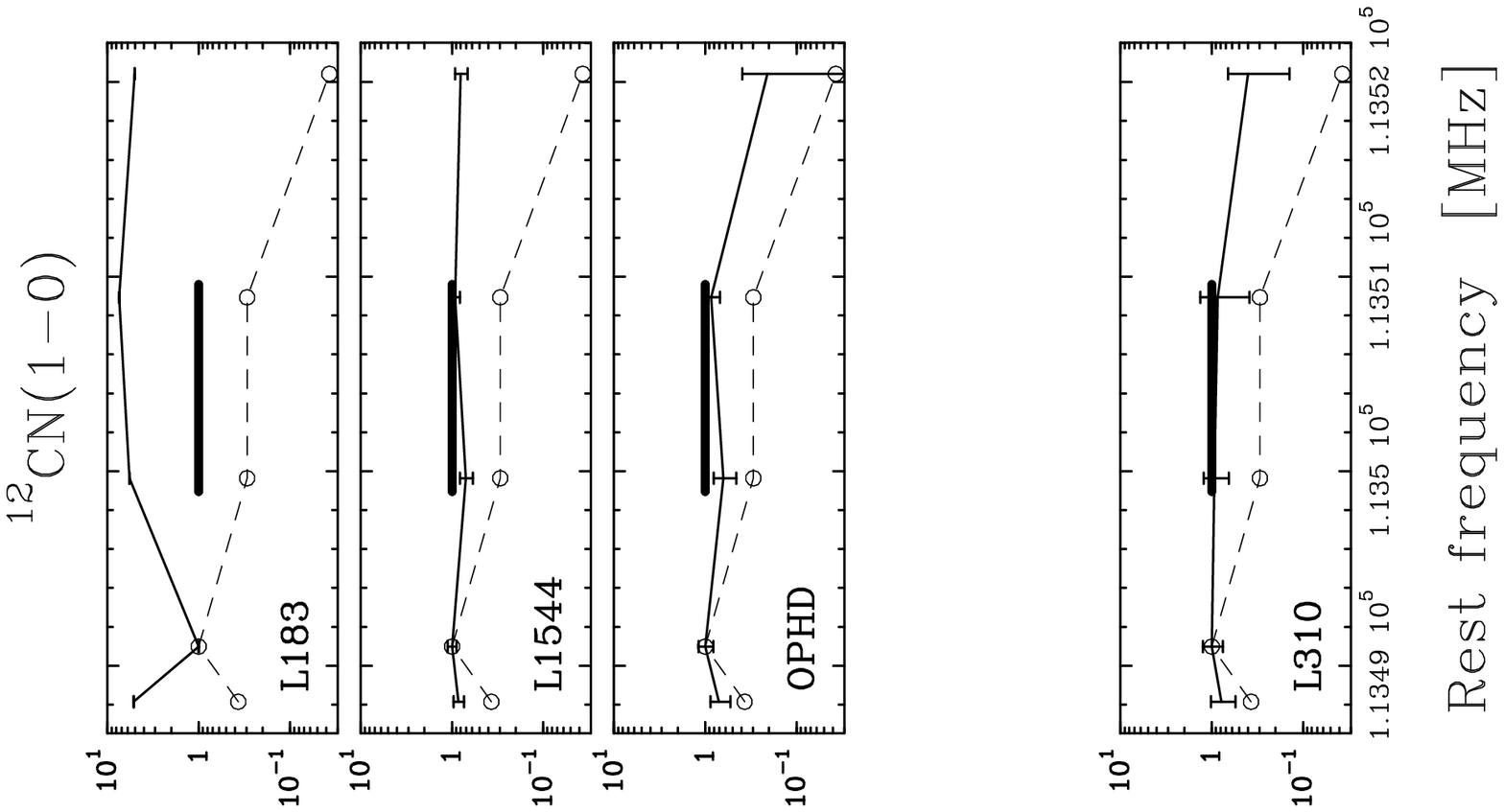}\hfill%
  \includegraphics[width=\wa,angle=-90]{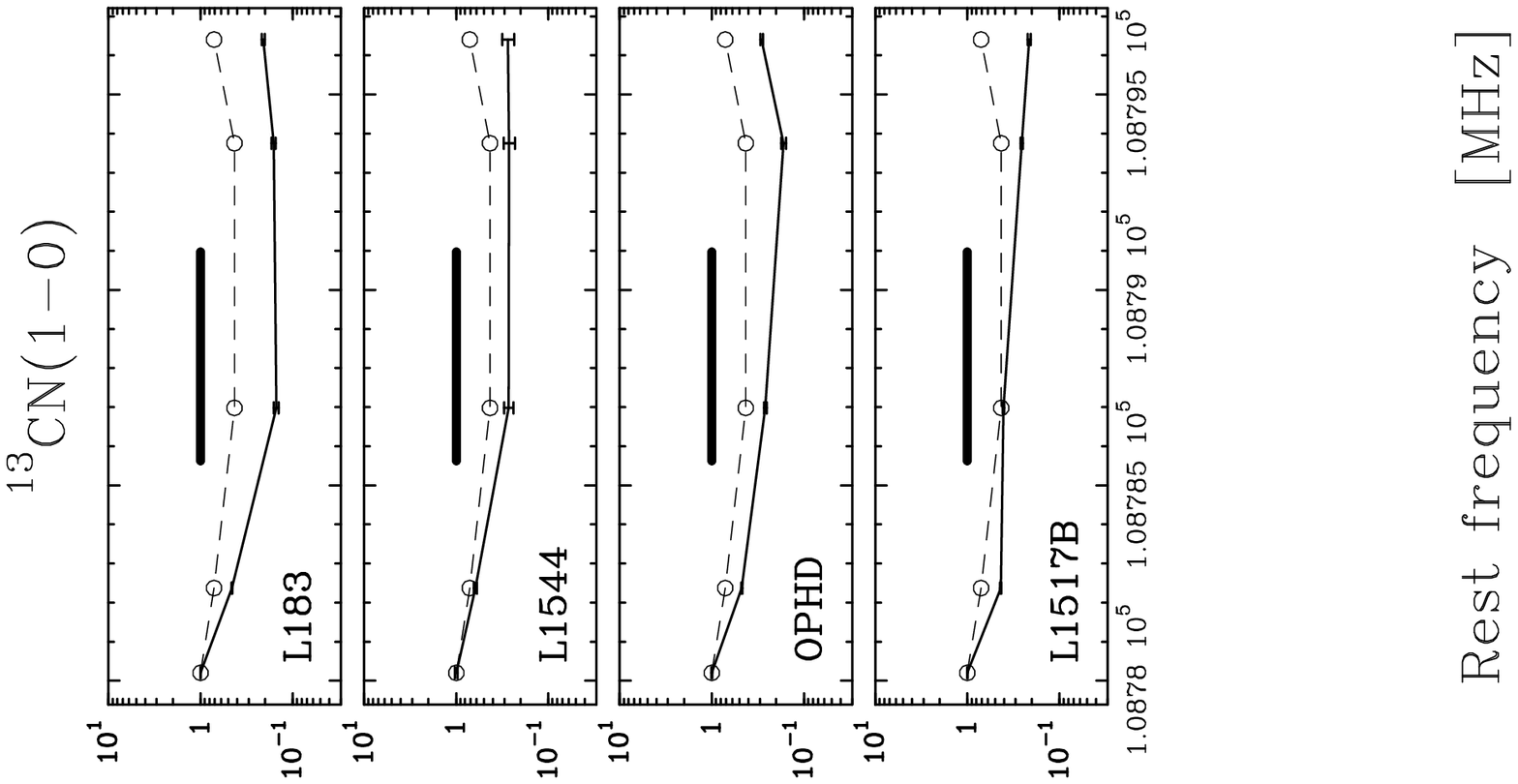}\hfill%
  \includegraphics[width=\wa,angle=-90]{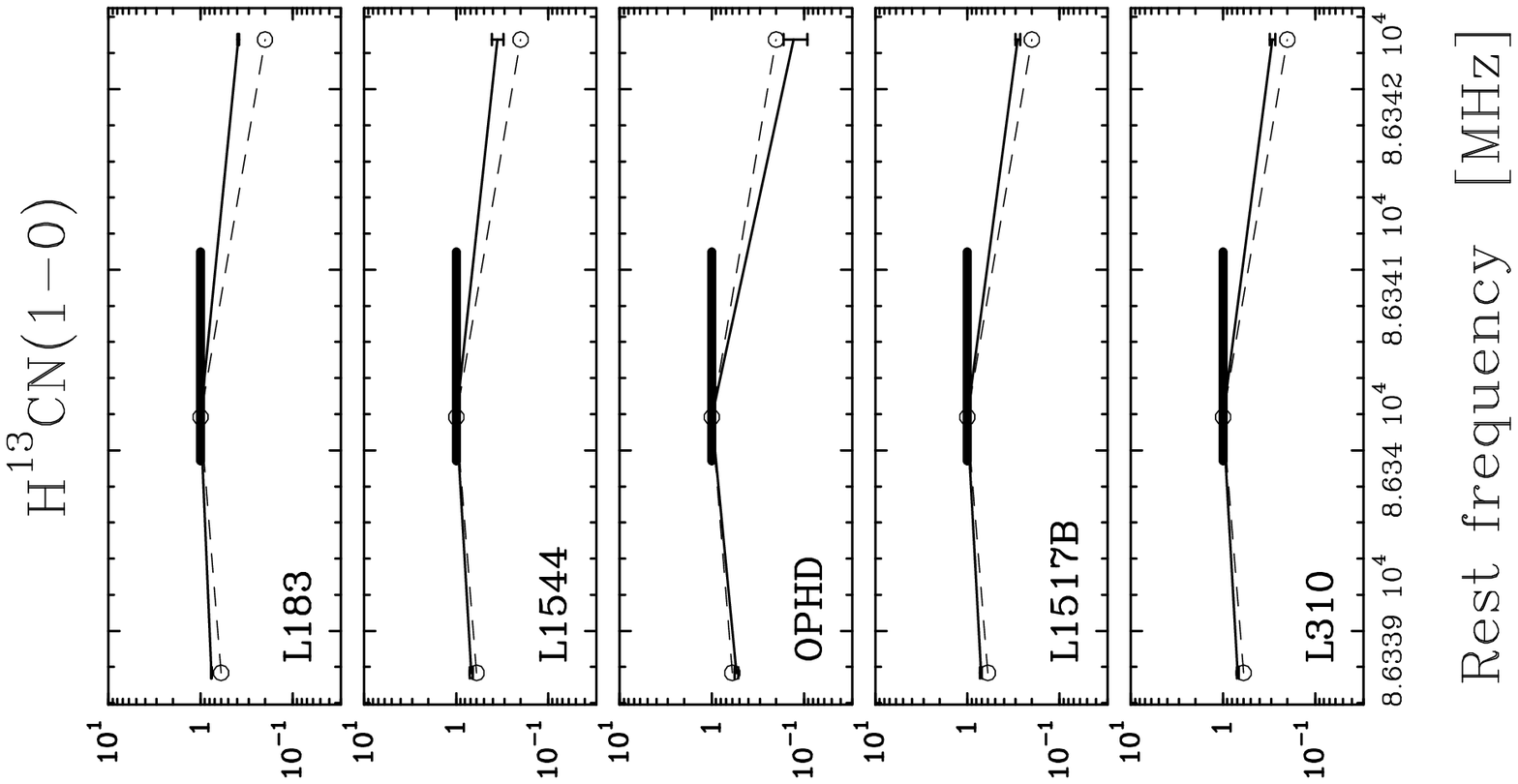}\hfill%
  \includegraphics[width=\wa,angle=-90]{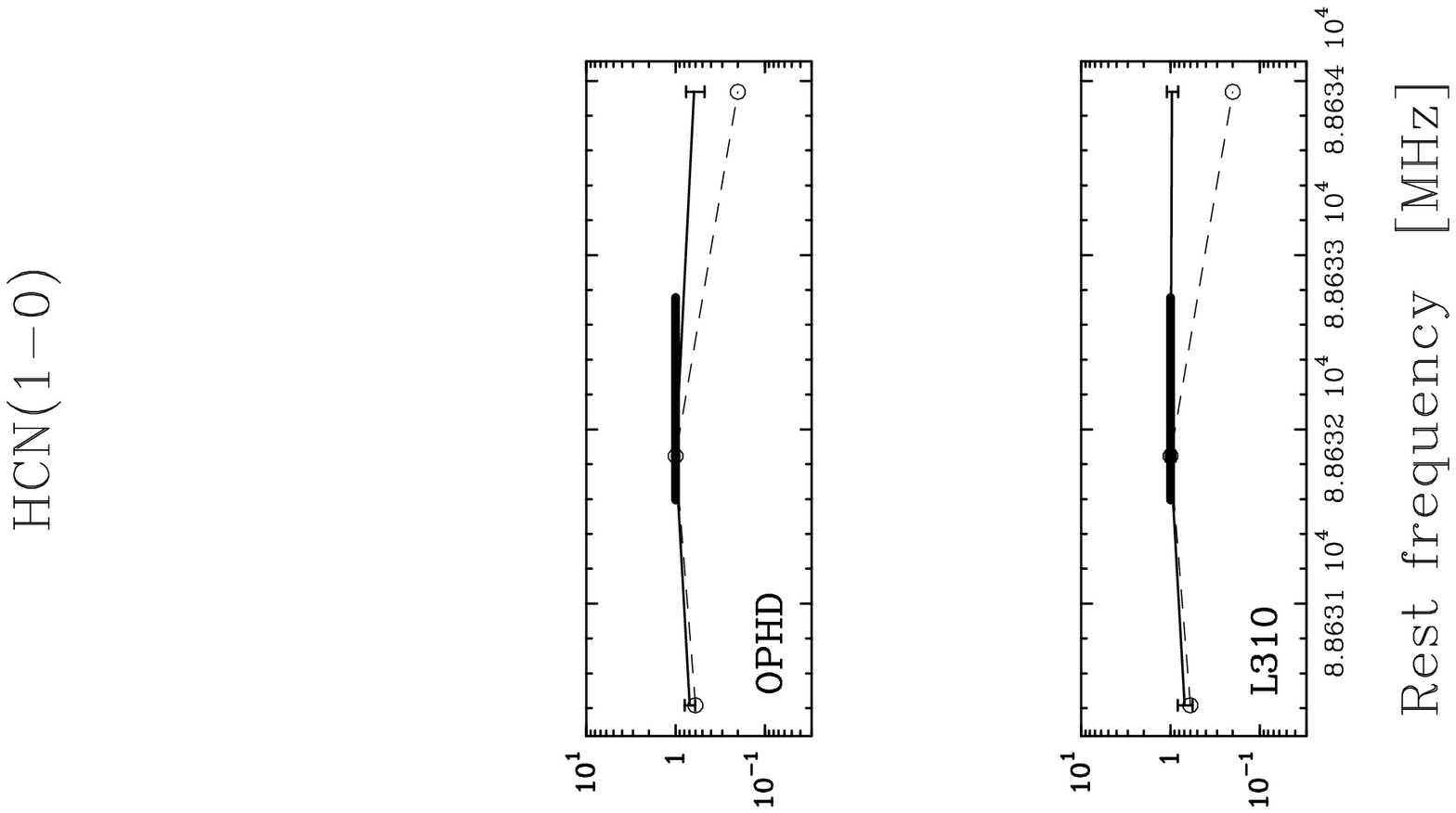}
  \caption{Relative integrated intensity of each HFS component for
    several species, at the central position for each source. In each
    panel, the dashed line indicates the relative intensities in LTE
    for optically thin emission. The thick line shows the optically
    thick limit. The abscissa is the rest line frequency. }
  \label{fig:hfs}
\end{figure*}

Most of the lines that have been observed are split by the hyperfine
interaction, and the relative intensities of the hyperfine components
can be used as a measure of optical depth. It is generally assumed
that the level populations of hyperfine states are in LTE and hence
proportional to the statistical weights, within a given rotational
level. However, it has been known for some time that this assumption
is often invalid \cite[see, for example, the discussion
of][]{walmsley1982}, and this is confirmed by our data. When the
populations are in LTE, one expects the satellite line intensity
ratios to lie between the ratios of the line strengths, in the
optically thin limit, and unity for high optical depths. As may be
seen from Fig.~\ref{fig:hfs}, this is usually but not always the case.
For example, it is clear that the CN\jone\ ratios towards L~183 are
inconsistent with this expectation, whereas the ratios observed
towards other sources suggest high optical depths and are broadly
consistent with equal excitation temperatures in the different
components. The observed \thcn\ ratios show clear signs of deviations
from LTE, but the effects are much less drastic than in the more
abundant isotopomer, and we suspect that optical depths are low. In
the case of H\thcn, departures from LTE appear to be minor.

The reasons for departures from LTE such as seen in Fig.~\ref{fig:hfs}
are presently unknown and need to be established. Such an
investigation would require calculations of the collisional rate
coefficients, analogous to those of \cite{monteiro1986}, as well as a
treatment of the radiative transfer; this is beyond the scope of the
current study. For the present, we use low abundance isotopomers such
as \thcn \ and H\thcn \ to trace abundance gradients, neglecting
collisional excitation and the possibility of fractionation of the
$^{13}$C and $^{15}$N isotopomers.


\begin{table*}[t]
  \centering
  \caption{Fractional abundances of CN, HCN, HNC and \nnhp\ relative
    to H towards the dust emission peaks of our source sample.
    Tables~\ref{tab:l183}--\ref{tab:l310} give the fractional
    abundances toward all the observed positions.}
  \begin{tabular}{l c c c c c c c} \hline\hline
    Source & [CN] & [HCN] & [HNC] & [\nnhp] & [NO]$^a$ & CN:HCN & HNC:HCN \\ 
    & \tdix{-9} & \tdix{-9} & \tdix{-9} & \tdix{-10} & \tdix{-9} & \\ 
    \hline
    L~183 & 0.40 & 0.17 & 0.45 & 0.78 & 10.0 & 2.4 & 2.6\\
    L~1544 & 1.11 & 0.56 & 1.12 & 1.20 & 4.0 & 2.0 & 2.0\\
    Oph~D & 0.37 & 0.23 & 0.38 & & & 1.6 & 1.7\\
    L~1517B& 1.12 & 0.48 & 1.50 & & & 2.3 & 3.1\\
    L~310 & 4.40 & 1.50 & 2.50 & & & 2.9 & 1.6\\
    \hline
  \end{tabular}
  \label{tab:peakab}
  \begin{list}{}{}
    \scriptsize
  \item $\rm [X]=n(X)/2n(\hh)$. An isotopic ratio $\twc:\thc=68$ was
    assumed when using the \thc\ isotopologues.
  \item A global uncertainty (including statistical and systematic
    errors) of 40\% is assumed.
  \item $^a$ The abundance of NO is taken from A07.
  \end{list}
\end{table*}

\begin{figure*}
  \centering
  \def\wa{.80\hsize}
  \def\wb{.465\hsize}
  \def\wy{\vspace{0.67cm}}
  \includegraphics[height=\wa,angle=-90]{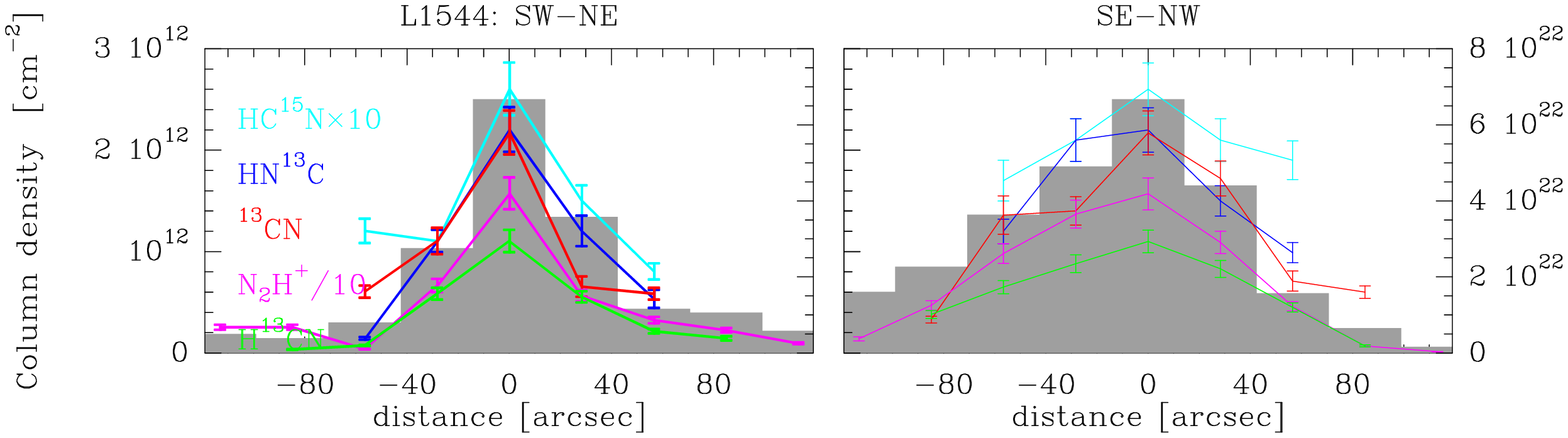}\bigskip\\
  \includegraphics[height=\wa,angle=-90]{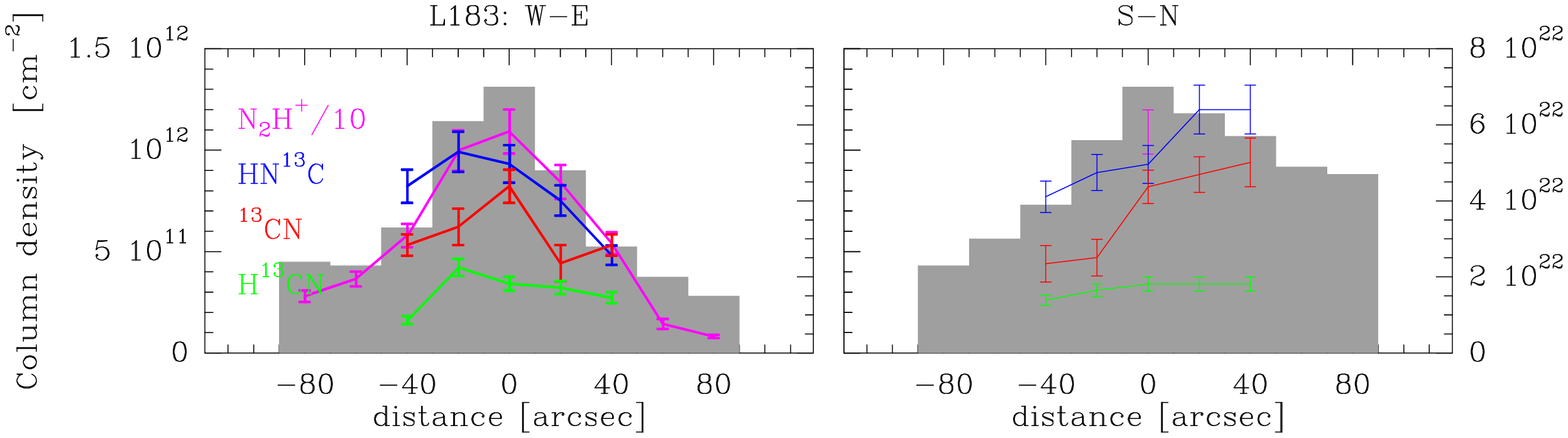}\bigskip\\
  \includegraphics[height=\wa,angle=-90]{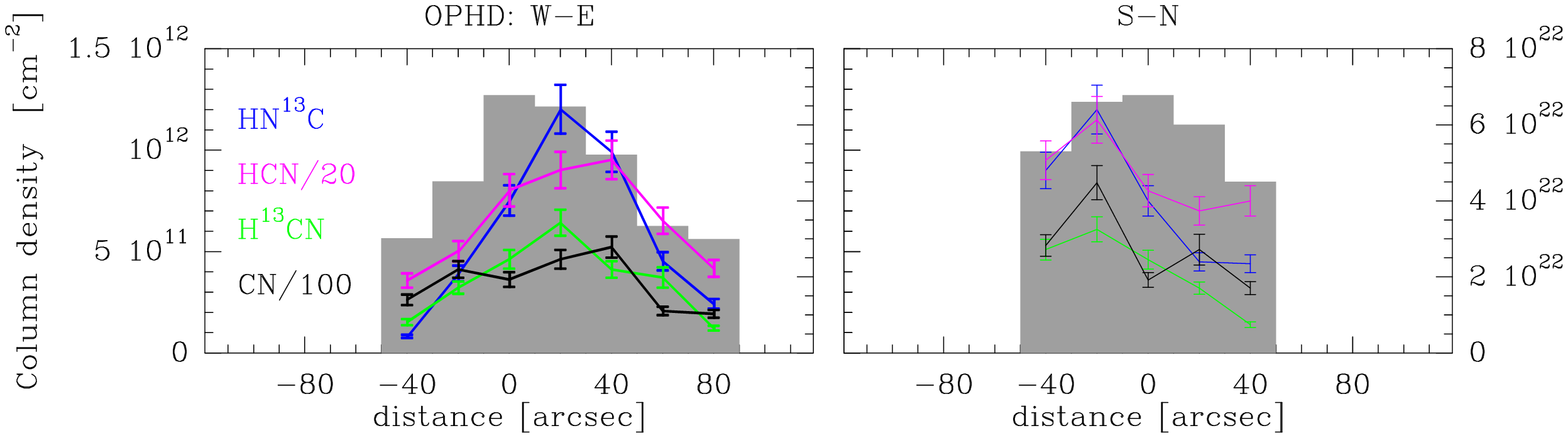}\bigskip\\
  \includegraphics[height=\wa,angle=-90]{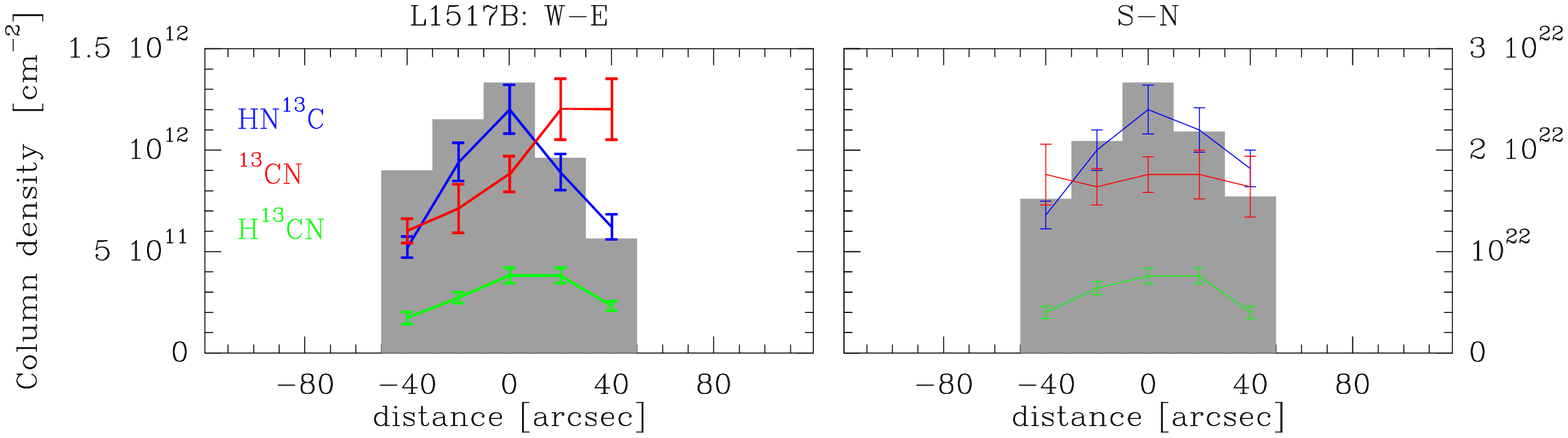}
  \caption{Derived column densities towards each source. From top to
    bottom: L~1544, L~183, Oph~D, and L~1517B. The column densities
    are plotted as a function of the distance from the dust emission
    peak, along both cuts.  Also plotted is the H$_{2}$ column density
    (grey histogram, right scale), as derived from the dust emission
    (assuming $\tdust=8$~K and $\kappa=0.01\,\rm cm^2\, g^{-1}$). The
    right panels show the corresponding abundances. \colfig}
  \label{fig:cdens}
\end{figure*}

\begin{figure*}
  \centering
  \def\wa{.80\hsize}
  \def\wb{.465\hsize}
  \def\wy{\vspace{0.67cm}}
  \includegraphics[height=\wa,angle=-90]{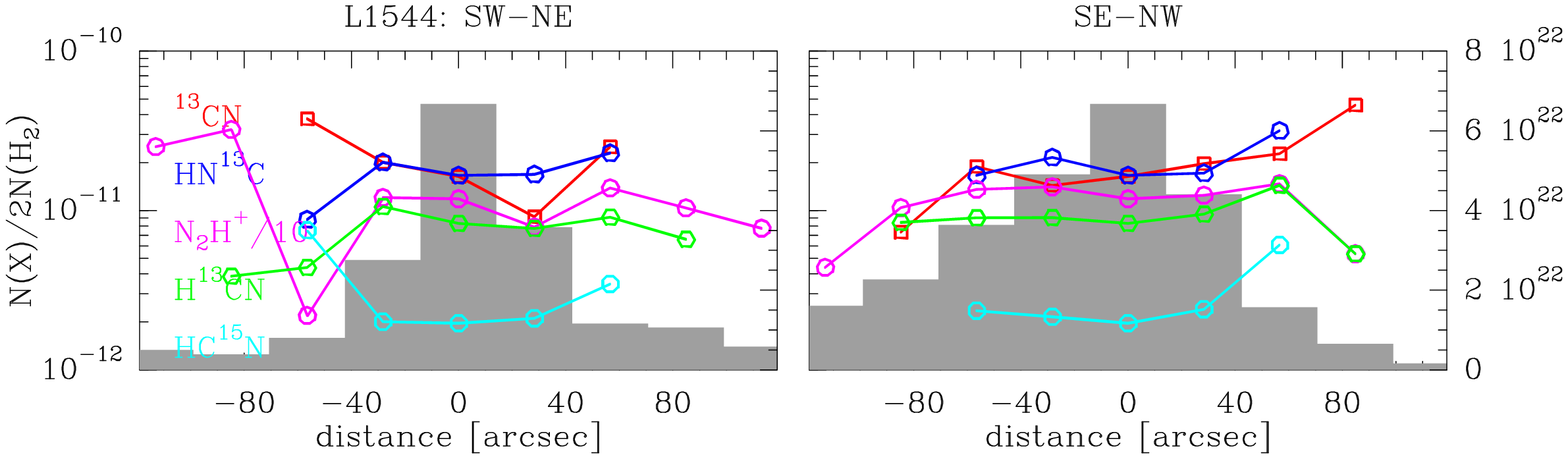}\bigskip\\
  \includegraphics[height=\wa,angle=-90]{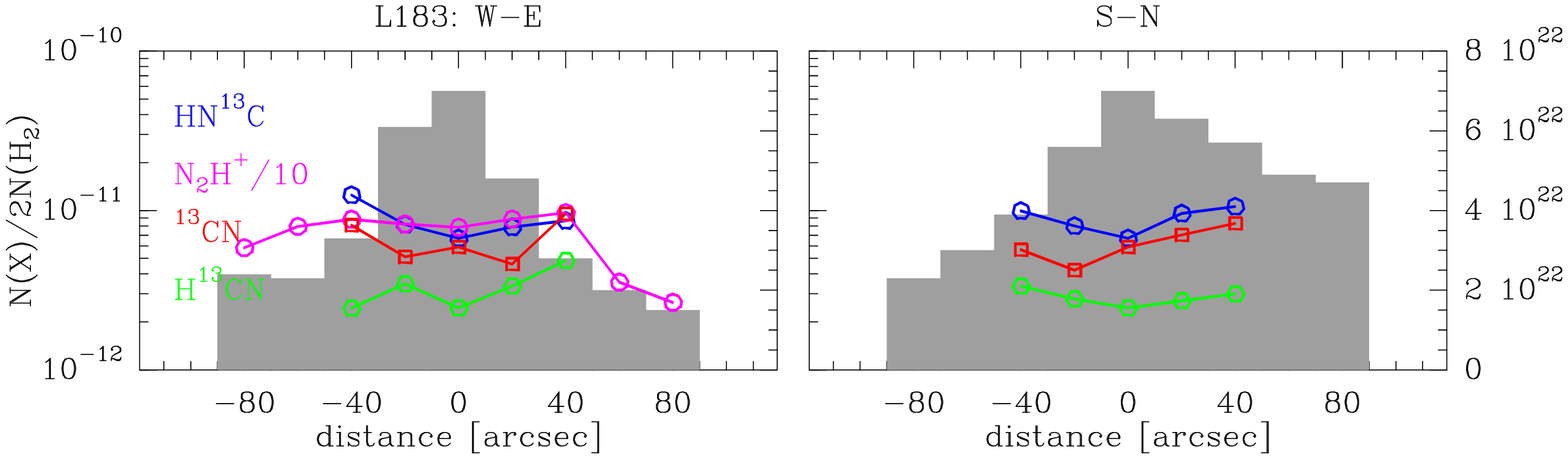}\bigskip\\
  \includegraphics[height=\wa,angle=-90]{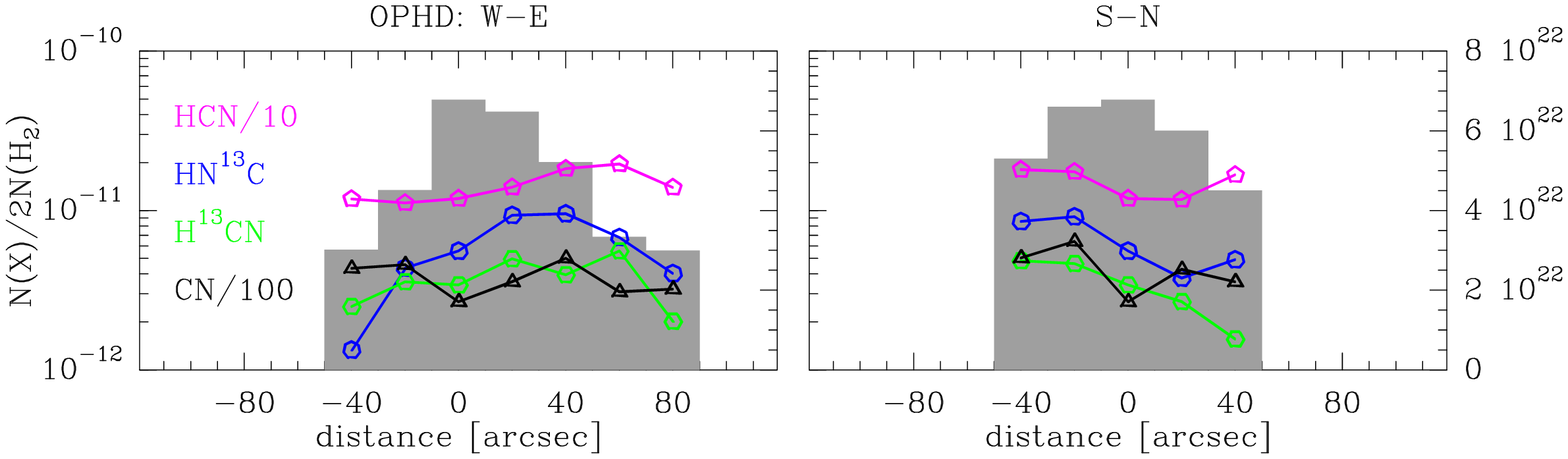}\bigskip\\
  \includegraphics[height=\wa,angle=-90]{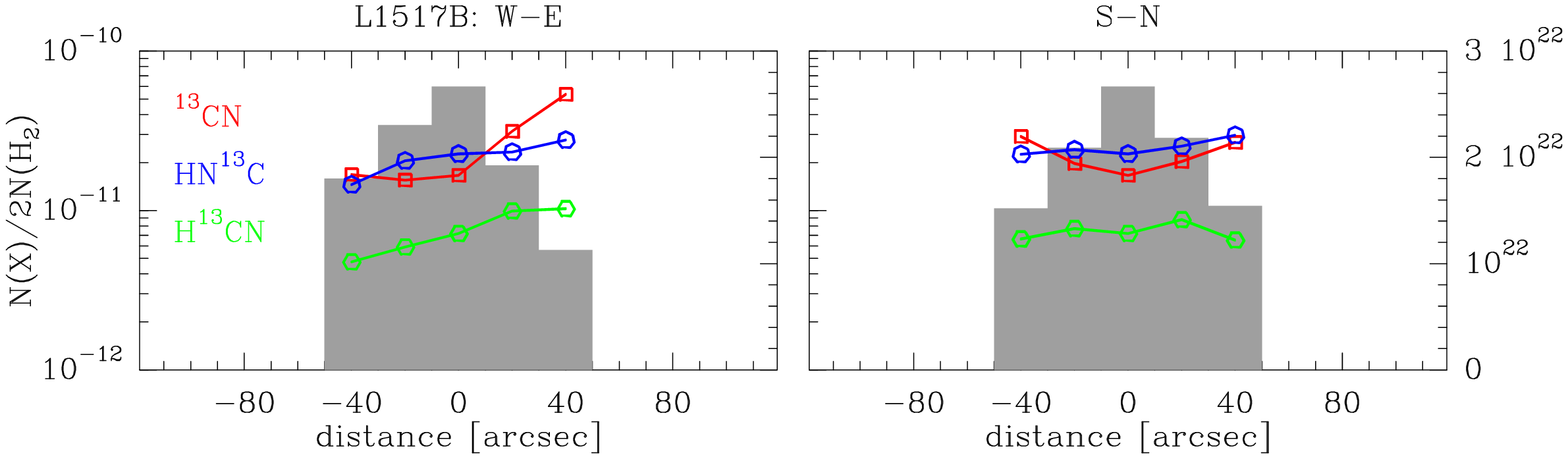}
  \caption{Same as Fig.~\ref{fig:cdens} for the derived fractional
    abundances.}
  \label{fig:xab}
\end{figure*}

\section{Column densities and abundance ratios}
\label{sec:cdens}

We determine column densities using the standard formalism described
in Appendix~\ref{app:cdens} (Eq.~\ref{eq:cdens1} and \ref{eq:N0}) and
assume the local (solar neighbourhood) $\rm ^{12}C:^{13}$C$ = 68$
ratio. It is instructive to consider also the abundance variations
from source to source. Converting column densities into relative
abundances requires the molecular hydrogen column density, \Nhh ,
which we have determined (indirectly) from measurements of the dust
emission, using bolometer maps available in the literature
\citep{ward1999, pagani2003, tafalla2004, bacmann2000} and smoothing
where necessary to a 20\arcsec\ beam. We assumed a dust temperature of
8~K and a 1.3~mm absorption coefficient of $\kappa=0.01~\rm
cm^2\,g^{-1}$ (see HWFP08). The results are shown in
Fig.~\ref{fig:xab} for L1544, L~183, Oph~D, and L~1517B.

We remark first that, towards L~1544 , the column densities of \thcn,
H\thcn, and \hnthc\ are all roughly proportional to the hydrogen
column density, as inferred from dust emission (see
Fig.~\ref{fig:cdens}); this has been noted already by HWPF08 for the
case of \thcn. The derived abundances do not change appreciably
towards the dust peak, in spite of a variation of almost an order of
magnitude in the hydrogen column density. Thus, in this source, and
with the current resolution, the CN--containing species do not appear
to be significantly depleted at high densities. On the other hand, the
abundances tend to increase to the NW of the dust emission peak (see
the SE--NW cut); we assume that this is related to asymmetry of the
source. It is interesting that \nnhp\ behaves in similar
fashion. Important for the later discussion is the fact that the
abundance ratios \hnthc:H\thcn\ and \thcn:H\thcn\ are approximately
equal to 2 (with variations of up to a factor of 2); we assume that
this reflects the ratios HNC:HCN and CN:HCN, respectively.

However, L~1544 is not typical. Towards L~183, for example (see the EW
cuts in Fig.~\ref{fig:cdens}), the peak H\thcn\ and \hnthc\ column
densities are offset to the east, relative to the dust emission,
whereas \nnhp\ and \thcn\ appear to follow the dust emission. The
situation is similar in Oph~D although we did not observe \thcn\ in
this source. Towards L~1517B, \hnthc\ correlates reasonably well with
$N(\hh)$ but this is not the case of \thcn\ nor, probably, of
H\thcn. Bearing in mind the inaccuracy of the abundance
determinations, and the possibility of $^{13}$C fractionation, we
conclude conservatively that there is no evidence for an order of
magnitude variation in the CN:HCN nor the HNC:HCN abundance ratios
between the dust emission peak and offset positions.

We give also in Table~\ref{tab:peakab} our estimates of the fractional
abundances, relative to H, of CN, HCN, and HNC towards the dust
emission peaks of our sample of sources; these abundances have been
derived assuming the local value of 68 for the $^{12}$C:$^{13}$C
ratio.  All the relative abundances are of order $\dix{-9}$, with CN
and HNC being more abundant than HCN by a factor of about 2. This
value is close to the ratio determined by \cite{irvine1984} toward
TMC-1. \corr{}{Values larger than 1 for the HNC:HCN abundance ratio
  were found towards a sample of 19 dark clouds by \cite{hirota1998}
  with an average ratio of $2.1\pm1.2$.} We do not see indications of
significant abundance differences between cores of high central
density (L~1544 and L~183) and cores of lower central density
(L~1517B, Oph~D). Whilst the complexities of the source structure and
of radiation transfer prevent our establishing the existence of small
abundance differences, we may conclude that there remain appreciable
amounts of CN, HCN, and HNC at densities above the typical density
(3\tdix{4}~\ccc) at which CO depletes on to grains.  Not surprisingly,
this effect is seen most readily in sources of high central column
density, like L~1544, in which emission from the low density envelope
is less important.

\section{Chemical considerations}
\label{chemistry}
\def\ch{\ensuremath{{\rm CH}}}
\def\chh{\ensuremath{{\rm CH_2}}}

In this Section, we seek to update and extend previous studies of the
interstellar chemistry of N--containing species \cite{pineau1990,
schilke1992}, with a view to providing a framework for the
interpretation of our observations of prestellar cores. We shall show
that it is possible to derive a simple expression for the CN:HCN
abundance ratio, in particular, by identifying the principal reactions
involved in the formation and destruction of these species.

\subsection{Main chemical reactions}
\label{reactions}

The fractional abundances of gas--phase species in prestellar cores
are determined by:
\begin{itemize}
\item the initial composition of the molecular gas which undergoes
  gravitational collapse;
\item \corr{}{variations of the density with time;}
\item the rates of gas--phase reactions at the low temperatures ($T
  \approx 10$~K) of the cores;
\item the rates of adsorption of the constituents of the gas on to
  grains.
\end{itemize}

The assumption of steady state, when computing the initial abundances,
is not crucial when dealing with species which are produced in
ion--neutral reactions. The timescales associated with ion--neutral
reactions are much smaller, in general, than dynamical timescales,
notably the gravitational free--fall time. Consequently, the evolution
of the fractional abundances of species which are formed in such
reactions soon becomes independent of the initial values. However, it
is believed that the timescales for producing N--containing species,
such as NH$_3$ and CN, are determined by slower neutral--neutral
reactions. It follows that their fractional abundances in the
subsequent gravitational collapse may depend significantly on the
initial values.

The conversion of atomic into molecular nitrogen in the gas phase is
believed to occur in the reactions
\begin{equation}
  \ce{N + OH -> NO + H}\label{equ1}
\end{equation}
\begin{equation}
  \ce{N + NO -> N2 + O}\label{equ2}
\end{equation}
and
\begin{equation}
  \ce{N + CH -> CN + H}\label{equ3}
\end{equation}
\begin{equation}
  \ce{N + CN -> N2 + C}\label{equ4}
\end{equation}

\corr{However, the reactions}{The reactions}
\begin{equation}
  \ce{C + NO -> CN + O}\label{equ5}
\end{equation}
\begin{equation}
  \ce{C + NO -> CO + N}\label{equ6}
\end{equation}
can also destroy NO, producing CN in the case of \cref{5}, and
\begin{equation}
  \ce{CN + O -> CO + N}\label{equ6bis}
\end{equation}
can destroy CN.

\corr{}{From the above, we see that NO forms from the reaction of N
  with OH, whereas CN forms from N and CH. It follows that the ratio
  of carbon to oxygen in the gas phase is a factor determining the
  relative abundance of NO and CN. The NO:CN abundance ratio will be
  lower in gas which is depleted of oxygen, either because of an
  intrinsically high C:O elemental abundance ratio, or due to the
  differential freeze--out of O and C on to the grains, where the
  oxygen is incorporated mainly as water ice.}

Once N$_2$ has formed, in \cref{2} and \cref{4}, N$_2$H$^+$ is
produced in the protonation reaction
\begin{equation}
  \ce{N2 + H3+ -> N2H+ + H2}\label{equ7}
\end{equation}
Dissociative ionization of N$_2$ by He$^+$ results in the production
of N$^+$:
\begin{equation}
  \ce{N2 + He+ -> N+ + N + He}\label{equ8}
\end{equation}

Whilst the reaction of N$^+$ with para--H$_2$ (in its ground
rotational state) is endothermic, by approximately 168~K, its reaction
with ortho--H$_2$ is slightly exothermic and occurs even at low
temperatures \cite{lebourlot1991}. Subsequent hydrogenation reactions
with H$_2$ lead to NH$_4^+$, which can dissociatively recombine to
produce NH$_3$. Thus, N$_2$ is a progenitor of both N$_2$H$^+$ and
NH$_3$, whilst NO and CN are intermediaries in the formation of
N$_2$.

HCN and HNC are produced principally in the reactions
\begin{equation}
  \ce{CH2 + N -> HCN + H}  \label{equ9}
\end{equation}
\begin{equation}
  \ce{NH2 + C -> HNC + H}  \label{equ10}
\end{equation}
in which the products have so much excess energy that rapid
isomerization is expected to yield practically equal amounts of HCN
and HNC \citep{herbst2000}. As CH and CH$_2$ are produced through the
dissociative recombination of hydrocarbon ions, notably CH$_3^+$, they
are expected to coexist in the medium. It follows that CN, HCN and HNC
should coexist also. HNC converts to HCN in the reaction
\begin{equation}
  \ce{HNC + H+ -> HCN + H+}  \label{equ11}
\end{equation}
The reverse reaction is endothermic and proceeds at a negligible rate
at low temperatures. HCN is destroyed principally in the charge
transfer reaction with H$^+$ \corr{}{and in the proton transfer
  reaction with \ce{H3+}}
\begin{equation}
  \ce{HCN + H+ -> HCN+ + H}  \label{equ12}
\end{equation}
\begin{equation}
  \ce{HCN + H3+ -> H2CN+ + H2}  \label{equ12b}
\end{equation}
\corr{and also in the proton transfer reaction with H$_3^+$, which
  yields H$_2$CN$^+$.}{} HCN$^+$ reacts rapidly with H$_2$, producing
H$_2$CN$^+$, which dissociatively recombines with electrons, producing
HCN and HNC; but there exists a branch to CN
\begin{equation}
  \ce{H2CN + e- -> CN + H2}  \label{equ13}
\end{equation}
for which the branching ratio $f_{\rm CN}$ was taken equal to 1/3, and
this represents a true destruction channel for HCN (and formation
channel for CN).

\begin{figure}[t]
  \centering
  \includegraphics[width=0.8\hsize]{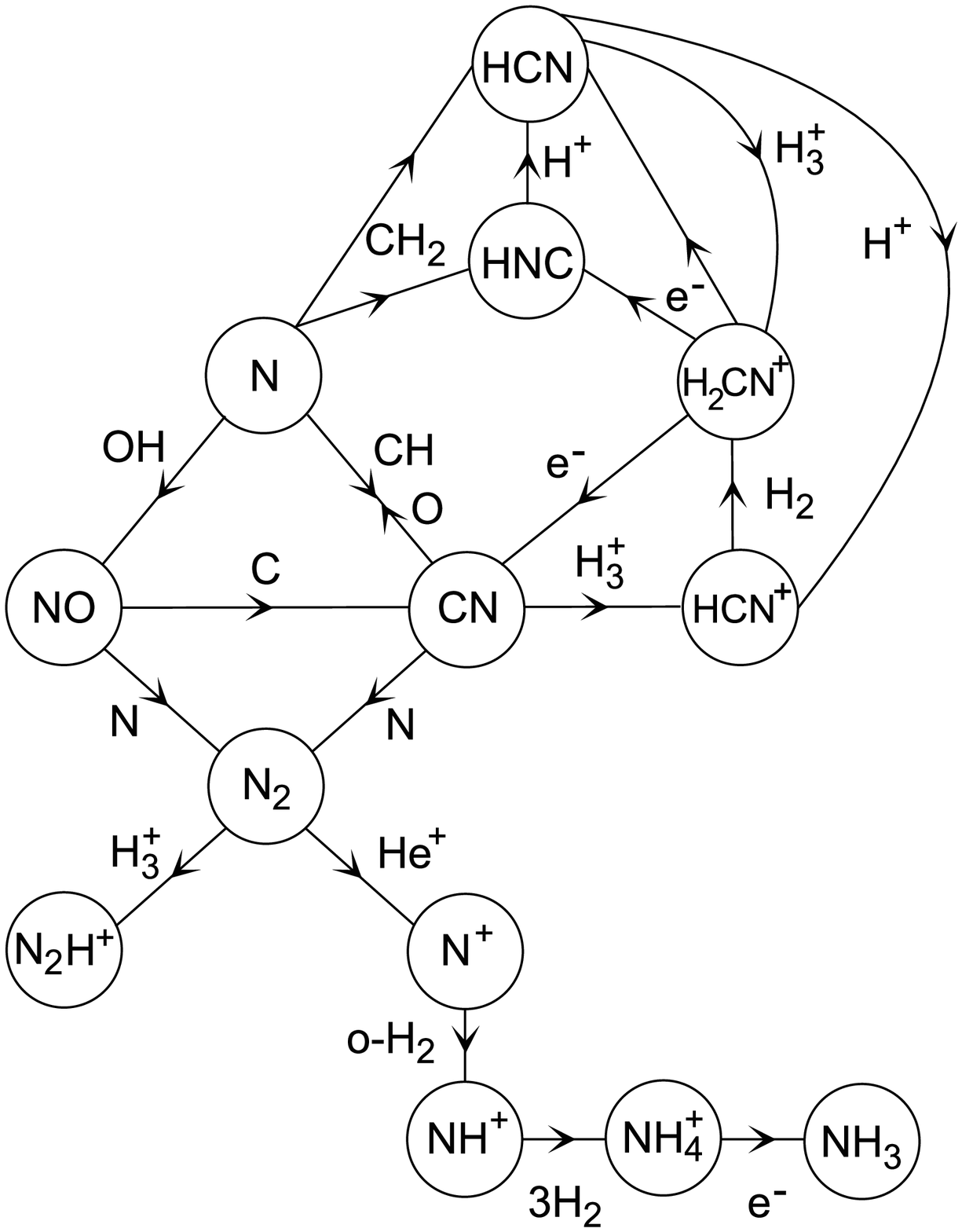}
  \caption{The principal reactions involved in the low--temperature
    chemistry of nitrogen--containing species, and specifically of CN,
    HCN and HNC.}
  \label{fig:network}
\end{figure}

\corr{}{Because CN but not HCN (nor HNC) is destroyed by O, the
  abundance ratio CN:HCN increases as the C:O ratio increases.  The
  chemical network is summarized in Fig.~\ref{fig:network}.}

\subsection{Simplified analysis of the chemistry}
\label{analysis}

We show now that an analysis of the principal reactions leading to the
formation and destruction of CN and HCN leads to a simple analytical
formula for the CN:HCN abundance ratio. \corr{As noted above,}{} When
reactions \cref{3} and \cref{4} determine the abundance of
CN, $$n({\rm CN}) = n({\rm CH}) k_{\ref{equ3}}/k_{\ref{equ4}} =
0.75n({\rm CH})$$ with the adopted values of the rate coefficients for
these reactions at $T = 10$~K.

We have seen already that HCN is formed in reaction \cref{9}, with a
rate coefficient $k_{\ref{equ9}} = 2.2\times \dix{-11}$~\cccs, and it
is removed principally by H$^+$ and H$_3^+$, forming H$_2$CN$^+$
(reactions \ref{equ12} and \ref{equ12b}). \corr{A fraction $f_{\rm
    CN} = 1/3$ of the dissociative recombinations of H$_2$CN$^+$ yield
  CN, and this is the true HCN destruction channel.}{} It follows
that $$k_{\ref{equ9}}n({\rm CH}_2)n({\rm N}) = k_{\rm L}n({\rm
  HCN})[n({\rm H}^+)+n({\rm H}_3^+)]f_{\rm CN},$$ where $k_{\rm L}$ is
the Langevin rate coefficient which characterizes the reactions of HCN
with H$^+$ and H$_3^+$.

\corr{}{Under the above assumptions, the CN:HCN ratio is proportional
  to the CH:\ce{CH2} ratio:
  $$ \frac {n({\rm CN})}{n({\rm HCN})} = \frac
  {k_{\ref{equ3}}}{k_{\ref{equ4}}}\frac {n({\rm CH})}{n({\rm
      CH}_2)}f_{\rm CN}\frac {k_{\rm L}n(\elec)}{k_{\ref{equ9}}n({\rm
      N})} \approx 100\frac {n({\rm CH})}{n({\rm CH}_2)}\frac
  {n(\elec)}{n({\rm N})}$$ where we have assumed that $n(\elec) =
  n(\ce{H+})+n(\ce{H3+})$.}

CH has its chemical origin in the reaction
of He$^+$ with CO
\begin{equation}
  \ce{He+ + CO -> C+ + O + He} \label{equ13a}
\end{equation}
for which the adopted value of the rate coefficient is
$k_{\ref{equ13a}} = 1.5\times \dix{-9}$~\cccs\ (essentially
Langevin). As CO depletes on to grains, there is competition with
reaction \cref{13a} from the reactions of He$^+$ with H$_2$,
\begin{equation}
  \ce{He+ + H2 -> H+ + H + He}\label{equ13b}
\end{equation}
\begin{equation*}
  \ce{He+ + H2 -> H2+ + He}
\end{equation*}
$k_{\ref{equ13b}} = 4.0\tdix{-14}$~\cccs\ \citep{schauer1989}. Thus,
the rate of production of carbon ions is $\zeta _{\rm He}n(\rm
{He})f_{\rm C}$, where $\zeta _{\rm He}=0.5\zeta _{{\rm H}_2}$ is the
rate of cosmic ray ionization of He and $f_{\rm C} =
k_{\ref{equ13a}}n({\rm CO})/[k_{\ref{equ13b}}n({\rm H}_2) +
  k_{\ref{equ13a}}n({\rm CO})].$

Most of the C$^+$ ions produced in reaction~\cref{13a} combine
radiatively with H$_2$ to form CH$_2^+$, which reacts rapidly with
H$_2$, forming CH$_3^+$ which then recombines dissociatively with
electrons, yielding CH
\begin{equation}
  \ce{CH3+ + e- -> CH + H2}  \label{equ13c}
\end{equation}
\begin{equation}
  \ce{CH3+ + e- -> CH + H +H}  \label{equ13d}
\end{equation}
but also C and CH$_2$. The ratio $f_{\rm CH} = 0.3$ is the fraction of
the dissociative recombinations of CH$_3^+$ which form CH and is
significant for the CN:HCN abundance ratio. CH and \ce{CH2} are
destroyed by atomic O:
\begin{equation}
  \ce{CH + O -> CO H}  \label{equ14}
\end{equation}
\begin{equation}
  \ce{CH2 + O -> CO + H + H}   \label{equ15}
\end{equation}
and
\begin{equation}
  \ce{CH2 + O -> CO + H2}    \label{equ16}
\end{equation}
forming CO, and hence the elemental C:O ratio is, once again, a
pertinent parameter for the CN:HCN ratio. So too are the rate
coefficients for reactions~\cref{14}, \cref{15} and \cref{16}. We
adopted the (temperature--independent) values of these rate
coefficients in the NIST chemical kinetics
database\footnote{\texttt{http://kinetics.nist.gov/kinetics/}}. All
three reactions are rapid, with rate coefficients of the order of
\dix{-10}\,\cccs. The number density of CH is predicted to be $n({\rm
  CH}) = {\zeta _{\rm He}n({\rm He})f_{\rm C}f_{\rm
    CH}}/[{k_{\ref{equ3}}n({\rm N})+k_{\ref{equ14}}n({\rm O})}]$,
Similarly that of CH$_{2}$ is predicted to be $ n({\rm CH}_{2}) =
{\zeta_{\rm He}n({\rm He})\, f_{\rm C}f_{\rm
    CH_2}/[k_{\ref{equ9}}n({\rm
      N})+(k_{\ref{equ15}}+k_{\ref{equ16}})n({\rm O})}]$.
\corr{At the same level of accuracy,}{Assuming that
  $k_{\ref{equ15}}=k_{\ref{equ16}}$, one can deduce from the above
  equations that the CH:\ce{CH2} ratio is given by}
$$
\frac{n({\rm CH})}{n({\rm CH_{2}})} \, = \, \frac{f_{\rm CH}}{f_{\rm
    CH_{2}}}\, \frac{2k_{\ref{equ15}}n({\rm O})+k_{\ref{equ9}}n({\rm
    N})}{k_{\ref{equ14}}n({\rm O})+k_{\ref{equ3}}n({\rm N})}
$$

It follows that $n(\ch)/n(\chh)$ depends on the ratio of the atomic N
and O abundances and varies between extreme values of approximately 2,
when $n({\rm O}) \gg n({\rm N})$, and 0.2, when $n({\rm N}) \gg n({\rm
  O})$. \corr{}{Model calculations (see Section~\ref{sec:gravcoll})
  confirmed that the above approximation reproduces the calculated
  CN:HCN abundance ratio to within a factor of approximately 2 over
  most of the density range $\dix{4} < \nh < \dix{6}$~\ccc.} In the
following Section~\ref{atomicN}, we consider the issue of the atomic
nitrogen abundance in prestellar cores in more detail.




\subsection{The abundance of atomic nitrogen in cores}
\label{atomicN}

Atomic abundances in prestellar cores are notoriously difficult to
determine. Although the atomic fine structure transitions are, in
principle, observable, it is difficult, in practice, to distinguish a
component corresponding to dense, cold molecular material from
emission arising from low density, hotter layers along the line of
sight. The emission from photon dominated regions (PDRs), for example,
tends to be stronger than that from cores.

The analysis in Section~\ref{analysis} suggests that the CH:CH$_2$
ratio is dependent on the abundances of atomic oxygen and
nitrogen. Combining the expressions for the CH:CH$_2$ and the CN:HCN
ratios, \corr{and assuming that $n({\rm N}) \gg n({\rm O})$,}{} there
follows the inequality
$$ x(\ce{N}) = n(\ce{N})/n(\ce{H2}) \leq 100 \frac {n(\ch)}{n(\chh)}
\frac {x({\rm e})}{R({\rm CN:HCN})},$$ where $x({\rm e})$ denotes
$n({\rm e})/n({\rm H}_{2})$ and $R({\rm CN:HCN})$ is the measured
value of the CN:HCN ratio. The inequality arises from neglecting
\corr{the term involving atomic oxygen}{the destruction of CN by
  oxygen in reaction \cref{6bis}}. {We adopt} $n({\rm CH})/n({\rm
  CH}_2) \approx 1$ and {estimate} $x({\rm e})$ from Fig.~5 of
\cite{walmsley2004} as $5\times 10^{-8} n_{5}^{-0.5}$, where $n_{5}$
is $n(\hh)$ in units of $10^5$~\ccc. From Table~\ref{tab:peakab}, we
infer a typical value of 2 for $R({\rm CN:HCN})$ and conclude that the
fractional atomic N abundance
$$n(\ce{N})/\nh \lesssim 1.25\tdix{-6}.$$ This value is considerably
less than the cosmic nitrogen abundance \citep[$n(\ce{N})/\nh =
  7.9\tdix{-5}$;][]{anders1989} and implies that there is only a small
fraction of elemental nitrogen in gaseous atomic form. Although some
approximations were made when deriving this value (such as $n({\rm
  CH})/n({\rm CH}_2) \approx 1$), our more precise and complete
chemical modelling has confirmed that this estimate is essentially
correct as long as the total gaseous nitrogen abundance is
$n(\ce{N})/\nh \le \dix{-5}$ (see Fig.~\ref{fig:fig7}). It
follows that most of the elemental nitrogen must be in the form of
gaseous N$_{2}$ or N-bearing solid compounds (\eg\ NH$_{3}$ or N$_{2}$
ices). Indeed, for $n(\ce{N})/\nh \le \dix{-5}$, steady--state
models suggest that most of the nitrogen is in ices (see
Section~\ref{models} as well as the discussion of Maret et al. 2006).

An analogous inequality, in terms of the measured CN:HCN ratio, can be
derived for the fractional abundance of atomic oxygen, $x({\rm
  O})$. However, it is a much weaker constraint than the limit on
$x({\rm N})$. Other observables, such as NO, provide stronger
constraints on the fractional abundance of atomic oxygen (see A07).

\section{Models}
\label{models}

\subsection{Steady state}
\label{steadystate}

\begin{figure}
  \begin{center}
    \includegraphics[width=0.7\hsize]{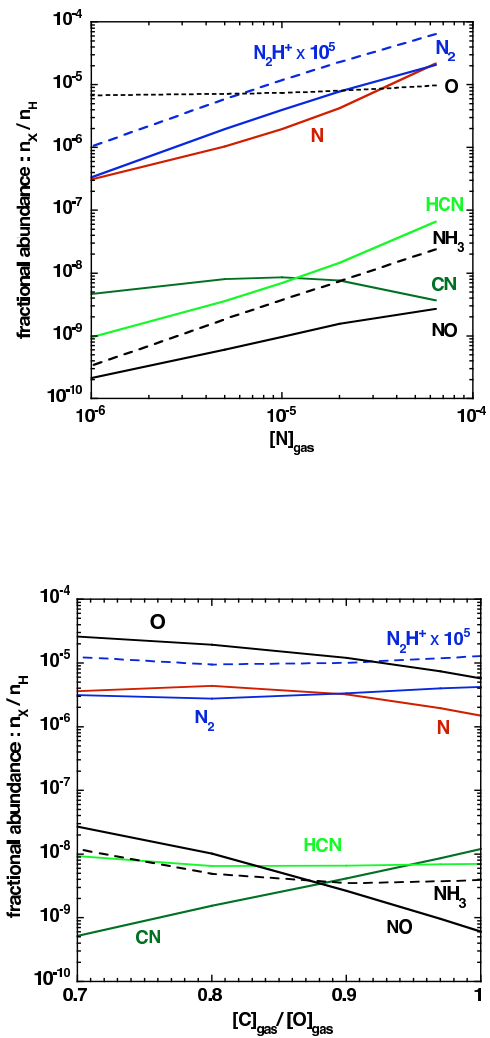}
    \caption{Steady--state fractional abundances of
      nitrogen--containing species for a density $\nh=10^4$~\ccc, a
      kinetic temperature $T = 10$~K, and a cosmic ray ionization rate
      of $\zeta = 10^{-17}$~s$^{-1}$. \textit{Upper panel}: C:O=0.97
      is held constant and the gas--phase nitrogen fractional
      abundance $\rm [N]_{gas}=n(\ce{N})/\nh$ is varied from \dix{-6}
      to 6.4\tdix{-5}.  \textit{Lower panel}: $\rm [N]_{gas}= 10^{-5}$
      is held constant and the gas--phase C:O ratio is varied by
      changing $\rm [O]_{gas}$. \colfig}
    \label{fig:fig7}
  \end{center}
\end{figure}

The timescale for the nitrogen chemistry to reach steady state is
known to be large, relative to the free--fall time in a prestellar
core, owing to the slow conversion of N to N$_2$ \citep[see, for
  example,][]{flower2006}. Our model calculations show that, with a
cosmic ray ionization rate $\zeta = 10^{-17}$~s$^{-1}$, this timescale
is of the order of $10^{6}$~yr. As a consequence, the results of the
time--dependent models of gravitational collapse depend on the initial
composition which is adopted and the rate of dynamical evolution.

Following the discussion in Section~\ref{chemistry}, it is
nevertheless instructive to examine the results of steady--state
calculations \corr{}{(that is to say time independent and only in the
  gas phase)}, as functions of the fractions of elemental oxygen and
nitrogen in the gas phase. In this way, an impression may be obtained
of the dependence of the observables on the degrees of depletion,
without the complications of the time dependence, which is considered
in Section~\ref{sec:gravcoll}.

In the upper panel of Fig.~\ref{fig:fig7}, we present the fractional
abundances of nitrogen--containing species in steady state for a
density $\nh=10^4$~\ccc, a kinetic temperature $T = 10$~K, and a
cosmic ray ionization rate of $\zeta = 10^{-17}$~s$^{-1}$. \corr{}{For
  reaction~\cref{6bis}, we adopted a temperature--independent rate
  coefficient\footnote{From the \texttt{osu\_03\_2008} rates of Eric
    Herbst's group
    (\texttt{http://www.physics.ohio-state.edu/{\~{}}eric})} ($4\times
  \dix{-11}\,\cccs$) although we note that there is some theoretical
  evidence that the rate of this reaction may decrease with
  temperature \citep{andersson2003}.}  The fraction of elemental
nitrogen in the gas phase varies from 0.017 to 1. We assume implicitly
that the `missing' nitrogen is in solid form. Following A07, we adopt
a relative abundance of elemental carbon to oxygen in the gas phase
$\rm C:O = 0.97$. In the lower panel, the fractional abundance of
elemental nitrogen in the gas phase is held constant at $n(\ce{N})/\nh
= 10^{-5}$ and the C:O ratio is varied by changing the fractional
abundance of oxygen in the gas phase $\rm [O]_{gas}$.

We see from Fig.~\ref{fig:fig7} that, in steady state, there tends to
be somewhat more molecular than atomic nitrogen in the gas
phase. Species such as NH$_3$ and \nnhp\ have abundances which are
roughly proportional to N$_{2}$. The HCN abundance is relatively
insensitive to changes in the gas phase C:O ratio but follows the
gas-phase nitrogen abundance. On the other hand, CN and NO are
sensitive to the C:O ratio. The net effect is that CN:HCN increases
with C:O and decreases with the fraction of nitrogen in the gas phase.

\begin{figure*}
  \begin{center}
    \includegraphics[width=0.43\hsize]{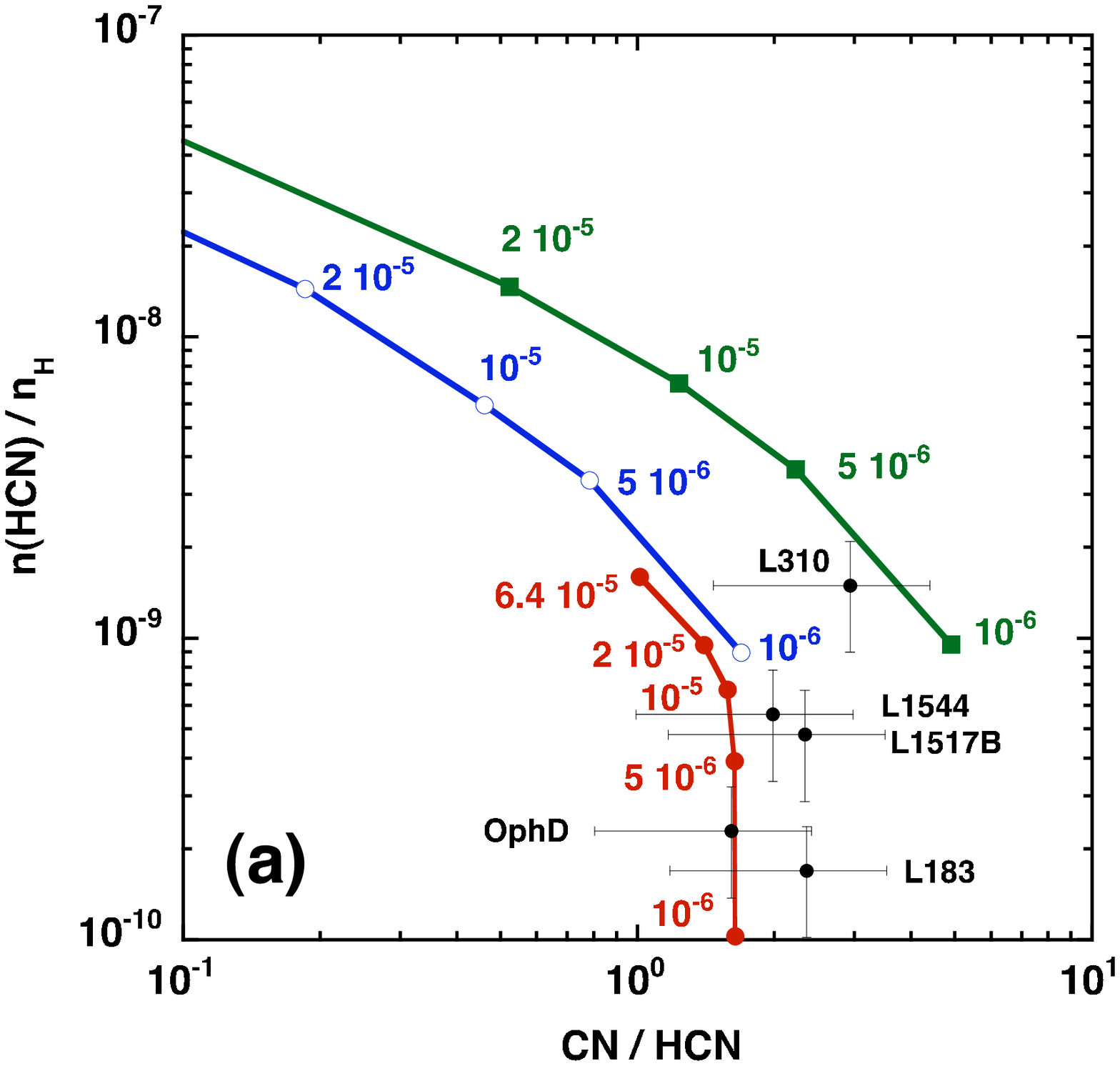}\hfill%
    \includegraphics[width=0.43\hsize]{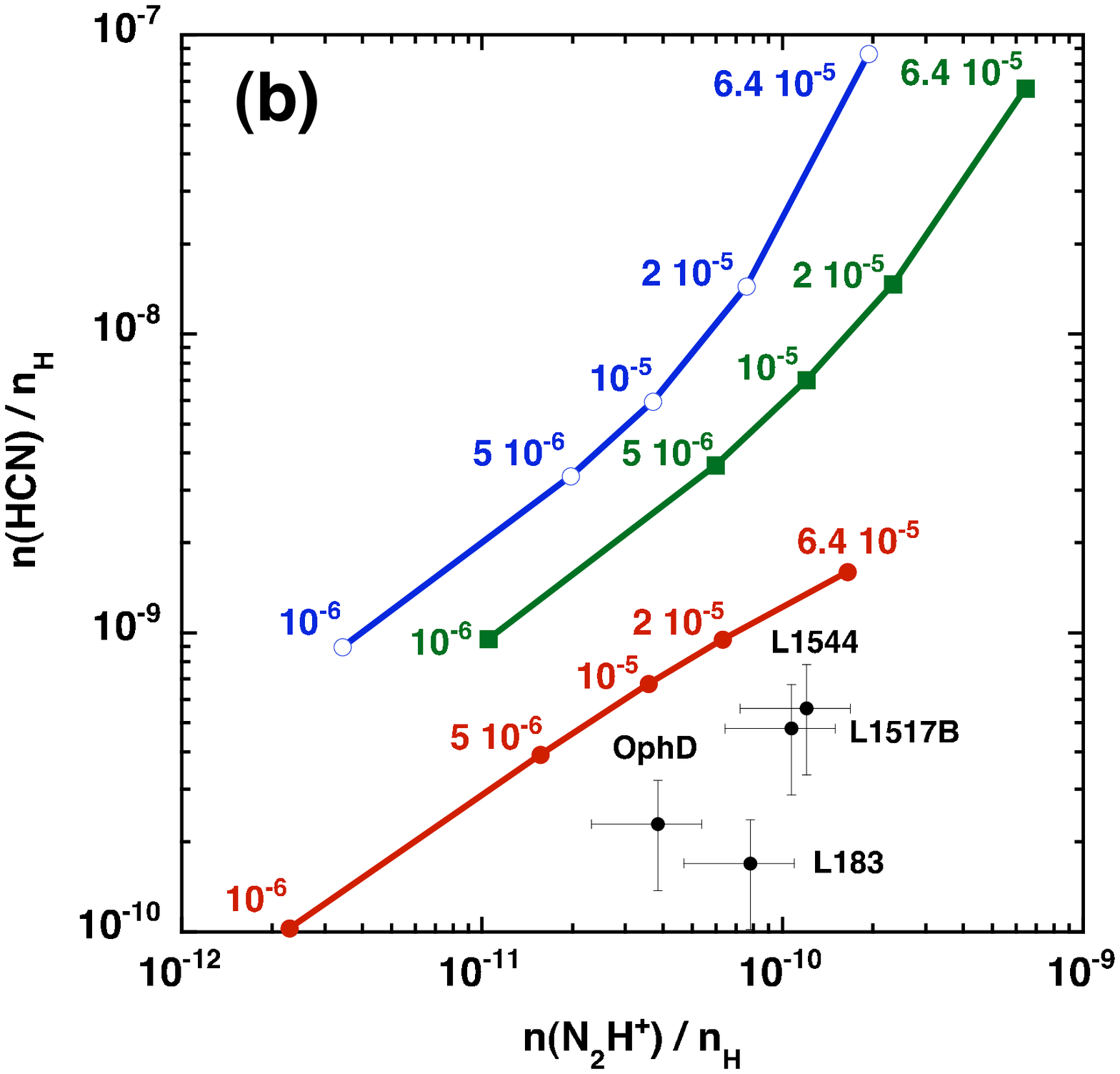}\bigskip\\
    \includegraphics[width=0.43\hsize]{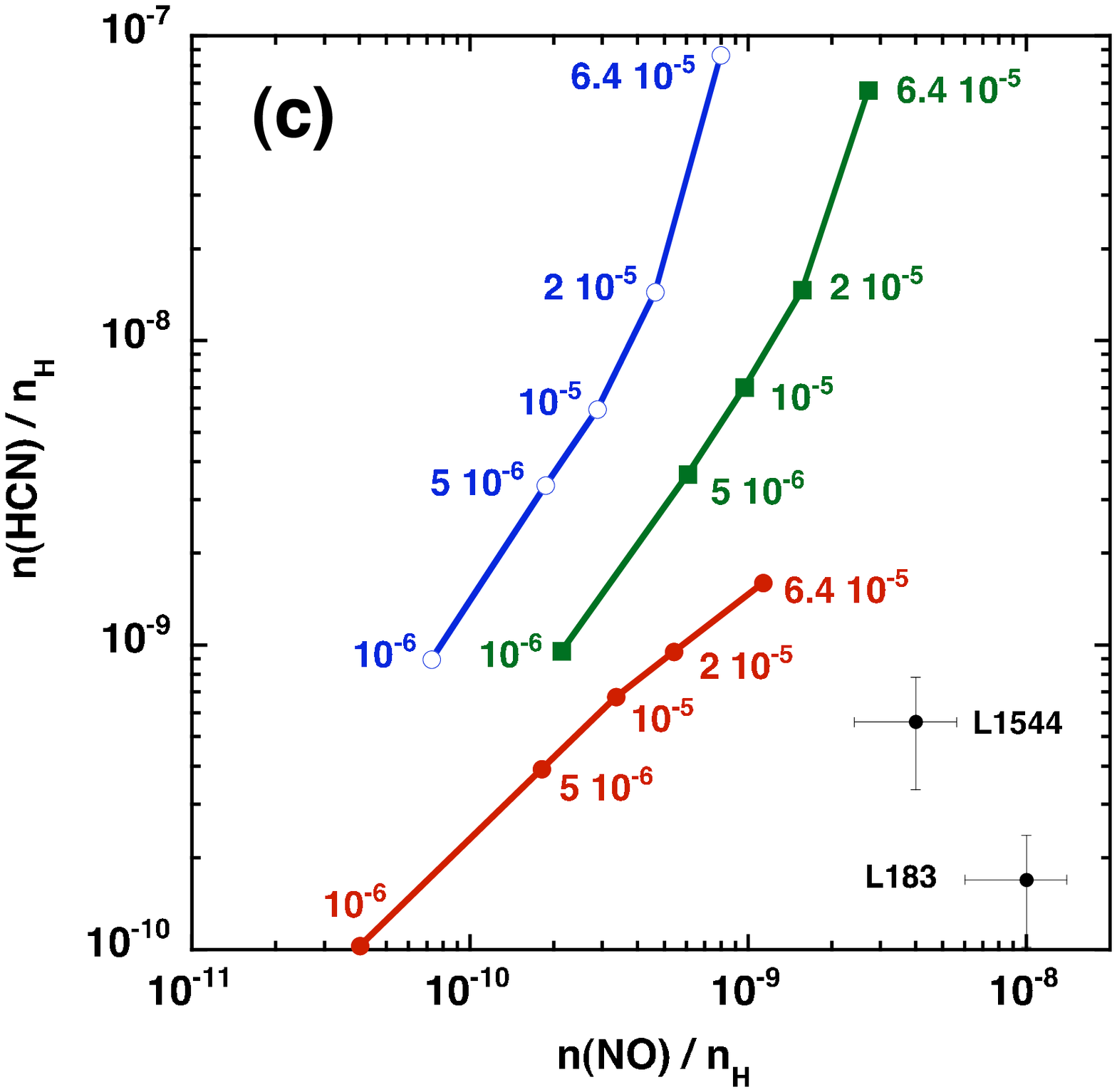}\hfill%
    \includegraphics[width=0.4\hsize]{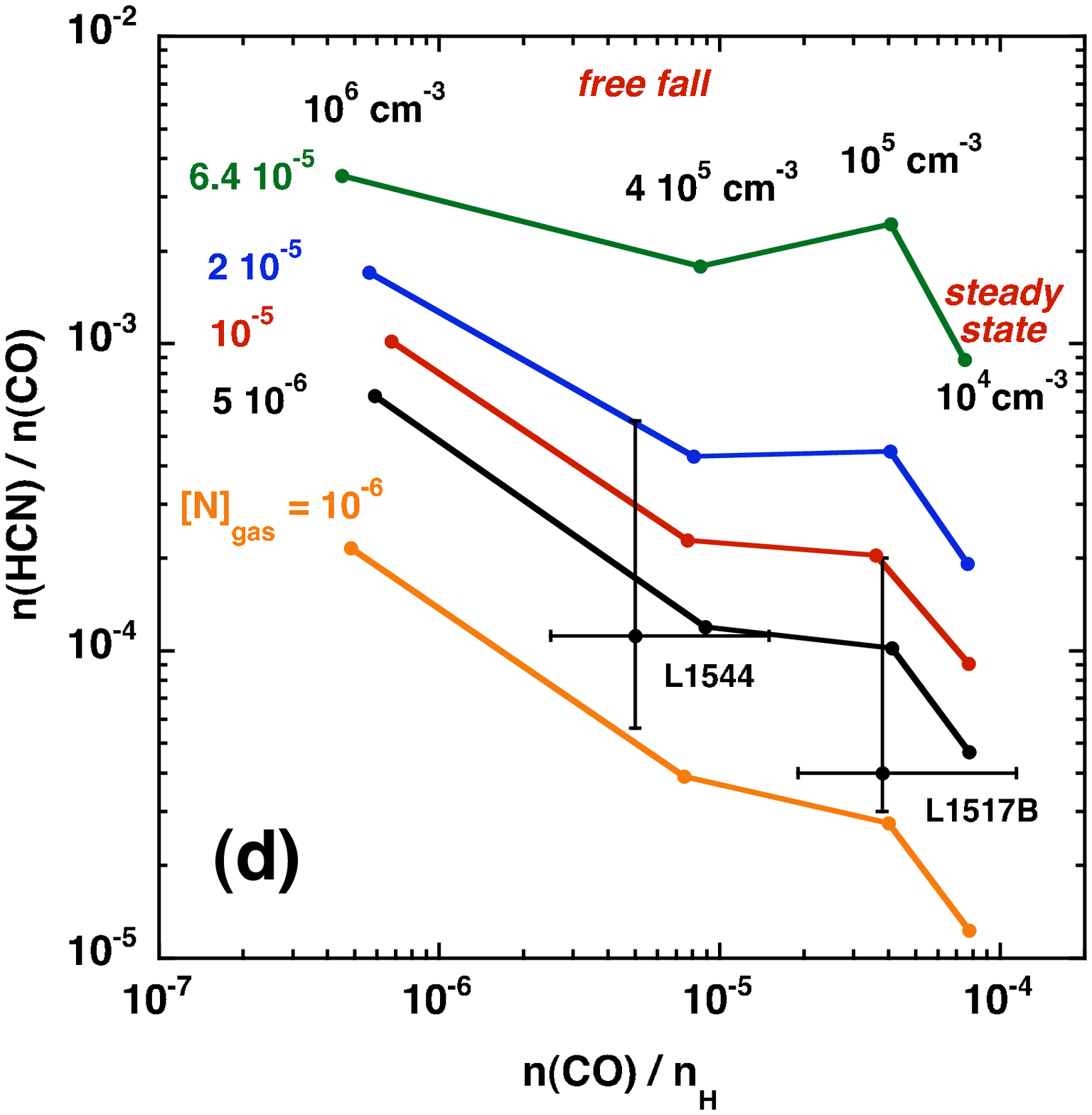}
    \caption{The fractional abundances computed by the gravitational
      collapse model. We show in panel (a) results for the HCN
      abundance as a function of the CN:HCN ratio. In panel (b) we
      show the HCN abundance as a function of the \ce{N2H+}
      abundance. In panel (c) the HCN abundance as a function of the
      NO abundance. The initial C:O ratio is assumed to be 0.97. Each
      point is labeled with the initial gas phase nitrogen abundance
      from \dix{-6} to 6.4\tdix{-5}. The initial steady state values
      (filled squares, green) are shown for comparison. Results for a
      density of $10^5$~\ccc\ are given as blue open circles and for
      $10^6$~\ccc\ as filled red circles. Data points are shown with
      black error bars. Data for \ce{N2H+} towards Oph~D and L~1517B
      are taken from \cite{crapsi2005}. \corr{}{In panel (d) we show
        the HCN:CO abundance ratio as a function of the CO fractional
        abundance, for different initial gas phase nitrogen
        abundances. For a given $\rm [N]_{gas}$, the ratios are
        followed during the collapse (snapshots at densities $10^5$,
        4\tdix{5} and $10^6$~\ccc\ are shown) starting from the
        initial steady state. The CO abundance was taken from
        \cite{caselli1999} for for L~1544, and for L~1517B both the
        HCN and CO abundances were taken from \cite{tafalla2006}.}}
    \label{fig:fig8}
  \end{center}
\end{figure*}

\subsection{Gravitational collapse}
\label{sec:gravcoll}

A more satisfactory approach to comparing observational results with
models is through a simulation of a gravitational collapse. In
Fig.~\ref{fig:fig8}, we compare the observed values of the HCN
abundance and the CN:HCN ratio with the predictions of models in which
the density and the chemistry evolve following free--fall
gravitational collapse. \corr{}{All neutral species are assumed to
  adsorb on to dust grains (of radius 0.5~\micr) with a sticking
  coefficient of unity and are desorbed by cosmic ray impacts
  \citep{flower2006}.} We assume that, initially, the chemical
composition of the gas has attained steady state at a density $\nh =
\dix{4}$~\ccc. We make various assumptions concerning the amount of
nitrogen initially in the gas phase (or, equivalently, the fraction
which is initially in the form of nitrogen--containing ices on grain
surfaces).

The fraction of elemental nitrogen in the ambient molecular medium
which is in solid form is poorly known. There is evidence for ammonia
ice in spectral profiles observed towards some young YSOs, with
perhaps 15\% of the abundance of water ice \citep{gibb2000}, but no
such evidence exists towards background stars; there is perhaps a
substantial fraction of the nitrogen in the form of N$_2$ ice also. In
prestellar cores, the abundance of \nnhp\ places lower limits on the
amount of gas--phase nitrogen, which we estimate conservatively to be
about $10^{-6}$. Accordingly, we have varied the initial gas phase
nitrogen abundance in our models in the range {$10^{-6} \le
  n(\ce{N})/\nh \le 6.4\times 10^{-5}$}, where the upper limit
corresponds to the value observed in diffuse interstellar gas
\citep{sofia2001}. Fig.~\ref{fig:fig8} displays the fractional
abundances computed in the course of the collapse, at densities $\nh =
10^5$~\ccc\ and $\nh = 10^6$~\ccc. We show also, for comparison, the
initial (steady state) values.

It may be seen from Fig.~\ref{fig:fig8} that, at a density of
$10^5$~\ccc\, which is the more relevant value for the purpose of
comparing with observations, reasonable agreement is obtained only for
initial gas--phase nitrogen abundances close to $10^{-6}$ -- in other
words, close to the lower limit. Even so, the computed abundances do
not fit well the observations of L~183 and Oph~D; but we note that the
density at the dust peak in L~183 approaches $10^6$~\ccc. Our model
results are dependent also on the fraction of oxygen locked in ices,
or, equivalently, on the initial gas--phase C:O ratio. It is possible
that this ratio varies considerably from source to source, resulting
in discrepancies when we compare observations with model predictions.

Figure~\ref{fig:fig8} (bottom right panel) shows the HCN:CO abundance
ratio as a function of the CO fractional abundance. For a given
initial abundance of gaseous nitrogen $[\rm N]_{gas}$
($=n(\ce{N})/\nh$), the abundance ratio is followed along the collapse
and values are shown at densities $\nh=10^5$, 4\tdix{5} and
\dix{6}~\ccc. The differential freeze-out of HCN and CO is evident. In
all these models, CO depletes by two orders of magnitude. The
behaviour of HCN regarding depletion is different in that it depends
on the initial $\rm [N]_{gas}$. For a large initial $\rm
[N]_{gas}=6.4\tdix{-5}$, HCN depletes only a factor of 3 less than
CO. However, at the other extreme value ($\rm [N]_{gas}=\dix{-6}$),
HCN depletes 10 times less than CO. Observational values towards
L~1544 \citep{caselli1999} and L~1517B \citep{tafalla2002} favour
differential freeze-out between HCN and CO and thus low initial $\rm
[N]_{gas}$.

We conclude from Fig.~\ref{fig:fig8} that the models fail to explain
the observations. One possible reason for this failure is our neglect
of line-of-sight effects in the models used to construct
Fig.~\ref{fig:fig8}. The observed quantities are column densities,
which are integrals along the line of sight over a range of densities;
our analysis neglects this effect. However, trial calculations for one
source (L~1544; see Appendix~\ref{app:model}) suggest that including
line-of-sight integration can reduce but not eliminate the
discrepancies between model predictions and
observations. \corr{}{Another possibility might be that the duration
  of the collapse is longer that the free-fall time. However, if this
  time is significantly increased, the abundance of gaseous CO drops
  too rapidly with increasing density \citep{flower2005}.} More
important may be errors in the rate coefficients that we have used for
some of the key reactions, discussed in Section~\ref{chemistry}. It is
clear, for example, that our predictions relating to CN are sensitive
to the rates of reactions (\ref{equ3}), (\ref{equ4}), and (\ref{equ9})
at temperatures of the order of 10~K.  Further progress in this field
will require reliable determinations of the rate coefficients of these
reactions at low temperatures.

We find that the values of the CN:HCN ratio observed in prestellar
cores indicate that the fraction of nitrogen in the gas phase is
likely to be considerably lower than the diffuse--gas value of $6.4
\times 10^{-5}$. Nitrogen (like oxygen) may deplete on to grain
surfaces at relatively low densities. Confirmation will require the
identification of nitrogen--containing ices and estimates of their
relative abundances. A rather similar conclusion has been reached by
\cite{maret2006} in a study of B68.

We finally note that our observations show HNC:HCN $\approx 2$,
whereas the exhaustive theoretical study of \cite{herbst2000}
predicted HNC:HCN $\approx 1$. It is possible that enhanced line
trapping in HNC, relative to HCN, results in our deriving an
anomalously high HNC:HCN abundance ratio; but it is unlikely that this
effect can explain fully the discrepancies with the model
predictions. \corr{}{Maybe more relevant is the apparent correlation
  between the HNC:HCN ratio and freeze-out, as suggested by the
  results from \cite{hirota1998} who show that the largest
  (resp. smallest) ratio is observed toward a strongly depleted core
  (resp. undepleted).}

\section{Concluding remarks}
\label{conclusions}

We have studied the behaviour of nitrogen--containing species,
principally CN, HCN, and HNC, in the pre-protostellar cores L~183,
L~1544, Oph~D, L~1517B, and L~310. Our main conclusions are as
follows.

\begin{itemize}
\item We observe that CN, HCN, and HNC remain present in the gas phase
  at densities above the typical density (3\tdix{4}~\ccc) at which CO
  depletes on to grains.
  
\item The CN:HCN and HNC:HCN ratios are larger than unity in all
  objects and do not vary much within in each core. Whilst the
  differential freeze--out of CN and CO can be understood, the
  approximate constancy of the CN:HCN ratio cannot.

\item The CN:HCN ratio puts upper limits on the abundance of atomic
  nitrogen in the gas phase, and the NO:HCN ratio constrains the C:O
  ratio. Though uncertain, the comparison between observations and
  models indicates that most of the nitrogen is locked into ices, even
  at densities probably as low as \dix{4}~\ccc.

\end{itemize}

Our current knowledge of the chemistry of nitrogen--containing species
is incompatible with the observed ratios. We recall that there exist
large uncertainties in the rate coefficients, at low temperatures, for
the key neutral--neutral reactions, discussed in
Section~\ref{chemistry}, including those with atomic
nitrogen. Essential to further progress in understanding the chemistry
of nitrogen--containing species in pre-protostellar cores are
measurements, at low temperatures, of at least some of these key
reactions.

\begin{acknowledgements}
We thank M.~Tafalla for providing us with the \nnhp\jone\ spectra
towards L~1517B and for his helpfull referee report. We also thank
Holger M\"uller of the CDMS for helpful comments on the
spectroscopy. This work has been been partially supported by the EC
Marie-Curie Research Training Network ``The Molecular Universe''
(MRTN-CT-2004-512302).
\end{acknowledgements}

\bibliographystyle{aa}
\bibliography{general,cores,phb,technic}

\appendix
\section{Spectroscopic data}

\begin{table}[!h]
  \centering
  \caption{Hyperfine structure in CN\jtwo\ ($NJF \rightarrow N'J'F'$);
    from \cite{skatrud1983}.}
  \begin{tabular}{c@{\,,\,}c@{\,,\,}c @{$\,\longrightarrow\,$} c@{\,,\,}c@{\,,\,}c c c}
    \hline\hline
    $N$ & $J$ & $F$ & $N'$ & $J'$ & $F'$ & $\nu^a$ & R.I.$^b$ \smallskip\\\hline
    2 &3/2 &1/2 & 1& 1/2& 1/2 & 226663.685 & 0.0494 \\
    2 &3/2 &3/2 & 1& 1/2& 1/2 & 226679.341 & 0.0617 \\
    2 &3/2 &1/2 & 1& 1/2& 3/2 & 226616.520 & 0.0062 \\
    2 &3/2 &3/2 & 1& 1/2& 3/2 & 226632.176 & 0.0494 \\
    2 &3/2 &5/2 & 1& 1/2& 3/2 & 226659.543 & 0.1667 \\
    2 &3/2 &1/2 & 1& 3/2& 1/2 & 226287.393 & 0.0062 \\
    2 &3/2 &3/2 & 1& 3/2& 1/2 & 226303.049 & 0.0049 \\
    2 &5/2 &3/2 & 1& 3/2& 1/2 & 226875.896 & 0.1000 \\
    2 &3/2 &1/2 & 1& 3/2& 3/2 & 226298.896 & 0.0049 \\
    2 &3/2 &3/2 & 1& 3/2& 3/2 & 226314.552 & 0.0120 \\
    2 &3/2 &5/2 & 1& 3/2& 3/2 & 226341.919 & 0.0053 \\
    2 &5/2 &3/2 & 1& 3/2& 3/2 & 226887.399 & 0.0320 \\
    2 &5/2 &5/2 & 1& 3/2& 3/2 & 226874.183 & 0.1680 \\
    2 &3/2 &3/2 & 1& 3/2& 5/2 & 226332.519 & 0.0053 \\
    2 &3/2 &5/2 & 1& 3/2& 5/2 & 226359.887 & 0.0280 \\
    2 &5/2 &3/2 & 1& 3/2& 5/2 & 226905.366 & 0.0013 \\
    2 &5/2 &5/2 & 1& 3/2& 5/2 & 226892.151 & 0.0320 \\
    2 &5/2 &7/2 & 1& 3/2& 5/2 & 226874.764 & 0.2667 \\
    \hline
  \end{tabular}
  \begin{list}{}{}
    \scriptsize
  \item $a$: Rest frequency in MHz.
  \item $b$: Relative intensities, with their sum normalized to
    unity.\\
  \end{list}
  \label{tab:12cn}
\end{table}

\section{Data reduction}
\label{AppA}

Data reduction was done with the CLASS90 software from the GILDAS
program suite\footnote{Available at
  \texttt{http://www.iram.fr/GILDAS}}. We summarize here the reduction
of the frequency--switched spectra obtained with the VESPA
autocorrelator facility at the IRAM~30~m radio telescope.

All spectra were corrected first from the instrumental spectral
transfer function. The resulting spectra were then folded and averaged
(each folded spectrum being weighted by the effective rms of the
residuals after baseline subtraction). A zero--order polynomial was
fitted to the resulting spectrum to compute the final rms. Whenever
platforming was present in the data, the spectrum was split into as
many parts as needed, and each part was treated individually with a
first--order polynomial in order to adjust the continuum level. The
concatenated sub-parts were then treated as a single spectrum. This
method proved to be robust. In some cases, the amplitude of ripples
was large enough to require special treatment. Ripples were subtracted
by fitting a sine wave to the spectrum, using an improved version of
sine fitting, as compared with the default CLASS90 procedure. In the
case of a spectrum presenting ripples, channels in the spectral
windows (where there is presumably some line emission) were replaced
by a sine wave, determined by the first--guess parameters (amplitude,
period and phase). As the critical parameter in the sine wave
minimization proved to be the period, the minimization was repeated
for several values of the period. In all cases, this algorithm
converged to an acceptable solution, as indicated by the residuals and
inspection by eye.

\section{Column density derivation}
\label{app:cdens}

\begin{table*}
  \centering
  \caption{Conversion factor, $N_0$ (in \dix{12}\cc/(\kkms)), at an
    excitation temperature $\texc=8$~K (see Eq.~\ref{eq:N0}).}
  \begin{tabular}{c c c c c c c c c c}\hline\hline
    Molecule & Transition & $\nu^a$ & $\delta v^b$ & $\beff^c$ & $B$ & $\mu_0$ & R.I.$^d$ &$Q^e$& $N_0/\rm R.I.$ \\
    & & MHz & \kms & & MHz & debye & & & \\\hline
    CN &\jone & 113520.414 & 0.052 & 0.74 & 56693.470 & 1.45 & 0.0184 & 3.3 & 293 \\
    CN &\jtwo & 226882.000 & 0.053 & 0.50 & & \\
    HCN &\jone & 88633.936 & 0.066 & 0.77 & 44315.976 & 2.9852 & 0.1111 & 4.1 & 16.4 \\
    \thcn &\jone & 108780.201 & 0.063 & 0.77 & 54353.130 & 1.45 & 0.194 & 3.4 & 29.5 \\
    H\thcn &\jone & 86342.251 & 0.054 & 0.75 & 43170.127 & 2.9852 & 0.5556 & 4.2 & 3.4 \\
    HN\thc &\jone & 87090.850 & 0.068 & 0.78 & 45331.980 & 3.05 & 1.000 & 4.0 & 1.7 \\
    HC\fifn &\jone & 86054.9664& 0.067 & 0.78 & 43027.648 & 2.9852 & 1.000 & 4.2 & 1.9 \\
    \nnhp &\jone & 93171.621 & 0.068 & 0.78 & 46586.880 & 3.40 & 0.037 & 3.9 & 35.2 \\
    \hline
  \end{tabular}
  \label{tab:summary}
  \begin{list}{}{}
    \scriptsize
  \item $^a$ Frequency of the HFS component considered for which we
    compute the flux $W$.
  \item $^b$ Spectral resolution in \kms.
  \item $^c$ Beam efficiency. The forward efficiency at 3~mm is
    $\feff=0.95$, and 0.91 at 1.3~mm.
  \item $^d$ Relative intensity of the HFS component. The total flux
    in Eq.~\ref{eq:N0} is $\wtot=W/$R.I.
  \item $^e$ Partition function computed as
    $Q=\dsum_{J=0}^{20}(2J+1)\exp^{-\frac{hBJ(J+1)}{k_{\rm B}\texc}}$,
    where $B$ is the rotational constant.
  \end{list}
\end{table*}

All column densities are derived assuming optically thin emission with
levels populated in LTE at the excitation temperature \texc. Under
these assumptions, the column density is directly proportional to the
integrated flux in the line $\wtot$. From a transition
$\jup\rightarrow\jlow$ between energy levels $E_{\rm u}$ and $E_{\rm
  l}$ (corresponding to an energy $T_0=(E_{\rm u}-E_{\rm l})/k_{\rm
  B}$), one can compute the total column density of the molecule as
\begin{equation}
  N_{\rm tot} = \frac{3\epsilon_0h}{2\pi^2}\,\frac{1}{\jup
    \mu_0^2}\,\frac{Q\,{\rm e}^{E_{\rm l}/k_{\rm
        B}\texc}\,\wtot}{\Delta J_\nu [1-\exp(-\frac{T_0}{\texc})]}
  \label{eq:cdens1}
\end{equation}
in SI units, where $\Delta J_\nu=J_\nu(\texc)-J_\nu(\tbg)$, with
$J_\nu(T)=T_0/[\exp(T_0/T) - 1]$, and $\mu_0$ is the dipole moment;
$\epsilon_0$ is the permittivity of free space. Numerically, we obtain
\begin{equation}
  N_{\rm tot} = \frac{8\tdix{12}}{\jup\mu_D^2}\,\frac{Q\,e^{E_{\rm
        l}/k_{\rm B}\texc}\,\wtot}{\Delta J_\nu
    [1-\exp(-\frac{T_0}{\texc})]} = N_0(\texc) \,\wtot
  \label{eq:N0}
\end{equation}
with $\wtot$ in \kkms and $N_{\rm tot}$ in cm$^{-2}$; $\mu_D$ is the
dipole moment in debye. We have introduced the conversion factor,
$N_0$, which depends on the molecular properties and the excitation
temperature. It is to be noted that, in the case of resolved hyperfine
structure, $\wtot$ is the total integrated intensity of the hyperfine
multiplet, i.e. of the rotational transition. Values of $N_0$ are
given in Table~\ref{tab:summary} for the observed molecules and at an
excitation temperature $\texc=8$~K (see HWFP08).

\begin{figure*}
  \centering
  \def\wa{0.7\hsize}
  \includegraphics[height=\wa,angle=-90]{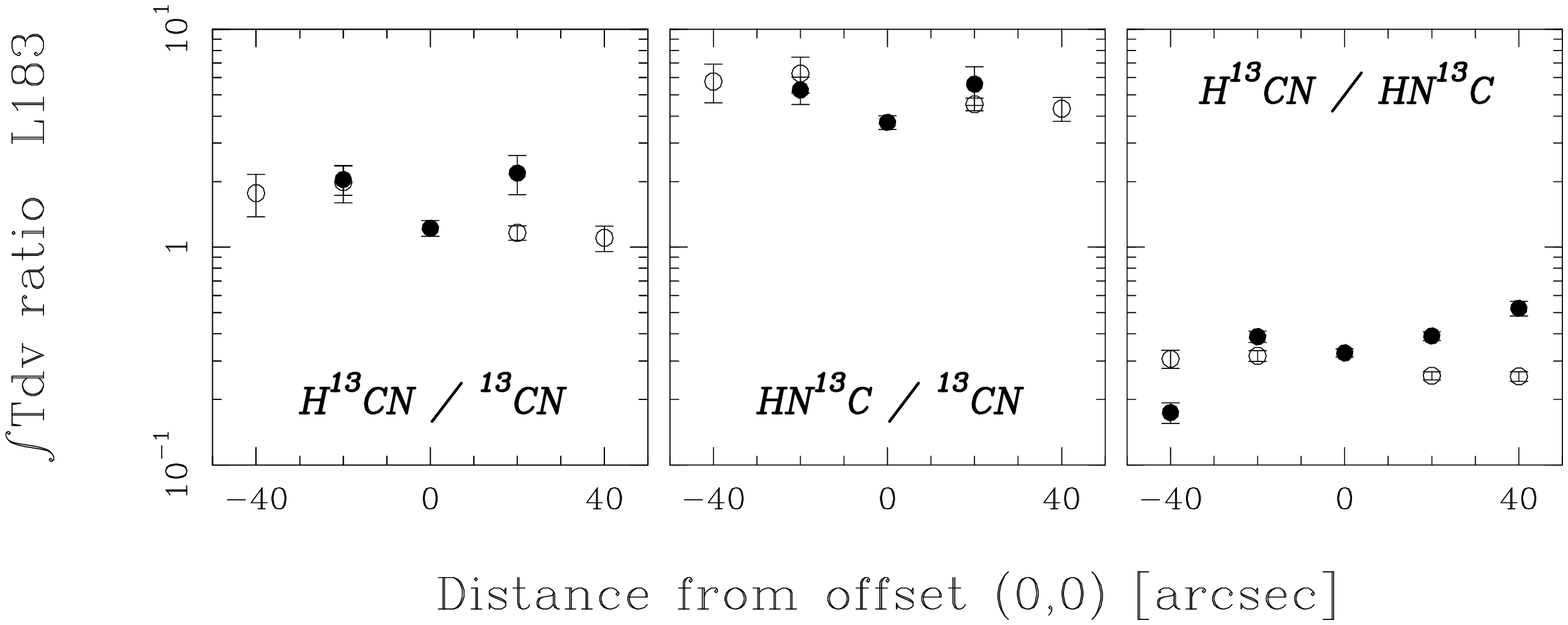}\smallskip\\
  \includegraphics[height=\wa,angle=-90]{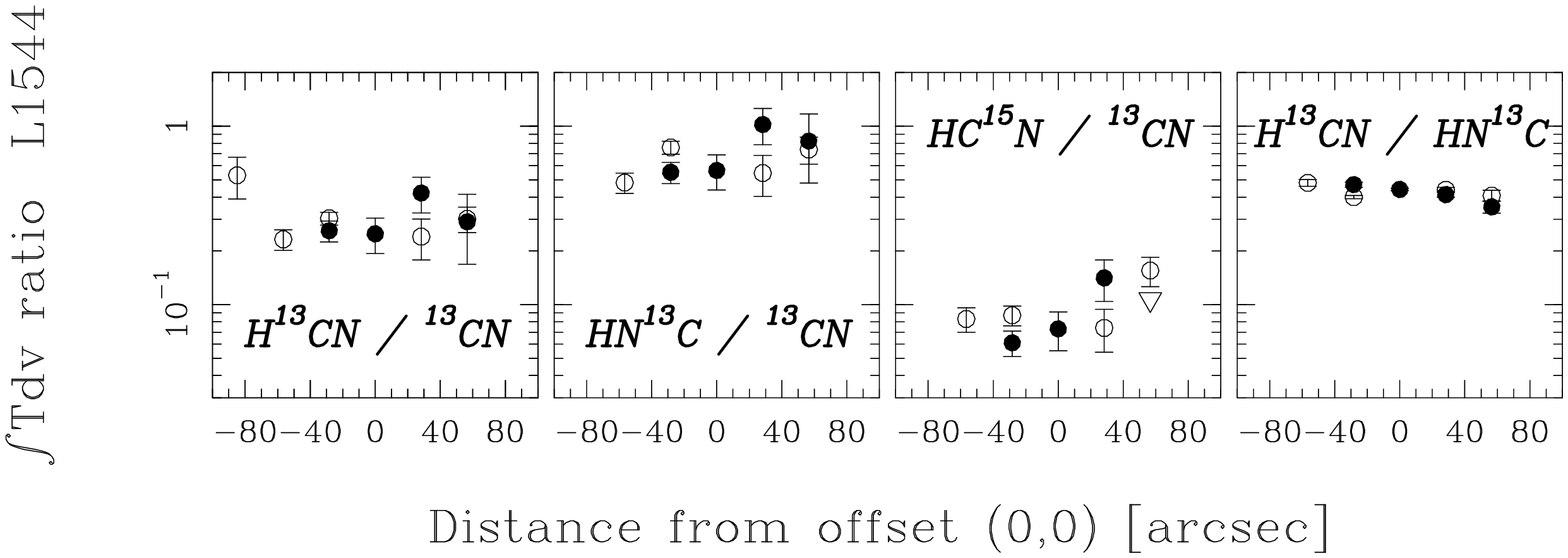}\smallskip\\
  \includegraphics[height=\wa,angle=-90]{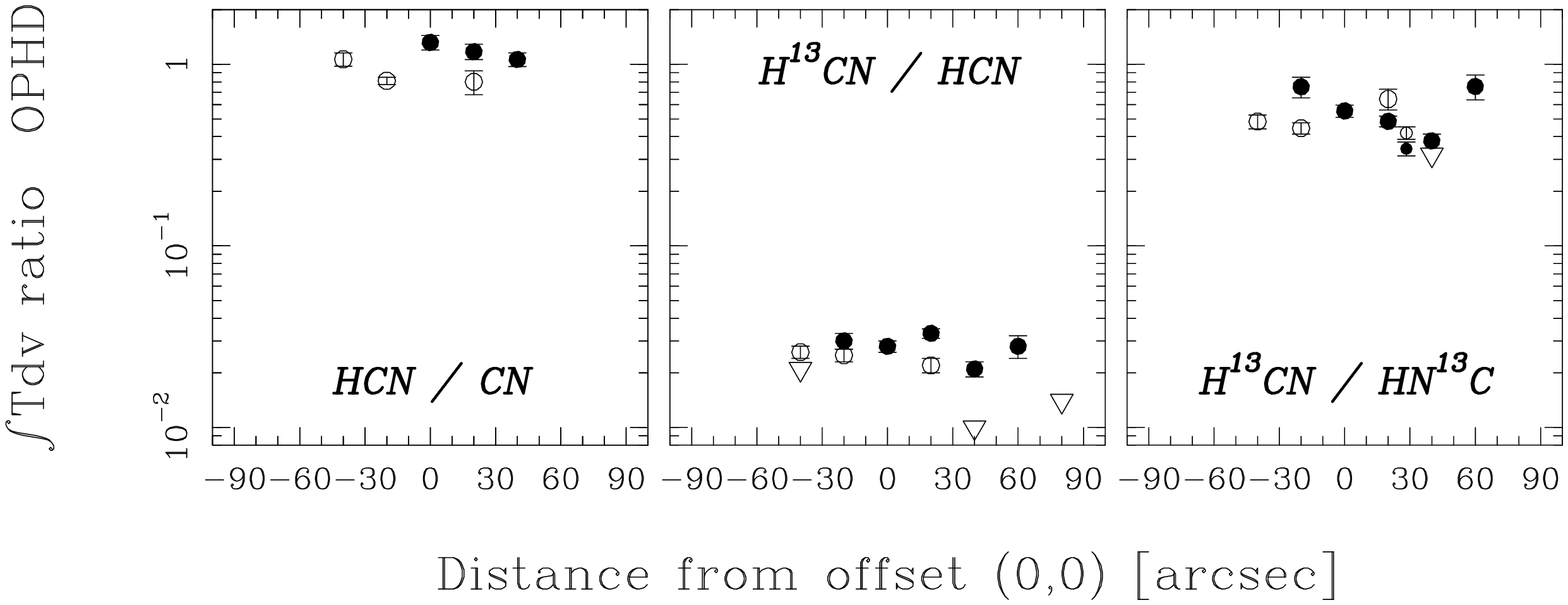}\smallskip\\
  \includegraphics[height=\wa,angle=-90]{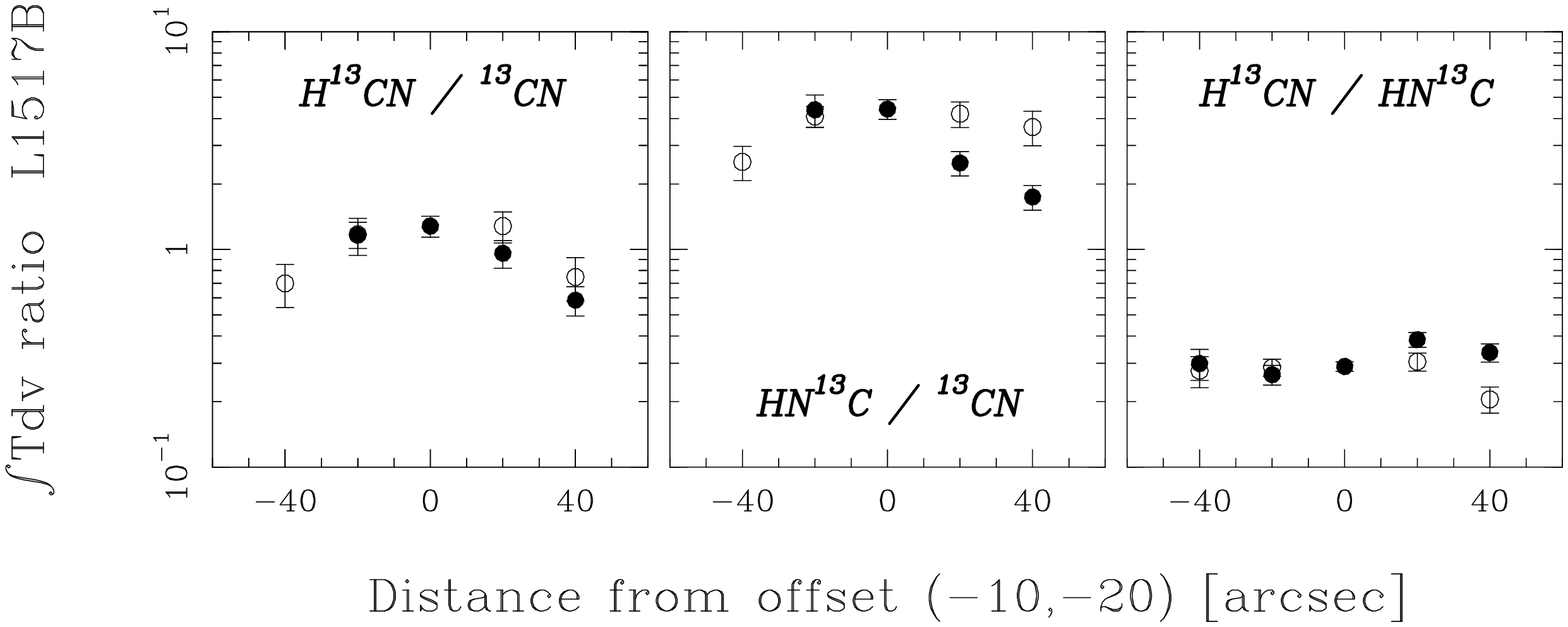}\smallskip\\
  \includegraphics[height=\wa,angle=-90]{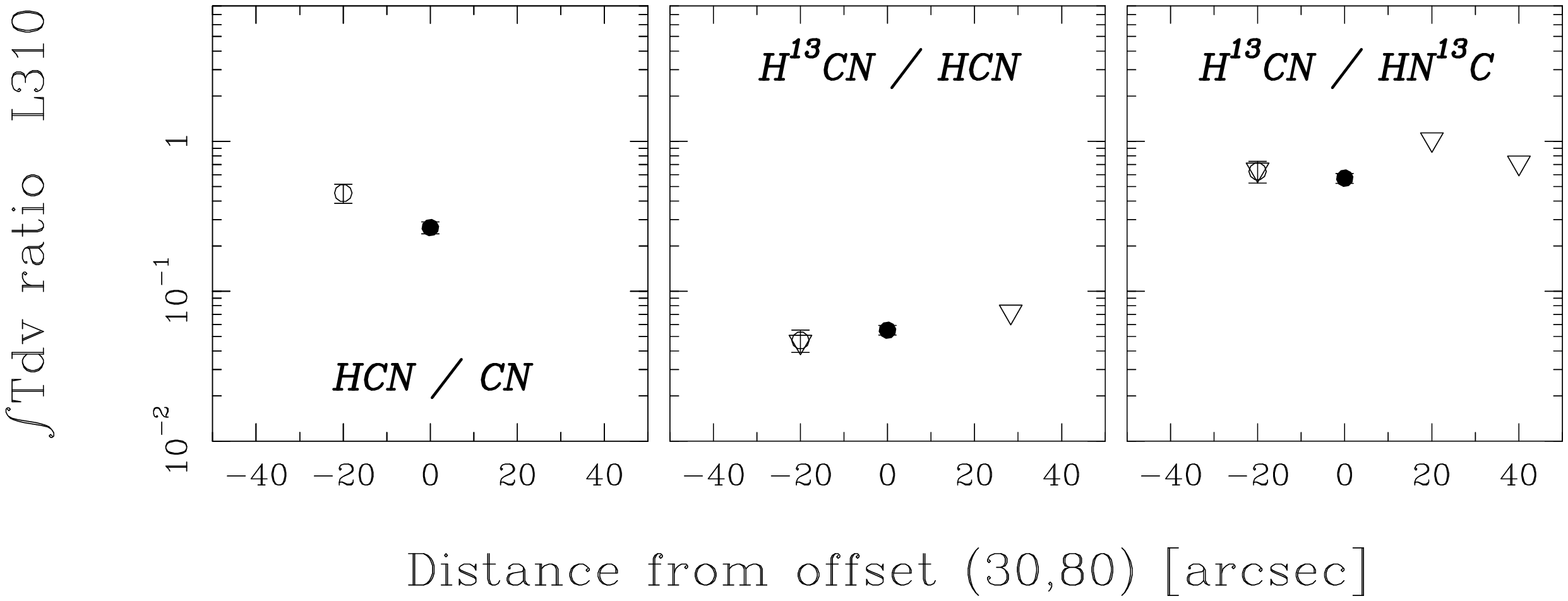}
  \caption{Ratio of the total integrated intensities $\wtot$ of the
    molecules indicated in each panel, as a function of the offset
    from the dust emission peak. For each molecule, $\wtot$ is derived
    from the values in the corresponding Table divided by the relative
    intensity of the HFS component, when resolved. Filled and open
    circles distinguish between the two cut directions. Open triangles
    are 5$\sigma$ upper limits. From top to bottom, L~183, L~1544,
    Oph~D, L~1517B, and L~310.}
  \label{fig:ratio-cnhcn-1}
\end{figure*}

\section{Modelling of the prestellar core L~1544}
\label{app:model}

In order to interpret our observations of L~1544, we make use of the
model of one--dimensional, free--fall gravitational collapse used in
our previous study (A07). This model incorporates dust grain
coagulation and a time--dependent chemistry, including the reactions
listed in Section~\ref{chemistry} above, which are directly relevant
to the present work. We assume a constant kinetic temperature, $T =
10$~K, and a cosmic ray ionization rate, $\zeta =
10^{-17}$~s$^{-1}$. Further information on the model may be found in
\cite{flower2005}.

An important aspect of the interpretation is to connect, as
realistically as possible, the abundance profiles (\ie\ the number
density of species X, $n$(X) as a function of the total density,
$\nh = 2n({\rm H}_2) + n({\rm H})$), which are the output of the
model, to the observed variations in L~1544 of column densities,
$N$(X), with impact parameter $r$. In order to make this connection,
we proceed as follows:
\begin{itemize}
\item we relate the gas density at $r$ to the central density by means
  of the relation $$\nh(r) = \nh(0)/[1+(r/r_2)^{\alpha }]$$
  \cite{tafalla2002}, where $r$ is the offset from centre, $r = 0$,
  and $r_2$ is the radial distance over which the density decreases to
  $\nh(0)/2$. Following \cite{tafalla2002}, we adopt $r_2 =
  20$\arcsec (equivalent to 0.014~pc at the distance of L~1544) and
  $\alpha = 2.5$; the central density $n(0) = \dix{6}$~\ccc\ (A07),
  which is somewhat smaller than the value reported in
  Table~\ref{tab:cores} for this object but within the probable
  uncertainties of its determination.

\item Using the computed values of $n$(X) \vs\ \nh, we calculate the
  corresponding column density, $N$(X), by integrating along the line
  of sight for any given value of $r$ in the adopted range $0 \le r
  \le 120$~arcsec, over which the density $\nh(r)$ decreases
  from \dix{6}~\ccc\ to $1.1\tdix{4}$~\ccc.

\item Finally, the column densities are convolved with a Gaussian
profile with a (1/e) radius of 15\arcsec , corresponding to a HPBW
of 25\arcsec, in order to simulate approximately the IRAM 30~m
telescope beam at 100~GHz.

\end{itemize}
As will be seen below, the consequence of this procedure is a
significant damping of the variations in the computed abundance
profiles, owing partly to the effect of integrating along the line of
sight and hence over a range of densities, and partly to the Gaussian
beam averaging.

We consider first the predictions of the chemical model, and
specifically the abundances of nitrogen--containing species. We turn
our attention then to the Gaussian--beam averaged column densities,
and their comparison with the observations.

\subsection{Abundance profiles}
\label{profiles}

\begin{figure}[t]
  \centering
  \includegraphics[width=0.6\hsize]{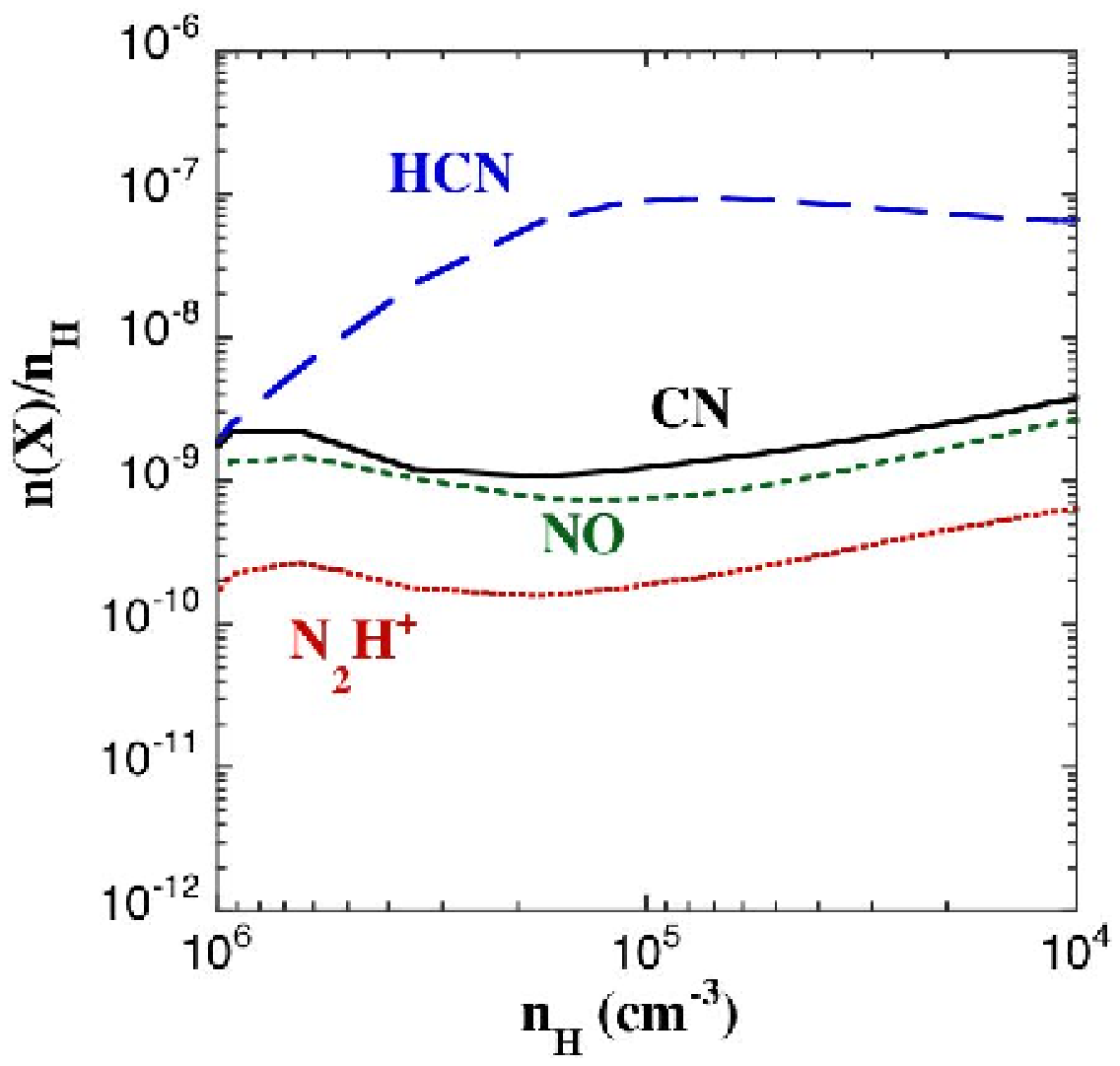}\\
  \includegraphics[width=0.6\hsize]{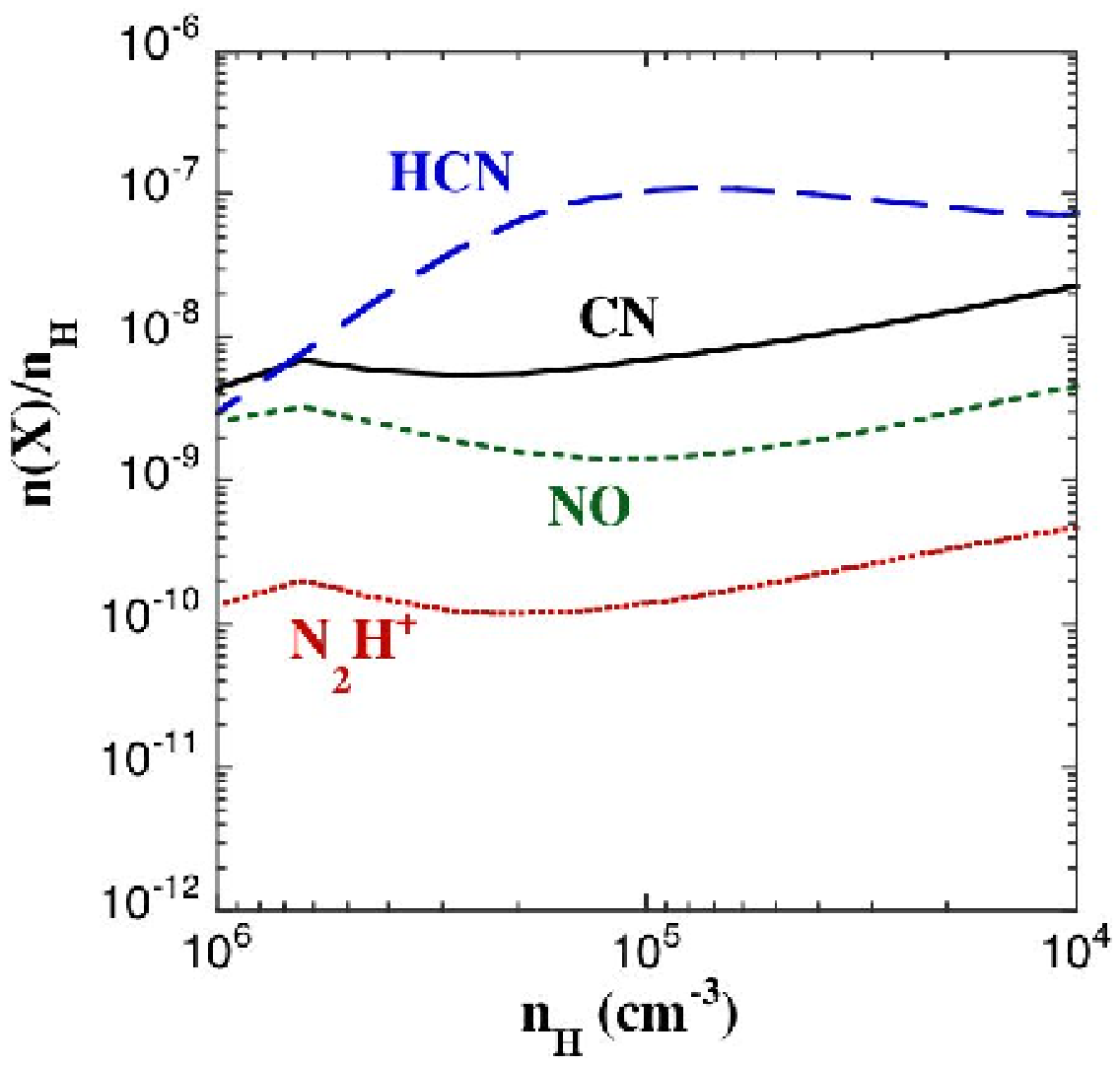}
  \caption{The number densities of CN, HCN, NO and N$_2$H$^+$,
    relative to $\nh = 2n({\rm H}_2) + n({\rm H})$, as predicted by a
    model of gravitational free--fall collapse. The lower panel
    illustrates the effects of introducing a barrier of 25~K to both
    reaction~(\ref{equ2}) and reaction~(\ref{equ4}). \colfig}
  \label{Figure2}
\end{figure}

In Fig.~\ref{Figure2} are plotted the fractional abundance profiles of
CN, HCN, NO and N$_2$H$^+$; note that the $x$-axis has been reversed
in order to facilitate the comparison with later Figures, in which the
$x$-coordinate is the offset from the centre, where the density of the
medium is highest. We see from Fig.~\ref{Figure2} that, at low
densities, the fractional abundance of HCN exceeds that of CN, by a
factor which approaches two orders of magnitude when $\nh \approx
\dix{5}$~\ccc. The fractional abundance of HCN decreases towards the
maximum density of \dix{6}~\ccc, where $n({\rm HCN}) \approx n({\rm
  CN})$. This behaviour can be understood by reference to the
discussion in Section~\ref{chemistry} above: CN is formed and
destroyed in reactions~(\ref{equ3}, \ref{equ4}) which involve atomic
nitrogen; HCN, on the other hand, is formed in reaction~(\ref{equ9})
with N but destroyed in reactions with H$^+$ and H$_3^+$ that
ultimately lead to CN. Consequently, as the density of the medium
increases, and neutral species begin to freeze on to the grains, the
fractional abundance of HCN falls, whereas the fractional abundance of
CN remains roughly constant until, finally, CN too freezes on to the
grains.

We have already seen in Section~\ref{observations} that the ratio of
the column densities of HCN and CN, which is of the order of 1 in
L~1544, shows little variation across this and the other sources in
our sample. Thus, it seems unlikely that the fractional abundance
profiles of HCN and CN seen in the upper panel of Fig.~\ref{Figure2}
are compatible with the observed column densities, given that the
ratio $n({\rm HCN})/n({\rm CN})$ varies by approximately 2 orders of
magnitude over the range of density $\dix{4} \le \nh \le
\dix{6}$~cm$^{-3}$; this tentative conclusion is confirmed by the
analysis in the following Section~\ref{columns}. It appears that the
observations are indicating that the natures of both the formation
\textit {and} the destruction processes are similar for both these
species. This realization led us to consider the possibility that
reaction~(\ref{equ4}), which destroys CN, may have a small barrier,
significant at the low temperatures relevant here ($T \lesssim 10$~K)
but difficult to detect by measurements at higher temperatures. A
similar situation may obtain for the analogous reaction~(\ref{equ2})
of NO with N. Accordingly, we show, in the lower panel of
Fig.~\ref{Figure2} the results which are obtained on introducing a
barrier of 25~K to both reaction~(\ref{equ2}) and
reaction~(\ref{equ4}); we note that the adopted barrier size is
arbitrary, the only requirement being that it should be small.

Small barriers can arise when the potential energy curves involved in
the atom--molecule reaction exhibit (much larger) barriers for certain
angles of approach but no barrier for others. In order to determine
the thermal rate coefficient, the probability of the reaction must be
averaged over the relative collision angle. If the rate coefficient is
then fitted to an Arrhenius form, $$k(T) = \gamma (T/300)^{\alpha }
\exp(-\beta /T),$$ small, positive values of $\beta $ may be
interpreted as the angle--averaged value of the reaction barrier; see
\cite{andersson2003}.

It is clear from Fig.~\ref{Figure2} that the small reaction barrier
has the effect of enhancing the fractional abundance of CN and
reducing the amplitude of the variation in the ratio $n({\rm
  HCN})/n({\rm CN})$. We shall see in the following
Section~\ref{columns} that this variation is damped further when the
ratio of the corresponding Gaussian--beam averaged column densities is
considered.

\begin{figure}[t]
  \centering
  \includegraphics[width=0.6\hsize]{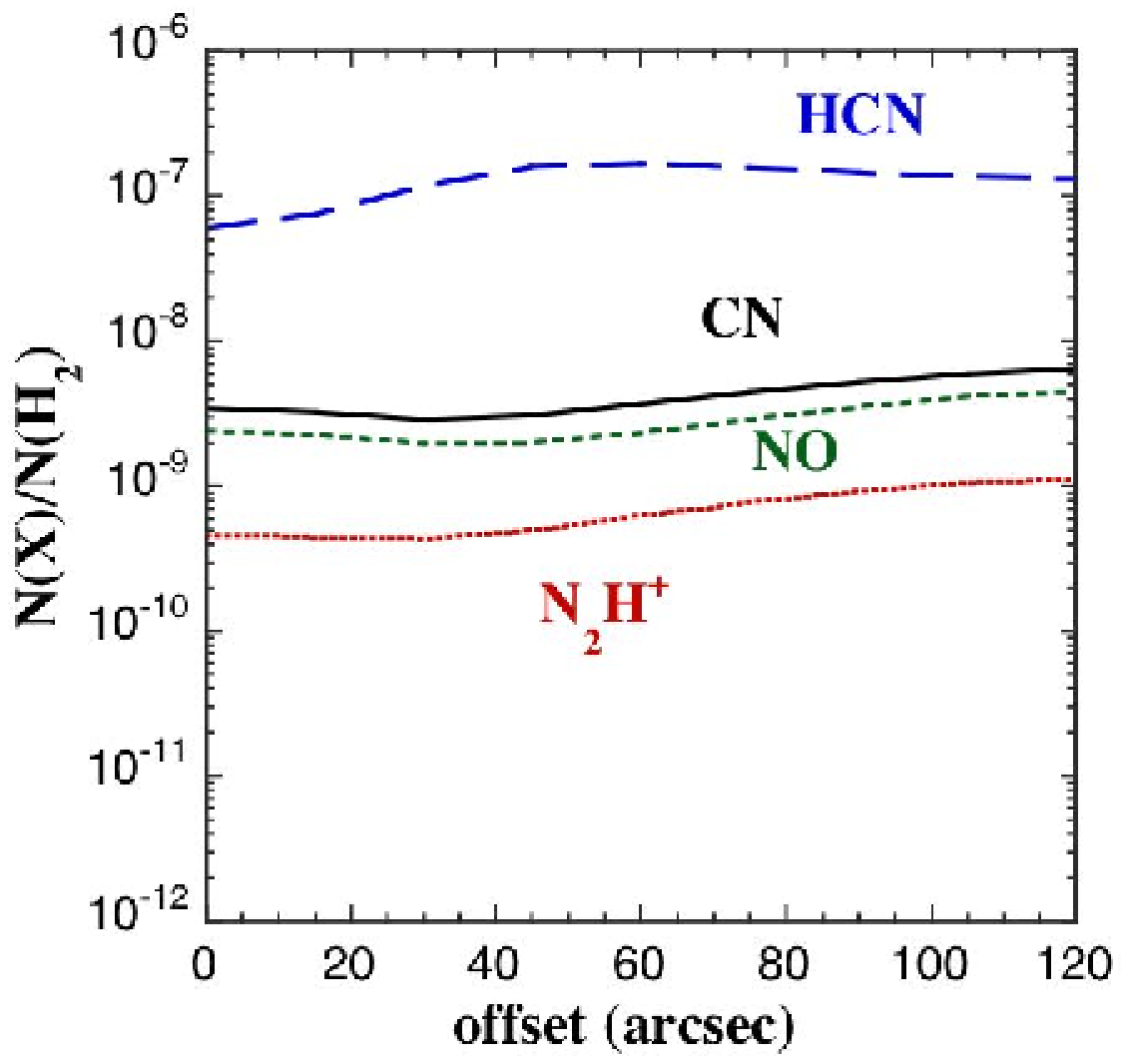}\\
  \includegraphics[width=0.6\hsize]{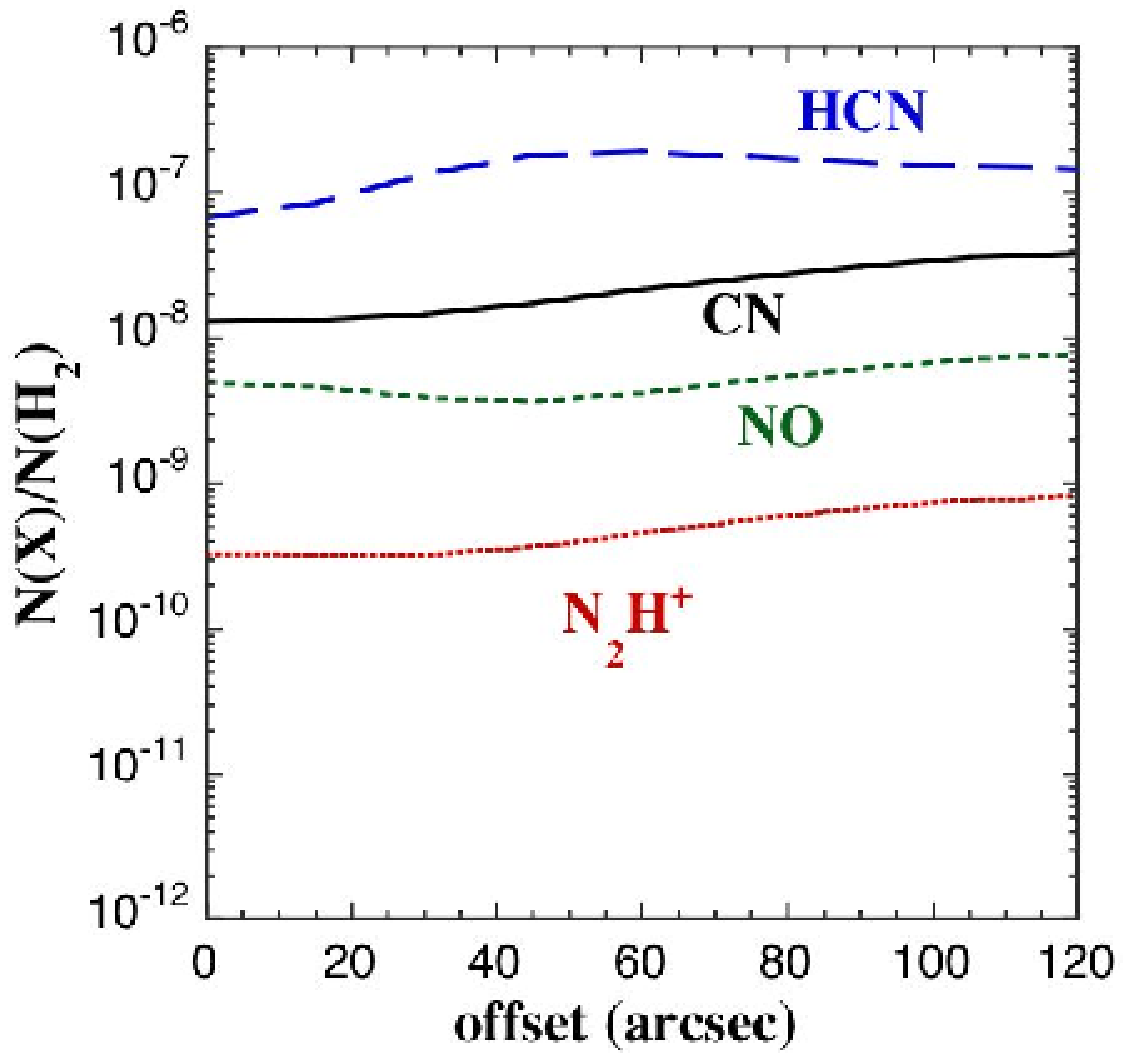}
  \caption{The column densities of CN, HCN, NO and N$_2$H$^+$,
    relative to H$_2$, as predicted by the model described in
    Section~\ref{app:model}. The lower panel illustrates the effects
    of introducing a barrier of 25~K to both reaction~(\ref{equ2}) and
    reaction~(\ref{equ4}). \colfig}
  \label{Figure3}
\end{figure}

The results in Fig.~\ref{Figure2} have been obtained assuming that the
grain--sticking probability was unity for all species, and that the
elemental abundance ratio $\rm C:O = 0.97$, \textit{i.e.}  a
marginally oxygen--rich medium. This value of the C:O ratio was an
outcome of the modelling by A07 of observations of NO and \ce{N2H+} in
L~1544. These authors investigated also the consequences of varying
the values of the sticking coefficient for atomic C, N and O, and the
initial value of the N:N$_2$ abundance ratio. We recall that
increasing the elemental C:O abundance ratio further has the
consequence of reducing the HCN:CN ratio, thereby improving the
agreement with the observations. On the other hand, the CN:NO ratio
also rises, and the values of this ratio in Fig.~\ref{Figure2}, where
$n({\rm CN}) > n({\rm NO})$, already exceed the values of the
corresponding column density ratio, observed in L~1544, where $N({\rm
  CN}) < N({\rm NO})$ by typically an order of magnitude. It is
possible that selective variations in the values of the sticking
probability, or in the initial N:N$_2$ abundance ratio\footnote{The
  equilibrium value of the N:N$_2$ ratio, adopted in the present
  models, is $n({\rm N})/n({\rm N}_2) = 1.0$, as compared with the
  (non-equilibrium) value of 18 adopted by A07.}, might alleviate some
of these discrepancies. However, whilst there remain such large
uncertainties in the values of the rate coefficients for the key
neutral--neutral reactions, discussed in Section~\ref{chemistry}, it
would perhaps be premature to investigate further the consequences of
modifying the values of other (and equally uncertain) parameters, in
an attempt to improve the agreement between the models and the
observations. Our aim here is to point to the discrepancies and
highlight the uncertainties; and it seems unlikely that further
progress can be made until the rates of at least some of the key
reactions have been measured at low temperatures.

\subsection{Column densities}
\label{columns}

In Fig.~\ref{Figure3} are shown the computed column densities of CN,
HCN, NO and \nnhp, relative to the column density of \hh. The
fractional abundances of these species, relative to $\nh$, derive from
the models discussed in the previous Section~\ref{profiles}.

At zero offset, the line of sight passes through regions with
densities covering the entire range of the model, $1.0\tdix{6}\ccc \ge
\nh \ge 1.1\tdix{4}$~\ccc. Consequently, the column density ratio,
$N({\rm HCN})/N({\rm H}_2) \gg n({\rm HCN})/\nh$, evaluated at the
peak density $\nh = \dix{6}$~\ccc; further smoothing is introduced by
the Gaussian--beam averaging. The overall effect of the
line--of--sight and Gaussian--beam averaging is a flattening of the
column density profiles (Fig.~\ref{Figure3}), compared with the
fractional abundance profiles (Fig.~\ref{Figure2}).  Comparing the two
panels of Fig.~\ref{Figure3}, we see that the introduction of the
small barriers to the reactions of CN and NO with N reduces
substantially the $N({\rm HCN})/N({\rm CN})$ column density
ratio. Although they do not attain the observed value, of the order of
$1$, the computed values of $N({\rm HCN})/N({\rm CN})$ in the lower
panel of Fig.~\ref{Figure3} are clearly more compatible with the
observations of L~1544 than are those in the upper panel.

\clearpage
\def\x{\quad\textbf{ZZZZZZ}}
\def\x{}
\begin{table*}[b]
  \centering
  \caption{Line properties (main beam temperature scale) and total
    column densities derived towards L~183. Numbers in parentheses are
    powers of 10.}  \scriptsize
  \begin{tabular}{l r r r r r c c r}\hline\hline
    Line & $\delta x$ & $\delta y$ & $T_{\rm mb}$ & $W$ & \dveq$^e$ & $N_{\rm tot}$$^f$ & $N(\hh)$ & $N_{\rm tot}/2N(\hh)$$^g$ \\
    & \arcsec & \arcsec & mK & m\kkms & \kms & \dix{12}\,\cc & \dix{22}\,\cc \smallskip\\\hline
    %
    CN
    & 0 & 0 & 700$\pm$ 60& 180$\pm$ 10& 0.26 & 53 & 7.0 & 3.8$(-$10) \\
    & -40 & 0 & 600$\pm$ 70& 140$\pm$ 12& 0.23 & 41 & 3.3 & 6.2$(-$10) \\
    & -20 & 0 & 650$\pm$ 70& 135$\pm$ 10& 0.21 & 40 & 6.1 & 3.3$(-$10) \\
    & 20 & 0 & 590$\pm$ 80& 110$\pm$ 10& 0.19 & 34 & 4.8 & 3.5$(-$10) \\
    & 40 & 0 & 455$\pm$ 80& 60$\pm$ 10& 0.13 & 18$\pm$2.9 & 2.8 & 3.2$(-$10) \\
    & 0 & -40 & 580$\pm$ 85& 130$\pm$ 10& 0.22 & 38 & 3.9 & 4.9$(-$10) \\
    & 0 & -20 & 755$\pm$ 60& 160$\pm$ 7& 0.21 & 47 & 5.6 & 4.2$(-$10) \\
    & 0 & 20 & 670$\pm$ 90& 140$\pm$ 10& 0.21 & 41 & 6.3 & 3.3$(-$10) \\
    & 0 & 40 & 660$\pm$ 80& 175$\pm$ 15& 0.26 & 48 & 5.7 & 4.2$(-$10) \\
    \hline
    \thcn$^a$ 
    & 0 & 0 & 115$\pm$ 12& 28$\pm$ 2& 0.24 & 0.82 & 7.0& 5.9$(-$12)\\
    & -40 & 0 & 75$\pm$ 25& $<$ 18& & $<$ 0.53 & 3.3& $<$8.0$(-$12)\\
    & -20 & 0 & 100$\pm$ 20& 21$\pm$ 3& 0.21 & 0.62 $\pm$ 0.09 & 6.1& 5.1$(-$12)\\
    & 20 & 0 & 65$\pm$ 20& 15$\pm$ 3& 0.23 & 0.44 $\pm$ 0.09 & 4.8& 4.6$(-$12)\\
    & 40 & 0 & 100$\pm$ 20& $<$ 18& & $<$ 0.53 & 2.8& 9.5$(-$12)\\
    & 0 & -40 & 65$\pm$ 20& 15$\pm$ 3& 0.23 & 0.44 $\pm$ 0.09 & 3.9& 5.6$(-$12)\\
    & 0 & -20 & 95$\pm$ 20& 16$\pm$ 3& 0.17 & 0.47 $\pm$ 0.09 & 5.6& 4.2$(-$12)\\
    & 0 & 20 & 115$\pm$ 20& 30$\pm$ 2& 0.26 & 0.88 & 6.3& 7.0$(-$12)\\
    & 0 & 40 & 115$\pm$ 20& 32$\pm$ 4& 0.28 & 0.94 $\pm$ 0.12 & 5.7& 8.2$(-$12)\\
    \hline
    H\thcn$^b$ 
    & 0 & 0 & 200$\pm$ 11& 98$\pm$ 4& 0.49 & 0.34 & 7.0 & 2.4$(-$12) \\
    & -40 & 0 & 135$\pm$ 17& 46$\pm$ 5& 0.34 & 0.16 $\pm$ 0.02 & 3.3 & 2.4$(-$12) \\
    & -20 & 0 & 250$\pm$ 20& 123$\pm$ 7& 0.49 & 0.42 & 6.1 & 3.4$(-$12) \\
    & 20 & 0 & 195$\pm$ 15& 94$\pm$ 4& 0.48 & 0.32 & 4.8 & 3.3$(-$12) \\
    & 40 & 0 & 160$\pm$ 15& 80$\pm$ 5& 0.50 & 0.27 & 2.8 & 4.8$(-$12) \\
    & 0 & -40 & 180$\pm$ 35& 76$\pm$ 7& 0.42 & 0.26 & 3.9 & 3.3$(-$12) \\
    & 0 & -20 & 245$\pm$ 18& 91$\pm$ 5& 0.37 & 0.31 & 5.6 & 2.8$(-$12) \\
    & 0 & 20 & 230$\pm$ 17& 100$\pm$ 4& 0.43 & 0.34 & 6.3 & 2.7$(-$12) \\
    & 0 & 40 & 230$\pm$ 18& 101$\pm$ 5& 0.44 & 0.34 & 5.7 & 3.0$(-$12) \\
    \hline
    HN\thc$^c$ 
    & 0 & 0 & 935$\pm$ 20& 540$\pm$ 4& 0.58 & 0.93 & 7.0 & 6.6$(-$12) \\
    & -40 & 0 & 788$\pm$ 35& 475$\pm$ 7& 0.60 & 0.82 & 3.3 & 1.2$(-$11) \\
    & -20 & 0 & 930$\pm$ 35& 570$\pm$ 7& 0.61 & 0.99 & 6.1 & 8.1$(-$12) \\
    & 20 & 0 & 710$\pm$ 33& 433$\pm$ 7& 0.61 & 0.75 & 4.8 & 7.8$(-$12) \\
    & 40 & 0 & 440$\pm$ 35& 275$\pm$ 13& 0.62 & 0.48 & 2.8 & 8.6$(-$12) \\
    & 0 & -40 & 840$\pm$ 35& 445$\pm$ 6& 0.53 & 0.77 & 3.9 & 9.9$(-$12) \\
    & 0 & -20 & 955$\pm$ 30& 517$\pm$ 6& 0.54 & 0.89 & 5.6 & 7.9$(-$12) \\
    & 0 & 20 & 1030$\pm$ 35& 700$\pm$ 7& 0.58 & 1.2 & 6.3 & 9.5$(-$12) \\
    & 0 & 40 & 1115$\pm$ 30& 712$\pm$ 7& 0.64 & 1.2 & 5.7 & 1.1$(-$11) \\
    \hline
    \nnhp$^d$ 
    & 0 & 0 & 1356$\pm$ 20& 309$\pm$ 6 & 0.23 & 10.9 & 7.0 & 7.8$(-$11)\\
    & 80 & 0 & 264$\pm$ 26& 67$\pm$ 5 & 0.25 & 0.786 & 1.5 & 2.6$(-$11)\\ 
    & 60 & 0 & 160$\pm$ 24& 40$\pm$ 7 & 0.25 & 1.41$\pm$ 0.246 & 2.0 & 3.5$(-$11)\\
    & 40 & 0 & 564$\pm$ 37& 153$\pm$ 11 & 0.27 & 5.39 & 2.8 & 9.6$(-$11)\\
    & 20 & 0 & 1037$\pm$ 30& 239$\pm$ 9 & 0.23 & 8.41 & 4.8 & 8.8$(-$11)\\
    & -20 & 0 & 1209$\pm$ 29& 283$\pm$ 9 & 0.23 & 9.96 & 6.1 & 8.2$(-$11)\\
    & -40 & 0 & 709$\pm$ 21& 164$\pm$ 6 & 0.23 & 5.77 & 3.3 & 8.7$(-$11)\\
    & -60 & 0 & 408$\pm$ 19& 103$\pm$ 6 & 0.25 & 3.63 & 2.3 & 7.9$(-$11)\\
    & -80 & 0 & 327$\pm$ 26& 79$\pm$ 8 & 0.24 & 2.78$\pm$ 0.282 & 2.4 & 5.8$(-$11)\\
    \hline
  \end{tabular}
  \label{tab:l183}
  \begin{list}{}{}
    \scriptsize
  \item Error bars are 1$\sigma$, and upper limits on $W$ and $N_{\rm
    tot}$ are at the 5$\sigma$ level. A final calibration uncertainty
    on the column density of 10\% has been adopted, unless smaller
    than 1$\sigma$. We adopt a systematic uncertainty on the dust
    column density of 30\% which reflects the uncertainties on
    \tdust\ and $\kappa_{\nu}$.
  \item $^a$ Strongest component at 108780.2010~MHz, with R.I. = 0.194.
  \item $^b$ Strongest component at 86340.1840~MHz, with R.I. = 0.556.
  \item $^c$ The integrated intensity includes the three blended HFS
    components.
  \item $^d$ Weakest HFS component at 93171.6210~MHz with R.I. = 0.037
    unless specified. At offsets (80\arcsec,0\arcsec) we use the
    isolated HFS component at 93176.2650~MHz with R.I.=0.1111 (see
    Table~\ref{tab:summary}).
  \item $^e$ $\dveq=W/T_{\rm mb}$ is
    the equivalent width. Error bars are 1$\sigma$, and upper limits
    on $W$ and 
  \item $^f$ $N_{\rm tot}=W\times N_0/R.I.$
  \item $^g$ Fractional abundances with respect to H assuming
    $N(\h)=2N(\hh)$.
  \end{list}
\end{table*}

\clearpage

\begin{table*}
  \centering
  \caption{As Table~\ref{tab:l183} but for lines observed towards L~1544.}
  \tiny
  \begin{tabular}{l r r r r r c c r}\hline\hline
    Line & $\delta x$ & $\delta y$ & $T_{\rm mb}$ & $W$ & \dveq & $N_{\rm tot}$ & $N(\hh)$ & $N_{\rm tot}/2N(\hh)$ \\
    & \arcsec & \arcsec & mK & m\kkms & \kms & \dix{12}\,\cc & \dix{22}\,\cc \smallskip\\\hline
    %
    CN$^a$
    & 0 & 0 & 1658$\pm$ 65 & 688$\pm$ 35 & 0.41 & 202 & 6.7 & 1.5$(-$9)\\
    & -40 & -40 & 325$\pm$ 63 & $<$ 50 & & $<$15 & & \\
    & -20 & -20 & 1171$\pm$ 103 & 422$\pm$ 27 & 0.36 & 124 & 2.8 & 2.2$(-$9)\\
    & 20 & 20 & 771$\pm$ 67 & 304$\pm$ 52 & 0.39 & 89.1$\pm$15.2 & 3.6 & 1.2$(-$9)\\
    & 40 & 40 & 729$\pm$ 94 & 254$\pm$ 20 & 0.35 & 74.4 & 1.2 & 3.2$(-$9)\\
    & -40 & 40 & 1257$\pm$ 105 & 565$\pm$ 62 & 0.45 & 166$\pm$18 & 3.6 & 2.3$(-$9)\\
    & -20 & 20 & 1641$\pm$ 66 & 745$\pm$ 106 & 0.45 & 218$\pm$31 & 4.9 & 2.2$(-$9)\\
    & 20 & -20 & 1569$\pm$ 93 & 587$\pm$ 52 & 0.37 & 172 & 4.4 & 2.0$(-$9)\\
    & 40 & -40 & 1124$\pm$ 80 & 296$\pm$ 30 & 0.26 & 86.7 & 1.6 & 2.8$(-$9)\\
    \hline
    \thcn 
    & 0 & 0 & 166$\pm$15 & 73$\pm$4 & 0.44 & 2.17 & 6.7 & 1.6$(-$11)\\
    &-40 & -40 & $<$ 60 & $<$ 20 & & $<$0.60\x & & \\
    &-20 & -20 & 108$\pm$19 & 37$\pm$4 & 0.34 & 1.10$\pm$0.13 & 2.8 & 2.0$(-$11)\\
    & 20 & 20 & 85$\pm$19 & 21$\pm$3 & 0.26 & 0.65$\pm$0.10 & 3.6 & 9.1$(-$12)\\
    & 40 & 40 & 55$\pm$20 & 20$\pm$5 & 0.35 & 0.58 & 1.2 & 2.5$(-$11)\\
    &-60 & 60 & 79$\pm$19 & 11 $\pm$3 & 0.14 & 0.33 & 2.3 & 7.3$(-$12)\\
    &-40 & 40 & 118$\pm$20 & 46$\pm$6 & 0.39 & 1.36$\pm$0.19 & 3.6 & 1.9$(-$11)\\
    &-20 & 20 & 166$\pm$17 & 47$\pm$3 & 0.29 & 1.40 & 4.9 & 1.4$(-$11)\\
    & 20 & -20 & 165$\pm$18 & 58$\pm$4 & 0.35 & 1.72 & 4.4 & 2.0$(-$11)\\
    & 40 & -40 & 94$\pm$20 & 23$\pm$3 & 0.25 & 0.71$\pm$0.10 & 1.6 & 2.3$(-$11)\\
    & 60 & -60 & $<$ 60 & $<$ 20 & & $<$0.60\x & & \\
    \hline
    H\thcn
    & 0 & 0 & 746$\pm$ 12 & 315$\pm$ 5 & 0.42 & 1.1 & 6.7 & 8.2$(-$12) \\
    & -60 & -60 & $<$ 55 & $<$ 10 & &$<$0.03\x & & \\
    & -40 & -40 & $<$ 57 & $<$ 20 & &$<$0.07\x & & \\
    & -20 & -20 & 432$\pm$ 20 & 172$\pm$ 10 & 0.40 & 0.58 & 2.8 & 1.1$(-$11) \\
    & 20 & 20 & 405$\pm$ 18 & 163$\pm$ 10 & 0.40 & 0.55 & 3.6 & 7.7$(-$12) \\
    & 40 & 40 & 207$\pm$ 19 & 61$\pm$ 3 & 0.29 & 0.21 & 1.2 & 9.0$(-$12) \\
    & 60 & 60 & 117$\pm$ 23 & 41$\pm$ 6 & 0.35 & 0.14$\pm$0.02& $<0.75$ & \\
    & -60 & 60 & 270$\pm$ 25 & 111$\pm$ 9 & 0.41 & 0.38 & 2.3 & 8.4$(-$12) \\
    & -40 & 40 & 396$\pm$ 20 & 191$\pm$ 6 & 0.48 & 0.65 & 3.6 & 8.9$(-$12) \\
    & -20 & 20 & 574$\pm$ 18 & 260$\pm$ 8 & 0.45 & 0.88 & 4.9 & 9.0$(-$12) \\
    & 20 & -20 & 591$\pm$ 18 & 243$\pm$ 7 & 0.41 & 0.83 & 4.4 & 9.4$(-$12) \\
    & 40 & -40 & 313$\pm$ 19 & 132$\pm$ 12 & 0.42 & 0.45 & 1.6 & 1.4$(-$11) \\
    & 60 & -60 & 61$\pm$ 19 & 22$\pm$ 4 & 0.37 & 0.07$\pm$0.01& & \\
    \hline
    HN\thc
    & 0 & 0 & 1758$\pm$ 29 & 1282$\pm$ 49 & 0.73 & 2.2 & 6.7 & 1.6$(-$11) \\
    & -40 & -40 & $<$ 204 & $<$ 80 & & $<$0.14\x & & \\
    & -20 & -20 & 945$\pm$ 67 & 661$\pm$ 58 & 0.70 & 1.1 & 2.8 & 2.0$(-$11) \\
    & 20 & 20 & 1066$\pm$ 67 & 710$\pm$ 90 & 0.67 & 1.2$\pm$0.15 & 3.6 & 1.7$(-$11) \\
    & 40 & 40 & 563$\pm$ 65 & 312$\pm$ 54 & 0.56 & 0.53$\pm$0.09 & $<0.75$ & \\
    & -40 & 40 & 914$\pm$ 67 & 716$\pm$ 58 & 0.78 & 1.2 & 3.6 & 1.6$(-$11) \\
    & -20 & 20 & 1577$\pm$ 66 & 1223$\pm$ 77 & 0.78 & 2.1 & 4.9 & 2.1$(-$11) \\
    & 20 & -20 & 1536$\pm$ 68 & 853$\pm$ 69 & 0.56 & 1.5 & 4.4 & 1.7$(-$11) \\
    & 40 & -40 & 908$\pm$ 71 & 585$\pm$ 38 & 0.64 & 0.99 & 1.6 & 3.1$(-$11) \\
    \hline
    HC\fifn
    & 0 & 0 & 258$\pm$ 19 & 136$\pm$ 13 & 0.53 & 0.26 & 6.7 & 1.9$(-$12)\\
    & -40 & -40 & $<$ 108 & $<$ 60 & &$<$ 0.12\x & & \\
    & -20 & -20 & 142$\pm$ 35 & 58$\pm$ 7 & 0.41 & 0.11 & 2.8 & 2.0$(-$12)\\
    & 20 & 20 & 163$\pm$ 34 & 80$\pm$ 7 & 0.49 & 0.15 & 3.6 & 2.1$(-$12)\\
    & 40 & 40 & 130$\pm$ 38 & $<$ 40 & &$<$ 0.08\x & 1.2 & 3.4$(-$12)\\
    & -40 & 40 & 217$\pm$ 36 & 92$\pm$ 10 & 0.42 & 0.17$\pm$0.02 & 3.6 & 2.3$(-$12)\\
    & -20 & 20 & 235$\pm$ 36 & 110$\pm$ 22 & 0.46 & 0.21\x & 4.9 & 2.1$(-$12)\\
    & 20 & -20 & 246$\pm$ 34 & 110$\pm$ 10 & 0.44 & 0.21\x & 4.4 & 2.4$(-$12)\\
    & 40 & -40 & 246$\pm$ 38 & 100$\pm$ 10 & 0.41 & 0.19 & 1.6 & 6.0$(-$12)\\
    \hline
    \nnhp 
    & 0 & 0 & 1134$\pm$ 11 & 445$\pm$ 3 & 0.39 & 15.7 & 6.7 & 1.2$(-$10) \\
    & -80 & -80 & $<$ 54 & $<$ 70 & 0.00 & 2.5 & & \\
    & -60 & -60 & $<$ 39 & $<$ 70 & 0.00 & 2.5 & & \\
    & -40 & -40 & 58$\pm$ 17 & $<$ 10 & 0.12 & 0.35 & 0.8 & \\
    & -20 & -20 & 556$\pm$ 16 & 188$\pm$ 5 & 0.34 & 6.6 & 2.8 & 1.2$(-$10) \\
    & 20 & 20 & 585$\pm$ 16 & 160$\pm$ 5 & 0.27 & 5.6 & 3.6 & 7.8$(-$11) \\
    & 40 & 40 & 414$\pm$ 15 & 90$\pm$ 5 & 0.22 & 3.2 & 1.2 & 1.4$(-$10) \\
    & 60 & 60 & 285$\pm$ 15 & 63$\pm$ 2 & 0.22 & 2.2 & 1.1 & 1.0$(-$10) \\
    & 80 & 80 & 134$\pm$ 17 & $<$ 25 & 0.18 & 0.9 & & \\
    & 80 & -80$^b$ & $<$ 57 & $<$ 25 & 0.50 & 0.13 & & \\ 
    & 60 & -60 & 72$\pm$ 17 & 20$\pm$ 3 & 0.27 & 0.7$\pm$0.1 & 0.7 & 5.3$(-$11) \\
    & 40 & -40 & 456$\pm$ 18 & 131$\pm$ 5 & 0.29 & 4.6 & 1.6 & 1.5$(-$10) \\
    & 20 & -20 & 911$\pm$ 18 & 311$\pm$ 6 & 0.34 & 10.9 & 4.4 & 1.2$(-$10) \\
    & -20 & 20 & 1032$\pm$ 16 & 389$\pm$ 5 & 0.38 & 13.7 & 4.9 & 1.4$(-$10) \\
    & -40 & 40 & 616$\pm$ 17 & 277$\pm$ 5 & 0.45 & 9.8 & 3.6 & 1.3$(-$10) \\
    & -60 & 60 & 433$\pm$ 17 & 133$\pm$ 5 & 0.31 & 4.7 & 2.3 & 1.0$(-$10) \\
    & -80 & 80 & 164$\pm$ 17 & 40$\pm$ 6 & 0.25 & 1.4$\pm$0.2 & 1.6 & 4.4$(-$11) \\
    \hline
  \end{tabular}
  \begin{list}{}{}
  \item $^a$ Weakest HFS component at 113520.414~MHz with R.I. = 0.0184
  \item $^b$ For this offset, the strongest HFS component at
    93173.7767~MHz with R.I.=0.2593 was used instead of the weakest.
  \end{list}
  \label{tab:l1544}
\end{table*}

\begin{table*}
  \centering
  \caption{As Table~\ref{tab:l183} but for lines observed towards Oph~D.}
  \scriptsize
  \begin{tabular}{l r r r r r c c r}\hline\hline
    Line & $\delta x$ & $\delta y$ & $T_{\rm mb}$ & $W$ & \dveq & $N_{\rm tot}$ & $N(\hh)$ & $N_{\rm tot}/2N(\hh)$ \\
    & \arcsec & \arcsec & mK & m\kkms & \kms & \dix{12}\,\cc & \dix{22}\,\cc \smallskip\\\hline
    %
    CN
    & 0 & 0 & 480$\pm$ 65& 121$\pm$ 11& 0.25 & 36 & 6.8 & 2.7$(-$10) \\
    & -40 & 0 & $<$ 320& $<$ 90& & $<$ 26 & 3.0 & $<$4.3$(-$10) \\
    & -20 & 0 & $<$ 320& $<$ 140& & $<$ 41 & 4.5 & $<$4.5$(-$10) \\
    & 20 & 0 & 750$\pm$ 105& 157$\pm$ 15& 0.21 & 46 & 6.5 & 3.6$(-$10) \\
    & 40 & 0 & 745$\pm$ 105& 176$\pm$ 15& 0.24 & 52 & 5.2 & 5.0$(-$10) \\
    & 60 & 0 & 435$\pm$ 110& $<$ 70& & $<$ 20.50 & 3.3 & $<$3.1$(-$10) \\
    & 80 & 0 & $<$ 340& $<$ 65& & $<$ 19.10 & 3.0 & $<$3.2$(-$10) \\
    & 0 & -40 & 670$\pm$ 110& 180$\pm$ 15& 0.27 & 53 & 5.3 & 5.0$(-$10) \\
    & 0 & -20 & 990$\pm$ 110& 287$\pm$ 13& 0.29 & 84 & 6.6 & 6.4$(-$10) \\
    & 0 & 20 & 500$\pm$ 105& 174$\pm$ 26& 0.35 & 51 $\pm$7.6 & 6.0 & 4.2$(-$10) \\
    & 0 & 40 & $<$ 310& $<$ 110& & $<$ 32 & 4.5 & $<$3.6$(-$10) \\
    \hline
    \thcn$^a$
    & 0 & 0 & 80$\pm$ 20& 25$\pm$ 7& 0.32 & 0.74$\pm$0.21 & 6.8 & 5.4$(-12)$\\
    & -20 & 0 & 110$\pm$ 35& 40$\pm$ 9& 0.36 & 1.2$\pm$0.3 & 4.5 & 1.3$(-11)$\\
    & 20 & 0 & 125$\pm$ 35& 57$\pm$ 10& 0.46 & 1.7$\pm$0.3 & 6.5 & 1.3$(-11)$\\
    & 0 & -20 & 150$\pm$ 35& 36$\pm$ 8& 0.24 & 1.1$\pm$0.2 & 6.6 & 8.3$(-12)$\\
    & 0 & 20 & $< $ 100& $< $ 25& & $<$0.75 & 6.0 & $<6.3(-12)$\\
    \hline
    HCN$^b$
    & 0 & 0 & 2140$\pm$ 35& 964$\pm$ 6& 0.45 & 16 & 6.8 & 1.2$(-$10) \\
    & -40 & 0 & 760$\pm$ 60& 435$\pm$ 20& 0.57 & 7.1 & 3.0 & 1.2$(-$10) \\
    & -20 & 0 & 1470$\pm$ 65& 629$\pm$ 11& 0.43 & 10 & 4.5 & 1.1$(-$10) \\
    & 20 & 0 & 2470$\pm$ 50& 1115$\pm$ 9& 0.45 & 18 & 6.5 & 1.4$(-$10) \\
    & 40 & 0 & 2490$\pm$ 65& 1131$\pm$ 11& 0.45 & 19 & 5.2 & 1.8$(-$10) \\
    & 60 & 0 & 1805$\pm$ 55& 781$\pm$ 10& 0.43 & 13 & 3.3 & 1.9$(-$10) \\
    & 80 & 0 & 1120$\pm$ 60& 504$\pm$ 10& 0.45 & 8.3 & 3.0 & 1.4$(-$10) \\
    & 0 & -40 & 2605$\pm$ 60& 1159$\pm$ 10& 0.44 & 19 & 5.3 & 1.8$(-$10) \\
    & 0 & -20 & 2720$\pm$ 60& 1408$\pm$ 10& 0.52 & 23 & 6.6 & 1.7$(-$10) \\
    & 0 & 20 & 2080$\pm$ 60& 842$\pm$ 10& 0.40 & 14 & 6.0 & 1.2$(-$10) \\
    & 0 & 40 & 1850$\pm$ 55& 906$\pm$ 10& 0.49 & 15 & 4.5 & 1.7$(-$10) \\
    \hline
    H\thcn
    & 0 & 0 & 439$\pm$ 35& 134$\pm$ 10& 0.31 & 0.46 & 6.8 & 3.4$(-$12) \\
    & -40 & 0 & $<$ 160& $<$ 45& & $<$ 0.15 & 3.0 & $<$2.5$(-$12) \\
    & -20 & 0 & 313$\pm$ 50& 94$\pm$ 10& 0.30 & 0.32$\pm$ 0.03 & 4.5 & 3.5$(-$12) \\
    & 20 & 0 & 551$\pm$ 50& 186$\pm$ 12& 0.34 & 0.64 & 6.5 & 4.9$(-$12) \\
    & 40 & 0 & 433$\pm$ 50& 120$\pm$ 10& 0.28 & 0.41 & 5.2 & 3.9$(-$12) \\
    & 60 & 0 & 283$\pm$ 55& 109$\pm$ 15& 0.39 & 0.37$\pm$ 0.05 & 3.3 & 5.5$(-$12) \\
    & 80 & 0 & $<$ 145& $<$ 35& & $<$ 0.12 & 3.0 & $<$2.0$(-$12) \\
    & 0 & -40 & 413$\pm$ 45& 148$\pm$ 12& 0.36 & 0.51 & 5.3 & 4.8$(-$12) \\
    & 0 & -20 & 649$\pm$ 50& 178$\pm$ 12& 0.27 & 0.61 & 6.6 & 4.6$(-$12) \\
    & 0 & 20 & 311$\pm$ 50& 93$\pm$ 10& 0.30 & 0.32$\pm$ 0.03 & 6.0 & 2.7$(-$12) \\
    & 0 & 40 & 220$\pm$ 50& 40$\pm$ 10& & 0.14\x & 4.5 & 1.6$(-$12) \\
    \hline
    HN\thc
    & 0 & 0 & 880$\pm$ 45& 435$\pm$ 10& 0.49 & 0.75 & 6.8 & 5.5$(-$12) \\
    & -40 & 0 & $<$ 170& 46$\pm$ 11& & 0.08 & 3.0 & 1.3$(-$12) \\
    & -20 & 0 & 440$\pm$ 65& 225$\pm$ 17& 0.51 & 0.39 & 4.5 & 4.3$(-$12) \\
    & 20 & 0 & 1170$\pm$ 60& 690$\pm$ 15& 0.59 & 1.2 & 6.5 & 9.3$(-$12) \\
    & 40 & 0 & 1045$\pm$ 60& 570$\pm$ 15& 0.55 & 0.99 & 5.2 & 9.5$(-$12) \\
    & 60 & 0 & 535$\pm$ 60& 260$\pm$ 20& 0.49 & 0.45 & 3.3 & 6.7$(-$12) \\
    & 80 & 0 & 255$\pm$ 60& 140$\pm$ 25& 0.99 & 0.24\x & 3.0 & 4.0$(-$12) \\
    & 0 & -40 & 920$\pm$ 60& 525$\pm$ 20& 0.57 & 0.90\x & 5.3 & 8.5$(-$12) \\
    & 0 & -20 & 1150$\pm$ 60& 720$\pm$ 20& 0.63 & 1.2 & 6.6 & 9.1$(-$12) \\
    & 0 & 20 & 660$\pm$ 65& 260$\pm$ 20& 0.39 & 0.45 & 6.0 & 3.7$(-$12) \\
    & 0 & 40 & 620$\pm$ 60& 255$\pm$ 15& 0.41 & 0.44 & 4.5 & 4.9$(-$12) \\
    \hline
  \end{tabular}
  \label{tab:ophd}
  \begin{list}{}{}
  \item $^a$ Detections are at the 4$\sigma$ level on $W$.
  \item $^b$ Weakest HFS component at 88633.9360~MHz with R.I. =
    0.111.
  \end{list}
\end{table*}

\begin{table*}
  \centering
  \caption{As Table~\ref{tab:l183} but for lines observed towards L~1517B.}
  \scriptsize
  \begin{tabular}{l r r r r r c c r}\hline\hline
    Line & $\delta x$ & $\delta y$ & $T_{\rm mb}$ & $W$ & \dveq & $N_{\rm tot}$ & $N(\hh)$ & $N_{\rm tot}/2N(\hh)$ \\
    & \arcsec & \arcsec & mK & m\kkms & \kms & \dix{12}\,\cc & \dix{22}\,\cc \smallskip\\\hline
    %
    \thcn
    & -10 & -20 & 115$\pm$ 15 & 30$\pm$ 3& 0.26 & 0.88 & 2.7 & 1.6$(-$11)\\
    & -50 & -20 & $<$ 70 & $<$ 20& & $<$ 0.60 & 1.8 & 1.7$(-$11)\\
    & -30 & -20 & 95$\pm$ 20 & 24$\pm$ 4& 0.25 & 0.71$\pm$ 0.12 & 2.3 & 1.5$(-$11)\\
    & 10 & -20 & 100$\pm$ 20 & 40$\pm$ 5& 0.40 & 1.2$\pm$ 0.15 & 1.9 & 3.1$(-$11)\\
    & 30 & -20 & 95$\pm$ 25 & 40$\pm$ 5& 0.42 & 1.2$\pm$ 0.15 & 1.1 & 5.3$(-$11)\\
    & -10 & -60 & 65$\pm$ 20 & 30$\pm$ 5& 0.46 & 0.88$\pm$ 0.15 & 1.5 & 2.9$(-$11)\\
    & -10 & -40 & 140$\pm$ 25 & 28$\pm$ 3& 0.20 & 0.82$\pm$ 0.09 & 2.1 & 2.0$(-$11)\\
    & -10 & 0 & 120$\pm$ 20 & 30$\pm$ 4& 0.25 & 0.88$\pm$ 0.12 & 2.2 & 2.0$(-$11)\\
    & -10 & 20 & 70$\pm$ 20 & 28$\pm$ 5& 0.40 & 0.82$\pm$ 0.15 & 1.5 & 2.7$(-$11)\\
    \hline
    H\thcn
    & -10 & -20 & 280$\pm$ 25 & 110$\pm$ 5& 0.39 & 0.38 & 2.7 & 7.1$(-$12)\\
    & -50 & -20 & 160$\pm$ 40 & 50$\pm$ 8& 0.31 & 0.17$\pm$ 0.03 & 1.8 & 4.7$(-$12)\\
    & -30 & -20 & 230$\pm$ 40 & 80$\pm$ 8& 0.35 & 0.27 & 2.3 & 5.9$(-$12)\\
    & 10 & -20 & 290$\pm$ 34 & 110$\pm$ 8& 0.38 & 0.38 & 1.9 & 9.9$(-$12)\\
    & 30 & -20 & 265$\pm$ 37 & 67$\pm$ 6& 0.25 & 0.23 & 1.1 & 1.0$(-$11)\\
    & -10 & -60 & 200$\pm$ 45 & 60$\pm$ 9& 0.30 & 0.20$\pm$ 0.03 & 1.5 & 6.6$(-$12)\\
    & -10 & -40 & 320$\pm$ 40 & 94$\pm$ 8& 0.29 & 0.32 & 2.1 & 7.7$(-$12)\\
    & -10 & 0 & 295$\pm$ 35 & 110$\pm$ 10& 0.37 & 0.38 & 2.2 & 8.7$(-$12)\\
    & -10 & 20 & 185$\pm$ 40 & 60$\pm$ 8& 0.32 & 0.20$\pm$ 0.03 & 1.5 & 6.5$(-$12)\\
    \hline
    HN\thc
    & -10 & -20 & 1105$\pm$ 80 & 683$\pm$ 16& 0.62 & 1.2 & 2.7 & 2.2$(-$11)\\
    & -50 & -20 & 753$\pm$ 71 & 301$\pm$ 12& 0.40 & 0.52 & 1.8 & 1.4$(-$11)\\
    & -30 & -20 & 896$\pm$ 76 & 542$\pm$ 15& 0.60 & 0.94 & 2.3 & 2.0$(-$11)\\
    & 10 & -20 & 944$\pm$ 71 & 514$\pm$ 14& 0.54 & 0.89 & 1.9 & 2.3$(-$11)\\
    & 30 & -20 & 774$\pm$ 73 & 359$\pm$ 13& 0.46 & 0.62 & 1.1 & 2.7$(-$11)\\
    & -10 & -60 & 661$\pm$ 71 & 390$\pm$ 25& 0.59 & 0.68 & 1.5 & 2.2$(-$11)\\
    & -10 & -40 & 982$\pm$ 75 & 589$\pm$ 15& 0.60 & 1.0 & 2.1 & 2.4$(-$11)\\
    & -10 & 0 & 1119$\pm$ 70 & 649$\pm$ 14& 0.58 & 1.1 & 2.2 & 2.5$(-$11)\\
    & -10 & 20 & 787$\pm$ 71 & 527$\pm$ 15& 0.67 & 0.91 & 1.5 & 2.9$(-$11)\\
    \hline
  \end{tabular}
  \label{tab:l1517b}
\end{table*}

\begin{table*}
  \centering
  \caption{As Table~\ref{tab:ophd} but for lines observed towards
    L~310. The dust and thus \hh\ column densities are only the
    3$\sigma$ level.}
  \scriptsize
  \begin{tabular}{l r r r r r c c r}\hline\hline
    Line & $\delta x$ & $\delta y$ & $T_{\rm mb}$ & $W$ & \dveq & $N_{\rm tot}$ & $N(\hh)$ & $N_{\rm tot}/2N(\hh)$ \\
    & \arcsec & \arcsec & mK & m\kkms & \kms & \dix{12}\,\cc & \dix{22}\,\cc \smallskip\\\hline
    %
    CN
    & 30 & 80 & 390$\pm$ 60& 230$\pm$ 20& 0.5 & 67 & 1.5 & $4.5(-9)$\\
    & 10 & 60 & $<$ 305 & $<$ 120 & & $<$ 35.2 &$<$0.7 & \\
    & 10 & 80 & $<$ 303 & $<$ 130 & & $<$ 38.1 &$<$0.7 & \\
    & 30 & 40 & $<$ 288 & $<$ 110 & & $<$ 32.2 &$<$0.7 & \\
    & 30 & 60 & 300$\pm$ 55& 110$\pm$ 15& 0.3 & 32$\pm$4.4 & 1.5 & $2.1(-9)$ \\
    & 50 & 60 & $<$ 280 & $<$ 110 & & $<$ 32.2 &$<$0.7 & \\
    \hline
    HCN
    & 30 & 80 & 615$\pm$ 35& 370$\pm$ 10& 0.60 & 6.1 & 1.5 & $4.1(-10)$ \\
    & 10 & 60 & 300$\pm$ 55& 230$\pm$ 20& 0.77 & 3.8 &$<$0.7 & \\
    & 10 & 80 & 290$\pm$ 60& 155$\pm$ 20& 0.53 & 2.5$\pm$0.33 &$<$0.7 & \\
    & 30 & 40 & 150$\pm$ 50& $<$ 70& & $<$ 1.15 &$<$0.7 & \\
    & 30 & 60 & 525$\pm$ 35& 300$\pm$ 15& 0.57 & 4.9 & 1.5 & $3.3(-10)$ \\
    & 50 & 60 & 280$\pm$ 55& 150$\pm$ 25& 0.54 & 2.5$\pm$0.41 &$<$0.7 & \\
    \hline
    H\thcn
    & 30 & 80 & 160$\pm$ 20& 101$\pm$ 7& 0.63 & 0.34 & 1.5 & $2.3(-11)$ \\
    & -10 & 80 & $<$ 90 & $<$ 36& & $<$ 0.12 &$<$0.7 & \\
    & 10 & 80 & $<$ 90 & $<$ 36& & $<$ 0.12 &$<$0.7 & \\
    & 30 & 40 & $<$ 105 & $<$ 42& & $<$ 0.14 &$<$0.7 & \\
    & 30 & 60 & 150$\pm$ 30& 70$\pm$ 11& 0.47 & 0.24$\pm$0.04 & 1.5 & $1.6(-11)$ \\
    & 30 & 100 & $<$ 95 & $<$ 40& & $<$ 0.14 &$<$0.7 & \\
    & 30 & 120 & $<$ 95 & $<$ 40& & $<$ 0.14 &$<$0.7 & \\
    & 50 & 60 & $<$ 135 & $<$ 55& & $<$ 0.19 &$<$0.7 & \\
    & 50 & 80 & 165$\pm$ 35& 65$\pm$ 8& 0.39 & 0.22$\pm$0.03 & 1.3 & $1.7(-11)$ \\
    & 50 & 100 & $<$ 140 & $<$ 60& & $<$ 0.20 & 1.5 & $<1.6(-11)$ \\
    & 70 & 80 & $<$ 90 & $<$ 40& & $<$ 0.14 &$<$0.7 & \\
    \hline
    HN\thc
    & 30 & 80 & 405$\pm$ 25& 320$\pm$ 10& 0.79 & 0.55 & 1.5 & $3.7(-11)$ \\
    & -10 & 80 & $<$ 125 & $<$ 55& & $<$ 0.10 &$<$0.7 & \\
    & 10 & 80 & 205$\pm$ 45& 100$\pm$ 15& 0.49 & 0.17$\pm$0.03 &$<$0.7 & \\
    & 30 & 40 & $<$ 120 & $<$ 50& & $<$ 0.09 &$<$0.7 & \\
    & 30 & 60 & 390$\pm$ 40& 200$\pm$ 10& 0.51 & 0.35 & 1.5 & $2.3(-11)$ \\
    & 30 & 100 & 150$\pm$ 40& 70$\pm$ 15& 0.47 & 0.12$\pm$0.03 &$<$0.7 & \\
    & 30 & 120 & $<$ 120 & $<$ 52& & $<$ 0.09 &$<$0.7 & \\
    & 50 & 60 & 230$\pm$ 60& $<$ 75& & $<$ 0.13 &$<$0.7 & \\
    & 50 & 80 & 465$\pm$ 40& $<$ 54& & $<$ 0.09 & 1.3 & $<6.9(-12)$ \\
    & 50 & 100 & 365$\pm$ 55& $<$ 73& & $<$ 0.13 & 1.5 & $<8.7(-12)$ \\
    & 70 & 80 & 165$\pm$ 40& 100$\pm$ 10& 0.61 & 0.17 &$<$0.7 & \\
    \hline
  \end{tabular}
  \label{tab:l310}
\end{table*}

\end{document}